\documentclass[fleqn,usenatbib]{mnras}

\usepackage{newtxtext,newtxmath}

\usepackage[T1]{fontenc}
\usepackage{ae,aecompl}

\usepackage{graphicx,epstopdf}	
\DeclareGraphicsExtensions{.ps}
\usepackage[utf8]{inputenc}
\usepackage{multirow}
\usepackage{hyperref}
\usepackage{natbib}
\usepackage{array}
\usepackage{verbatim}
\usepackage{float}
\usepackage{epsfig}
\usepackage{subfigure}
\usepackage{color}

\graphicspath{ {Figures/} }

\newcommand{\HII}{\textrm{H~{\textsc{ii}}}}

\newcommand{\OI}{\textrm{O~{\textsc{i}}}}
\newcommand{\CII}{\textrm{C~{\textsc{ii}}}}

\newcommand{\mum}{$\mu$m}

\title[PAHs in NGC~1333]{Characterizing Spatial Variations of PAH Emission in the Reflection Nebula NGC~1333}

\author[C. Knight et al.]{
C. Knight$^{1}$, E. Peeters$^{1,2,3}$, M. Wolfire$^{4}$, D.J. Stock$^{1}$
\\
$^{1}$Department of Physics and Astronomy, University of Western Ontario, London, ON N6A 3K7, Canada; collinknight11@gmail.com \\
$^{2}$ Institute for Earth and Space Exploration, University of Western Ontario, London, ON, N6A 3K7, Canada\\
$^{3}$Carl Sagan Center, SETI Institute, 189 N. Bernardo Avenue, Suite 100, Mountain View, CA 94043, USA \\
$^{4}$Department of Astronomy, University of Maryland, College Park, MD 20742, USA}

\date{Accepted XXX. Received YYY; in original form ZZZ}

\pubyear{2021}

\begin{document}
\label{firstpage}
\pagerange{\pageref{firstpage}--\pageref{lastpage}}
\maketitle

\begin{abstract}

Infrared emission features at 3.3, 6.2, 7.7, 8.6, and 11.2~$\mu$m, attributed to polycyclic aromatic hydrocarbons, show variations in relative intensity, shape, and peak position.
These variations depend on the physical conditions of the photodissociation region (PDR) in which strong PAH emission arises, but their relationship has yet to be fully quantified. We aim to better calibrate the response of PAH species to their environment using observations with matching apertures and spatial resolution. We present observations from the Field-Imaging Far-Infrared Line Spectrometer (FIFI-LS) on board the Stratospheric Observatory for Infrared Astronomy (SOFIA) of the gas cooling lines [OI] 63, 146~$\mu$m and [CII] 158~$\mu$m in the reflection nebula NGC~1333 and use archival dust continuum observations from the Photodetector Array Camera and Spectrometer (PACS) on board Herschel.  We employ PDR modelling to derive the physical conditions and compare these with the characteristics of the PAH emission as observed with the Infrared Spectrometer (IRS) on board Spitzer. We find distinct spatial characteristics for the various PAH spectral components. We conclude that the ionic bands (6.2, 7.7, 8.6, and 11.0) and the 7--9~$\mu$m emission are due to multiple PAH sub-populations and that the plateaus are distinct from the features perched on top. The 6--9~$\mu$m PAH emission exhibit a significant change in behaviour between the irradiated PDR and diffuse outskirts, confirming these bands arise from multiple PAH sub-populations with different underlying molecular properties. We find multiple promising relationships between PAH ratios and the FUV radiation field strength but no clear correlations with the PAH ionization parameter.

\end{abstract}




\section{Introduction}
\label{intro}

The mid-infrared (MIR) spectra of a vast number astronomical sources are dominated by prominent emission features at 3.3, 6.2, 7.7, 8.6, 11.2, and 12.7~$\mu$m  along with weaker associated bands found at 5.2, 5.7, 6.0, 11.0, 12.0, 13.5, and 14.2~$\mu$m. These features are widely attributed to the infrared fluorescence of polycyclic aromatic hydrocarbons \citep[PAHs,e.g.][]{leg84,all85,all89,pug89} and related carbonaceous species such as PAH clusters, polycyclic aromatic nitrogen heterocycles \citep[PANHs,][]{hud05,bau08} or PAHs with functional groups attached  \citep[e.g.][]{job96a,slo97,pil15,mal16,sha19}. These species are characterized by their shared molecular composition, made up primarily of a planar collection of fused benzene rings with hydrogen atoms located on the outer edges. These molecules are electronically excited by the absorption of far-ultraviolet (FUV) photons from a nearby stellar source. This energy is rapidly redistributed to lower lying vibrational states, where these molecules cascade back to their ground state by radiating MIR photons corresponding to vibration relaxation. Astronomical sources where these bands have been routinely observed include  H\,\textsc{ii} regions, young stellar objects (YSOs), post-AGB stars, planetary nebula(e) (PN(e)), reflection nebula(e) (RN(e)), external galaxies as well as the diffuse ISM \citep[e.g.][]{hon01,ver01,pee02,ber07,smi07a, gal08,boe12,sha16,sto16}.

The spectral fingerprints of these PAHs show significant variation in the relative intensities, profile shape, and peak positions between sources and spatially within extended sources. Intensity variations are mainly attributed to the molecular charge state, i.e. the PAH emission in the 6--9~$\mu$m  range increases in cationic species, whereas the 3.3 and 11.2~$\mu$m bands are more prominent in neutral species \citep{all89}, the 12.7~$\mu$m band has both cationic and neutral components \citep[e.g.]{pee12, boe13,sha16}. To a lesser extent, intensity variations are driven by the size distribution of a PAH population because smaller molecules tend to have less vibrational modes available, hence they have more energy per mode or reach a higher internal temperature upon the absorption of a FUV photon \citep[]{sch93, cro16, kni21}. Additionally, intensity variations in PAH bands in the 10--15~$\mu$m range are attributed to structural differences in their carriers, i.e. the 11.2~$\mu$m band is stronger in compact--symmetrical species whereas the 12.7~$\mu$m band becomes more prominent in species with an irregular edge structure \citep[e.g.][]{hon01}.

PAH emission is prominent in photodissociation regions (PDRs). These are regions where
FUV photons of energies between $>$~6~eV and $<$~13.6~eV (i.e. the ionization energy of hydrogen) control the physics and chemistry of the gas and which includes the neutral atomic and molecular hydrogen in IR-luminous regions around young massive stars \citep{hol97, hol99}. PDRs extend to a wide variety of neutral environments, in fact they account for all atomic and a minimum of 90\% of the molecular gas in the Galaxy \citep{hol99}. 
In addition to PAH features, observed emission features from PDRs include the far-infrared (FIR) dust continuum, H$_{2}$ emission lines, the atomic fine structure lines of species such as [\OI] and [\CII], and CO rotational lines. These emission features have all proven to be useful in predicting the physical conditions of PDR environments through the use of PDR modelling \citep[e.g.][]{tie85a,wol90,kau99,you02,kau06,wol10,neu16}. These models use underlying assumptions about thermal balance, gas chemistry, and the ionization balance \citep{tie85a} to derive how the emission lines change as function of environment, using such parameters as elemental abundances, gas density (n$_{\textrm{H}}$), and FUV radiation field strength (G$_{0}$). However, PAH emission features remain one avenue of measuring the physical conditions of PDR environments that is still in its infancy. 

Recent efforts have uncovered promising correlations between PAH emission features and PDR physical conditions. Based on three distinct PDR environments, \cite{gal08} have empirically determined a quantitative relationship between the observed 6.2/11.2 PAH emission and the PAH ionization rate to recombination rate ratio or the PAH ionization parameter $\gamma$~=~G$_{0}$~T$^{0.5}$/~n$_{\textrm{e}}$ \citep{bak94}, where n$_{\textrm{e}}$ is the electron density. Similarly, \cite{boe15} found a relationship between the 6.2/11.2 emission ratio and the ionization parameter in the RN NGC~7023, where they fit spectra with a collection of PAH species from the NASA Ames PAH IR Spectroscopic database \citep[PAHdb,][]{bau10,boe14a,bau18} to calculate the ionization parameter. Additionally, it has been shown that the 7.7~$\mu$m PAH complex can be decomposed into four Gaussian components \citep{pee17}. The ratio of two of these components, G7.8/G7.6, has been shown to have a linear relationship with G$_{0}$ over a wide range range of PDR environments \citep{sto17}. Finally, through the use of the spectral fitting program, PAHTAT, an empirical relationship between the fraction of carbons contained within very small grains (VSG) and G$_{0}$ has been established over a wide range of PDR environments \citep{pil12}.

Unfortunately each of the above relationships between PAH emission and environmental conditions have outstanding issues that must be resolved before they become more applicable to astronomy at large. For instance, the correlations between the 6.2/11.2 and $\gamma$ determined by \cite{gal08} and \cite{boe15} disagree with each other by an order of magnitude in $\gamma$. Another important issue with the latter relationships is the individual parameters that make up $\gamma$ (i.e. G$_{0}$, n$_{\textrm{e}}$, and T) cannot be isolated. Hence, this correlation alone does not provide enough information to properly characterize the PDR environment. Likewise, the \cite{sto17} relationship between G7.8/G7.6 and G$_{0}$ suffers from a bias in their data set towards higher values of G$_{0}$, with almost no representation from more quiescent regions. Moreover, due to a lack of FIR observations at a comparable resolution to the MIR observations, the \cite{sto17} relationship is currently forced to make simple assumptions to derive an estimate for G$_{0}$. The \cite{pil12} correlation between fraction of carbon locked in VSG and G$_{0}$ also has its limitations as this method is unable to properly measure the rising dust continuum found in spectral observations of H\,\textsc{ii} regions and star--forming galaxies.

In this work, we aim to further investigate the relationship between the PAH emission features and PDR physical conditions using matching apertures and spatial resolution for observations of the RN NGC~1333.
We obtain {\it SOFIA} FIR spectroscopic maps to measure three dominant PDR cooling lines: [\OI] 63, 146~$\mu$m and [\CII] 158~$\mu$m. We use archival {\it Herschel} FIR photometric observations to measure the FIR dust continuum and archival {\it Spitzer} MIR spectral maps to characterize the PAH emission features and the relevant emission ratios.
In Section \ref{ngc1333}, we introduce NGC~1333. Section \ref{data} presents the observations and data reduction applied. Section \ref{analysis} reviews the methodology used to measure the MIR and FIR emission involved in this study. We present our major results in terms of emission maps, correlations, cross cuts, and PDR modelling in Section \ref{results} and provide an interpretation of these results in Section \ref{discussion}. Finally, we give a summary of these findings in Section \ref{conclusion}.

\section{NGC 1333}
\label{ngc1333}
\begin{figure}
\begin{center}
\includegraphics[clip,trim =0.cm 0cm 0.cm 0cm,width=8.5cm]{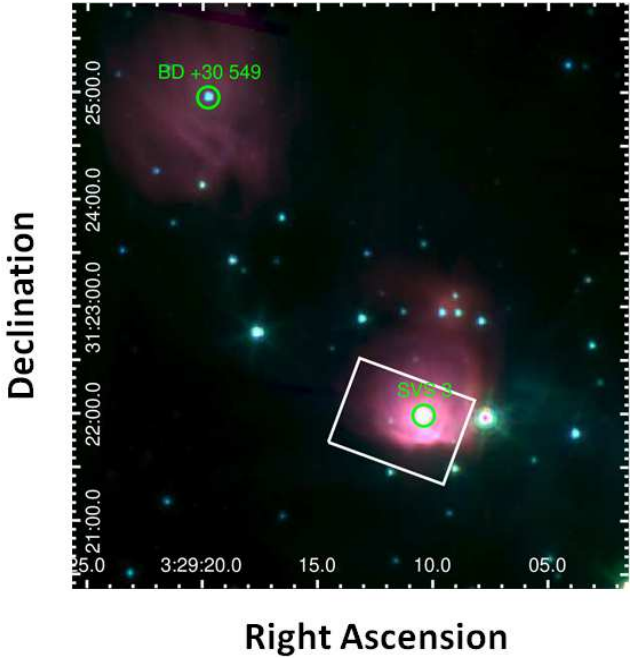}
\end{center}
\caption{Spitzer IRAC mosaic of NGC~1333 RN region \citep{gut08}. IRAC~3.6, 4.5, and 8.0~$\mu$m are shown in blue, green, and red respectively. Illuminating sources BD +30$^{\circ}$549 and SVS~3 are indicated by green circles and the region covered by Spitzer IRS~SL is shown as a white rectangle. Axes are given in right ascension and declination (J2000).}
\label{NGC1333 IRAC}
\end{figure}

The source, NGC~1333, primarily refers to a large star--forming region located within the Perseus cloud complex \citep[e.g.][]{lis88,gut08,wal08}, however it was originally used to refer to the RN at the core of this region (Figure \ref{NGC1333 IRAC}). In this paper we will use the original nomenclature when referring to NGC~1333 \citep{str76}. Astrometry of H$_{2}$O masers within the star--forming region of NGC~1333 has shown this complex is at a distance of 235~$\pm$~18~pc from the Sun \citep{hir08}. This RN is illuminated primarily by SVS~3, a B6 type star located at the center of the RN while the nebulous region at 3.5$^\prime$ to the northeast of SVS~3 is illuminated by the B6 star BD~+30$^{\circ}$~549 \citep[see Figures \ref{NGC1333 IRAC} and \ref{NGC1333 PACS},][]{str76,har84}. A study of the FIR cooling lines of the PDR gas within the RN by \citet{you02} found a far--ultraviolet radiation field strength of 4800 times the average interstellar field as well as gas density and temperature of 2~$\times$~10$^{4}$~cm$^{-3}$ and 690~K respectively. 

There have been several notable MIR studies of NGC~1333 which are focused on its PAH emission. 
\cite{bre93} used ground--based spectral imaging of the nebula around SVS~3 to show that the 3.3 and 11.2~$\mu$m bands are spatially distinct. In particular, the 3.3~$\mu$m emission peaks farther from SVS~3 than the 11.2~$\mu$m emission and has a distribution comparable to a limb--brightened shell. \cite{job96a} showed how the relative intensity of the aliphatic 3.4~$\mu$m band to the 3.3~$\mu$m band increased significantly in the diffuse ISM surrounding the RN with respect to pointings near SVS~3, suggesting that interstellar PAHs are highly methylated. Other ground--based MIR spectroscopic observations of NGC~1333 found the 8.6~$\mu$m band peaked at the location of SVS~3, while the 11.2~$\mu$m peaked 10$^{\prime\prime}$ south of SVS~3, suggesting this was evidence for the existence of ionized PAHs in the ISM \citep[e.g.][]{job96b, slo99}. \citet{roc94} determined that the relative intensities of the PAH emission features in the 8--13~$\mu$m range in NGC~1333 and another well--studied RN, NGC~2023, are comparable. More recently, \cite{sto16} did a survey of MIR--bright regions using {\it Spitzer} IRS~SL observations including both NGC~1333 and NGC~2023, where it was again found 
that correlations between the observed PAH emission features are similar in both sources \citep[for a recent overview of the observed PAH emission features in NGC~2023, see][]{pee17}. Moreover, using ISOCAM spectral imaging, \cite{bre05} observed a shift in the central wavelength of the 7.7~$\mu$m band from 7.75 to 7.65~$\mu$m with decreasing distance to SVS~3. These authors also found the 11.2/7.7 ratio decreases with proximity to the star which they relate to PAH ionization. 

\section{The Data}
\label{data}
\subsection{Observations}
\label{obs}

\subsubsection{Spitzer}
\label{spit}
We present observations obtained with the Infrared Spectrograph \citep[IRS,][]{hou04} on--board the {\it Spitzer} Space Telescope \citep{wer04a}.  We retrieved Spitzer-IRS 5-14~$\mu$m short--low (SL) spectral mapping observations from the Spitzer archive (AORKEY 14587648, PI T. Bergin). These observations have a spectral resolution ranging from 60 to 128 over three orders of diffraction: SL1, SL2, and SL3. The SL mode has a pixel size of 1.8$^{\prime\prime}$, with a slit width of 3.6$^{\prime\prime}$ and a slit length of 57$^{\prime\prime}$. The region covered by the IRS SL spectral data is shown in Figure \ref{NGC1333 IRAC}.

\subsubsection{SOFIA}
\label{sofia}
The Field--Imaging Far--Infrared Line Spectrometer (FIFI--LS) on--board SOFIA is an integral field FIR spectrometer which includes two independent grating spectrometers with wavelength ranges from 51--210~$\mu$m and 115--200~$\mu$m respectively \citep{col18,fis18}. Both channels have a 5~$\times$~5~pixel projection onto the sky, with centers offset by 10$^{\prime\prime}$. The short and long wavelength channels have a pixel size of 6$^{\prime\prime}$~$\times$~6$^{\prime\prime}$ and 12$^{\prime\prime}$~$\times$~12$^{\prime\prime}$ respectively which translates to a 0.5 and 1.0~arcminute--squared FOV respectively. The 5~$\times$~5~pixels are re--organized along a 25~$\times$~1 line and subsequently dispersed into 16~pixels in the spectral dimension. Spectral resolution ranges from 600--2000 depending on the observed wavelength, which tends to be higher towards the longer wavelengths in both spectrometers.

We obtained FIFI--LS spectral observations of NGC~1333 for the following cooling lines: [\OI] 63, 145~$\mu$m and [\CII] 158~$\mu$m (PID: 05\_0110, PI: E. Peeters). These data cubes have a PSF FWHM of 6.4$^{\prime\prime}$, 14.6$^{\prime\prime}$, and 15.9$^{\prime\prime}$ respectively. Our observations are centered at 3:29:09.3, +31:21:47.2 (J2000) and are comprised of 4 parallel pointings with a stepsize of 30$^{\prime\prime}$.

\subsubsection{Herschel}
\label{herschel}

We obtained FIR photometric observations taken with the Photodetector Array Camera and Spectrometer \citep[PACS;][]{pog10} on--board the {\it Herschel} Space Observatory \citep{pil10} from the Herschel Science archive. PACS includes a dual--band photometer with an instantaneous FOV of 3.5$^\prime$~$\times$~1.75$^\prime$ that has bandpass combinations of either 60--85~$\mu$m and 125--210~$\mu$m (70/160~$\mu$m filter) or 85--125~$\mu$m and 125--210~$\mu$m (100/160~$\mu$m filter). The 70 and 100~$\mu$m bands have a pixel scale of 3.2$^{\prime\prime}$ while the 160~$\mu$m band has a pixel scale of 6.4$^{\prime\prime}$. The absolute flux uncertainties for the 70, 100, 160~$\mu$m filters are $\pm$~10, 10, and 20\% respectively \citep{pog10}.

\begin{figure}
\begin{center}
\includegraphics[clip,trim =0.cm 0cm 0.cm 0cm,width=8.5cm]{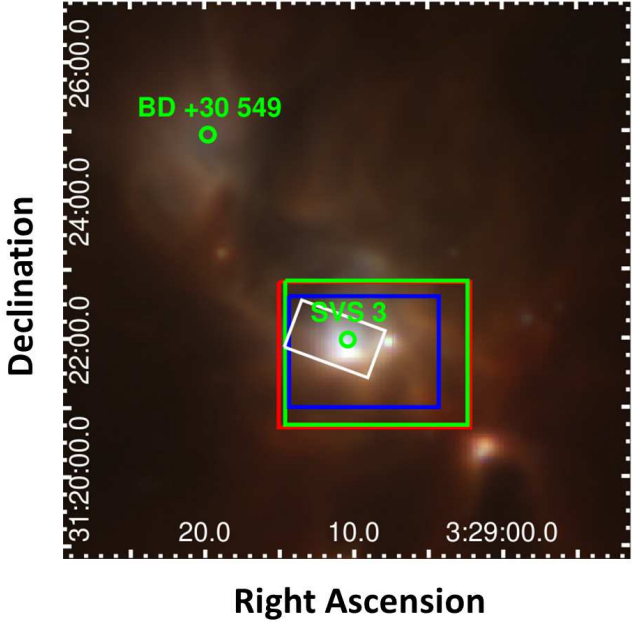}
\end{center}
\caption{Hershel PACS mosaic of NGC~1333 RN region. PACS 70, 100, and 160~$\mu$m data are shown in blue, green and red respectively. All images are shown on a square root scale for clarity. Illuminating sources BD +30$^{\circ}$549 and SVS~3 are indicated by green circles and the region covered by Spitzer IRS~SL, FIFI--LS [\OI] 63~$\mu$m, 146~$\mu$m, and [\CII] 158~$\mu$m are shown as white, blue, green, and red rectangles respectively. Axes are given in right ascension and declination (J2000).}
\label{NGC1333 PACS}
\end{figure}

These observations consist of imaging data taken of the Perseus Complex in the 70, 100 and 160~$\mu$m bands (PID: KPGT\_pandre\_1, OBS\_ID: 1342190326 and 1342227103) created with the JScanam task (Figure \ref{NGC1333 PACS}). Notably the 70~$\mu$m map was taken in the SPIRE/PACS parallel mode with a nominal scan velocity of 60$^{\prime\prime}$~s$^{-1}$ corresponding to a PSF FWHM of 5.86$^{\prime\prime}$~$\times$~12.16$^{\prime\prime}$. The 100 and 160~$\mu$m maps were taken in the PACS photo mode with a nominal scan speed of 20$^{\prime\prime}$~s$^{-1}$ corresponding to PSF FWHMs of 6.89$^{\prime\prime}$ $\times$ 9.74$^{\prime\prime}$ and 11.31$^{\prime\prime}$~$\times$~13.32$^{\prime\prime}$ respectively.

\subsection{IRS Reduction}
\label{IRS red}
The SL raw data were processed with the S18.18 pipeline by the Spitzer Science Center. The resulting bcd data products were further processed using cubism \citep{smi07a}. We applied a {\it wavesamp} of 0.04--0.96 to exclude spurious data at the extremities of the SL slit and cubism's automatic bad pixel generation ($\sigma_{TRIM}$~=~7 and Minbad-fraction~=~0.5 and 0.75 for respectively global and record bad pixels). Remaining bad pixels were removed manually. The resulting SL2 and SL3 data cubes were regridded to the SL1 grid. We clip the original SL1 data cube to a 24~$\times$~41~pixel aperture around the nebula surrounding SVS~3 where we find appreciable PAH emission. 

We find mismatches between the different orders of the SL module (SL1, SL2, and SL3). To remedy this issue, scaling factors of $<$~20\% were applied to the SL2 data and $<$~10\% to the SL3 data to scale these orders to the SL1  and to the combined SL1 and scaled SL3 data respectively, with the exception of outlier pixels in the southwest and southeastern corners of the FOV considered which are masked out. We also masked out two prominent YSOs in the western edge of the final FOV, as they exhibit strong silicate absorption and extinction. The SL1 and scaled SL2 data are combined into a single spectrum for each pixel.

\section{Analysis}
\label{analysis}

\subsection{IRS~SL Continuum and Extinction}
\label{IRS cont}

\begin{figure}
\begin{center}
\includegraphics[clip,trim =0cm 2.5cm 1.cm 3.7cm,width=4cm]{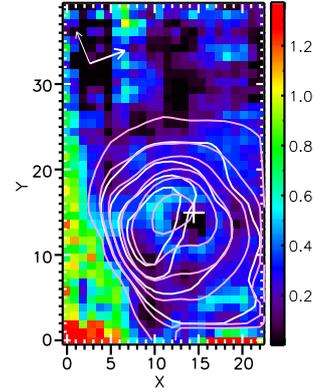}
\end{center}
\caption{NGC~1333 extinction map in units of A$_{k}$. North and East are indicated by the thick and thin white arrows respectively in the upper left corner of the map. Contours of the 11.2 and 7.7~$\mu$m emission are shown respectively in white (1.0, 1.4, 1.8, and 2.2~$\times$~10$^{-5}$~W~m$^{-2}$~sr$^{-1}$) and pink (1.0, 2.0, 3.0, 4.5, 6.0 and 9.5~$\times$~10$^{-5}$~W~m$^{-2}$~sr$^{-1}$). The position of SVS~3 is indicated by a white cross. Axes are given in IRS~SL pixel units.  }
\label{sil_extinct}
\end{figure}

To correct our IRS~SL spectra for extinction, we apply the `modified Spoon method' \citep{sto13,sto14,sto16}. The Spoon method \citep{spo07} is an iterative fitting procedure that interpolates a power law continuum with anchor points at 5.5 and 14.5~$\mu$m of the form y~=~a x$^{k}$ and then calculates the natural log of the ratio of this continuum at 9.8~$\mu$m to the observed flux at 9.8~$\mu$m. This is referred to as the optical depth of the 9.8~$\mu$m silicate absorption feature, $\tau_{9.8}$. In some cases the 14.5~$\mu$m is affected by silicate absorption, leading to an underestimation of $\tau_{9.8}$. The spectrum is then dereddened using the derived $\tau_{9.8}$ and the Spoon method is reapplied to get a new value for $\tau_{9.8}$. This process is reiterated until the anchor point at 14.5~$\mu$m continuum does not change, at which point the final continuum at 9.8~$\mu$m is compared to the observed flux at 9.8~$\mu$m in the original spectrum. However, RN typically have a more linear shaped continua \citep[i.e.][]{ber07,sto16, pee17}. Thus we use a linear fit to the continua for this procedure as done by \cite{sto16}, referred to as the modified Spoon method. We define a linear continuum of the form, y~=~a~+~b~x, where b is the slope. 
We then deredden our spectral data by dividing by our final silicate extinction, A$_{k}$~=~1.079~$\tau_{9.8}$ \citep{sto13}. We assume an uncertainty in silicate extinction of 10\%. We show a map of the extinction in Figure \ref{sil_extinct}. We find that the extinction is very high (A$_{k}$~$>$~1) in the southwestern edge of the map and the western edge near the two YSOs. 

\begin{figure}
\begin{center}
\includegraphics[clip,trim =0.5cm 0.5cm 0.5cm 0.5cm,width=8.cm]{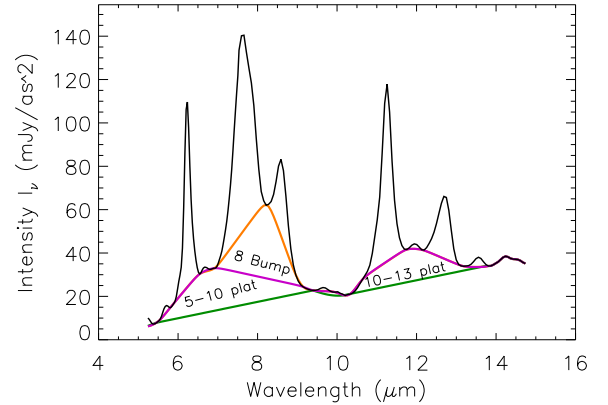}
\end{center}
\caption{Typical IRS~SL spectrum from the RN within NGC~1333 is shown. The orange line traces the local spline continuum (LS), the magenta line traces the global spline continuum (GS), and the green line traces the underlying dust continuum (PL). We label the different PAH emission plateaus defined between the aforementioned continuum components as follows: `8 bump' for the 8 \mum\ bump between the LS and GS continua, `5--10 plat' for the 5--10 \mum\ plateau emission between GS and PL continua, and `10--13 plat' for the 10--13 \mum\ plateau emission between the GS and PL continua.  }
\label{cont_ex}
\end{figure} 

We make use of the spline decomposition method \citep[e.g.][]{hon01,pee02,van04,boe12,sto14,sha15,sha16,sto16,pee17} to fit the continua in each of the spectra as local spline (LS), global spline (GS), and the underlying dust continuum (PL). An example of the typical spectrum and its continua is shown in Figure \ref{cont_ex}. For the LS continuum, we apply anchor points at 5.36, 5.46, 5.86, 6.58, 6.92, 8.28, 9.15, 9.40, 9.89, 10.14, 10.51, 10.76, 11.8,12.13, 13.18, and 13.49~$\mu$m. A spectral artefact arises beyond 14~$\mu$m, therefore we exclude this wavelength range from further analysis. The GS continuum uses all the same anchor points as the LS continuum except for the removal of 8.28 $\mu$m point. The PL continuum is fit using straight lines between anchor points at 5.46 and 9.40~$\mu$m and 10.14 to 13.8~$\mu$m. We refer to the broad emission component in the 7--9~$\mu$m range between the LS and GS as the 8~$\mu$m bump. Likewise, we define the broad emission components between GS and the PL continua as the 5--10~$\mu$m and 10--15~$\mu$m plateaus. We use the term `5--10 plat' to refer to the 5--10~$\mu$m plateau under the GS continuum and bounded below by the PL continuum. Likewise, we use the term `10--13 plat' to refer to the 10-13~$\mu$m plateau under the GS continuum and bounded below by the PL continuum.  

\subsection{IRS~SL Flux measurement}
\label{IRS Flux}

The fluxes of the major PAH bands are in general derived by integrating the continuum--subtracted spectra. In the case of the 6.2 and 11.2~$\mu$m bands, there is significant blending with the weaker 6.0 and 11.0~$\mu$m bands respectively. We fit Gaussians to the 6.0 and the (blue part of the) 6.2~$\mu$m simultaneously with $\lambda$ (FWHM) of 6.027 (0.10) and 6.228 (0.18)~$\mu$m respectively. We obtain these parameters by taking the average peak position for both Gaussians in pixels with sufficiently high 7.7~$\mu$m flux ($\ge$~1.0~$\times$~10$^{-5}$~~W~m$^{-2}$ ~sr$^{-1}$). The peak positions are fixed and the average FWHM over the same set of pixels is determined and subsequently fixed as well. We then subtract the flux of 6.0~$\mu$m Gaussian from the integrated 6.0 + 6.2~$\mu$m flux. The same treatment is used to disentangle the 11.0 and 11.2~$\mu$m bands where we use Gaussians with $\lambda$ (FWHM) of 10.979 (0.21) and 11.261 (0.258)~$\mu$m respectively.

The 12.7~$\mu$m feature blends with the 12.3~$\mu$m H$_{2}$ line. Therefore, we use a decomposition method similar to that used in previous studies \citep{sto14,sto16} to isolate the 12.7~$\mu$m emission band. In short, we use the NGC~2023 12.7~$\mu$m profile at the PDR front from \cite{pee17} as a template and scale it to the 12.7~$\mu$m band between 12.4--12.8~$\mu$m. We subtract this scaled profile from the spectra and a Gaussian is used to fit the 12.3~$\mu$m H$_{2}$ line. We then integrate each spectra from 12.15--13.1~$\mu$m and subsequently subtract the 12.3~$\mu$m H$_{2}$ Gaussian component to measure the 12.7~$\mu$m flux. 

Following \citet{pee17} and \citet{sto17}, we decompose the PAH emission in the 7--9~$\mu$m range using four Gaussians, referred to as G7.6, G7.8, G8.2, and G8.6 components. We begin with a constrained fit to each of these component where the peak positions and FWHMs are allowed to vary within a 0.2 and 0.25~$\mu$m range respectively. The starting values for these parameters are set to the average RN values found by \citet[][their Table 1]{sto17}. Each Gaussian is then fixed to the average peak position found for the pixels where the 7.7~LS integrated flux is above a threshold value of 1.0 $\times$ 10$^{-5}$~W~m$^{-2}$ ~sr$^{-1}$. The average FWHMs over this set of pixels is determined and fixed as well. The fitting procedure is run once more with the following fixed $\lambda$ (FWHM) values for G7.6, G7.8, G8.2, and G8.6~$\mu$m: 7.58 (0.497), 7.90 (0.443), 8.25 (0.29), and 8.57 (0.41)~$\mu$m respectively. For additional details and example fits of the above decompositions, see \cite{sto17, pee17}.

We estimate the signal--to--noise ratio of the PAH emission features as $ \textrm{SNR} = \textrm{F}/ ( \textrm{rms} \times \sqrt{\textrm{N}} \times \Delta \lambda) $ where F is the feature flux (in W~m$^{-2}$~sr$^{-1}$), rms the rms noise, N the number of spectral wavelength bins within the feature, and $\Delta \lambda$ is the wavelength bin size determined from the spectral resolution. The rms noise is determined from featureless portions of the spectra between 9.3--9.5, 13.3--13.5, and 13.7--13.9~$\mu$m. For atomic and H$_{2}$ lines, the signal-to-noise is the ratio of the peak line flux to the underlying rms noise.

\subsection{FIFI-LS Flux measurement}
\label{FIFI measure}

We make use of the SOSPEX spectral cube analysis software \citep{fad18} to measure the FIR cooling line fluxes. We apply an atmospheric correction to all spectra. Specifically, for the [\OI] 63~$\mu$m line, we apply an atmospheric correction equal to the median value of the atmospheric transmission, $\sim$~0.56, due to the atmospheric transmission curve having a large drop towards the red end of this line. For the [\OI] 146~$\mu$m line, we apply an atmospheric correction equal to the median value of the atmospheric transmission, over the wavelength range of the spectra, $\sim$~0.8, due to a significant broad absorption band corresponding with the reference wavelength of the line. For the [\CII] 158~$\mu$m line, we apply an atmospheric correction equal to the value of the atmospheric transmission at the reference wavelength of the line, $\sim$~0.85, to avoid a broad absorption band at 157.5~$\mu$m \citep{lor92}.

We fit a straight line to anchor points at both ends of each spectral line to define the underlying continuum over the entire spectral cube with the specification that the continuum must have positive values. 
The continuum for each pixel is set to the median value within the surrounding pixels. To measure line intensities, we use a Gaussian model for [\OI] 63~$\mu$m and a Voigt model for [\OI] 146 and [\CII] 158~$\mu$m (Dario Fadda, private communication).

We convolve the [\OI] 63 and 146~$\mu$m flux maps to match the larger PSF of the [\CII] 158~$\mu$m map. For example, the width of the Gaussian kernel for the [\OI] 63~$\mu$m to [\CII] 158~$\mu$m convolution is $ w = \sqrt{(15.9)^{2} - (6.4)^{2}}$, where 6.4$^{\prime\prime}$ and 15.9$^{\prime\prime}$ are the [\OI] 63~$\mu$m and [\CII] 158~$\mu$m PSFs respectively \citep[e.g.][]{hou07}. The FIR line maps are then converted to units of W~m$^{-2}$~sr$^{-1}$ and regridded to the largest spatial resampling size of 2$^{\prime\prime}$ for the red spectrometer of FIFI--LS and clipped to the FOV of the [\OI] 63~$\mu$m map allowing us to compare both data sets at a matching spatial resolution and FOVs.

\subsection{FIR SED Fitting}
\label{SED}
\begin{figure}
\begin{center}
\includegraphics[clip,trim =0.5cm 0.5cm 0.2cm 0.5cm,width=8.cm]{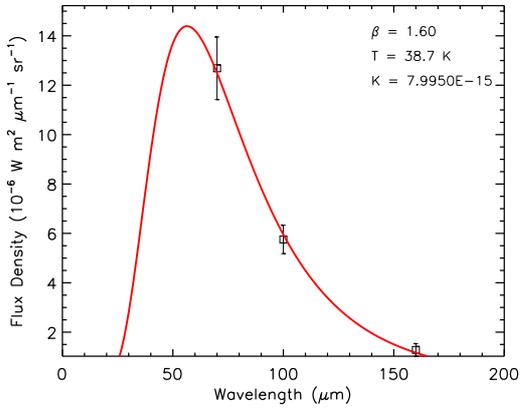}
\end{center}
\caption{Example of the SED fitting procedure used to fit the PACS data. The data is shown as black squares and the modified blackbody fit is shown in red. The fit parameters for this pixel are given in the upper right.}
\label{sed_ex}
\end{figure}

We use PACS images at 70, 100, and 160~$\mu$m to measure the FIR dust continuum emission. We first convolve each of these images to the largest PSF in our data set, specifically the [\CII] 158~$\mu$m PSF, using a 2D Gaussian kernel to account for the differnt PSFs of the PACS data. We extract a sub image of the convolved PACS maps equal to the FOV of the [\CII] 158~$\mu$m map and regrid to the PACS 160~$\mu$m pixel scale. We then fit the data with a modified blackbody function of the form:

\begin{equation}
I(\lambda, T) = \frac{K}{\lambda^{\,\beta}}B(\lambda,T)  
\end{equation}

\noindent where K is a scaling parameter, $\beta$ is the spectral index, T is the dust temperature and B($\lambda$,T) is the Planck function \citep[e.g.][]{abe10,ber12,and18}. We use starting values of $\beta$~=~1.8, T~=~20~K and K~=~1.0~$\times$~10$^{-15}$ and modified $\beta$, T, and K until a best fit was achieved. The obtained modified blackbody fits to the PACS data points agrees with the observations within the absolute flux uncertainties for each pixel (Figure \ref{sed_ex}). The FIR flux is determined by integrating the area under the fitted function for each pixel.

\subsection{Matching Apertures}
\label{aper}

In order to compare the MIR IRS~SL spectral maps with the FIR FIFI-LS and PACS maps, we convolve all the maps to the lowest common resolution as was discussed in previous sections. Because the [\CII] 158~$\mu$m observation has the largest PSF in this data set, all of the other maps are convolved to this spatial resolution. In our comparison of the MIR and FIR maps in Section \ref{PAHs and PDRs}, subimages of the FIR maps are extracted to match the IRS~SL FOV, the smallest FOV within our data set.

\section{Results}
\label{results}

\subsection{IRS Results}
\label{IRS results}

In this section, we probe the relationships between the PAH emission features, the underlying plateaus, the H$_{2}$ emission, and the dust continuum emission obtained from the IRS~SL data. We present our maps at the native pixel scale of the SL cube to explore this data at the highest resolution possible. 

\subsubsection{SL Maps}
\label{slmaps}

\begin{figure*}
\begin{center}
\resizebox{\hsize}{!}{%
\includegraphics[clip,trim =0.cm 2.5cm 0.8cm 2cm,width=0.20\textwidth]{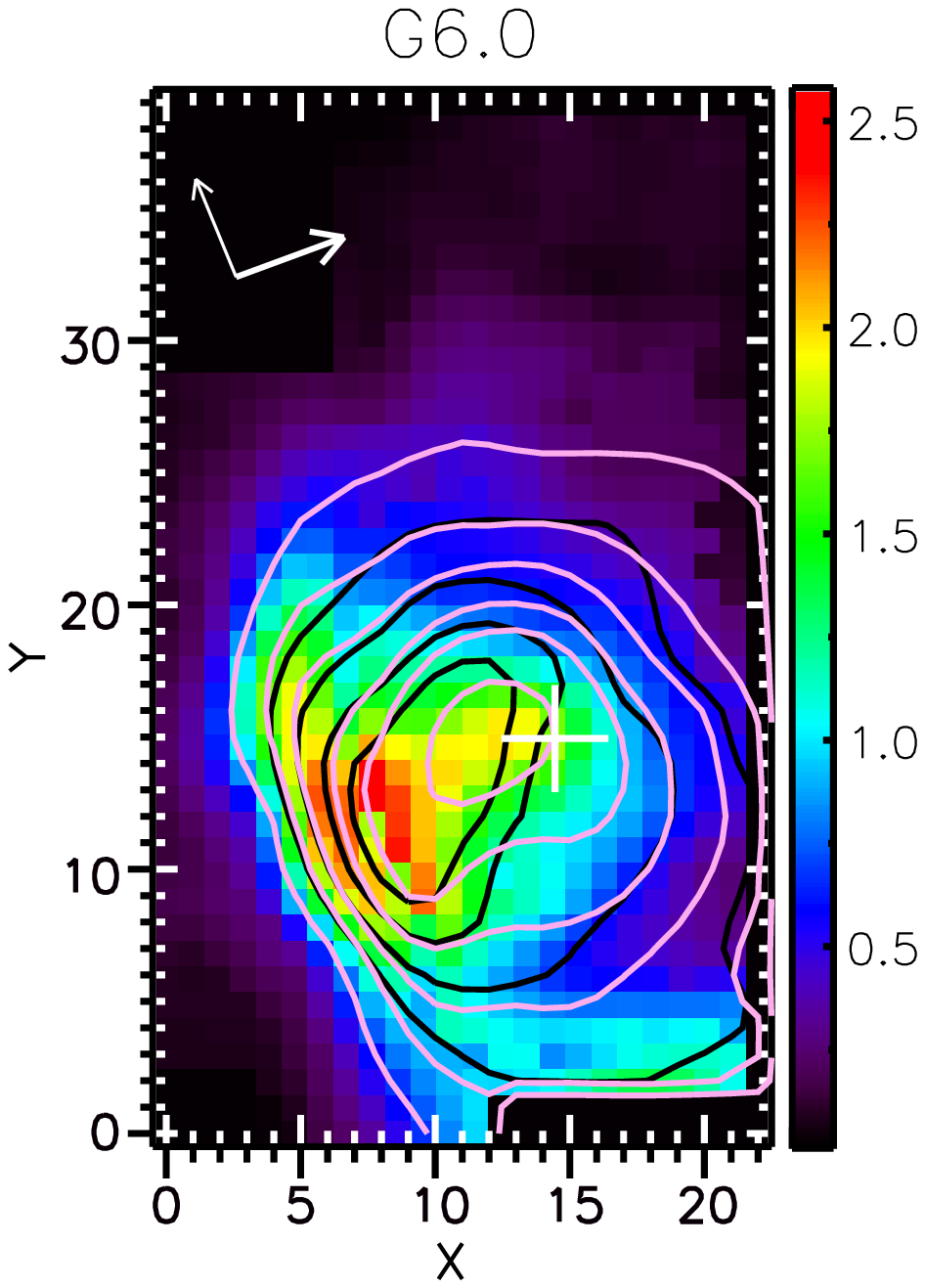}
\includegraphics[clip,trim =0.cm 2.5cm 0.8cm 2cm,width=0.20\textwidth]{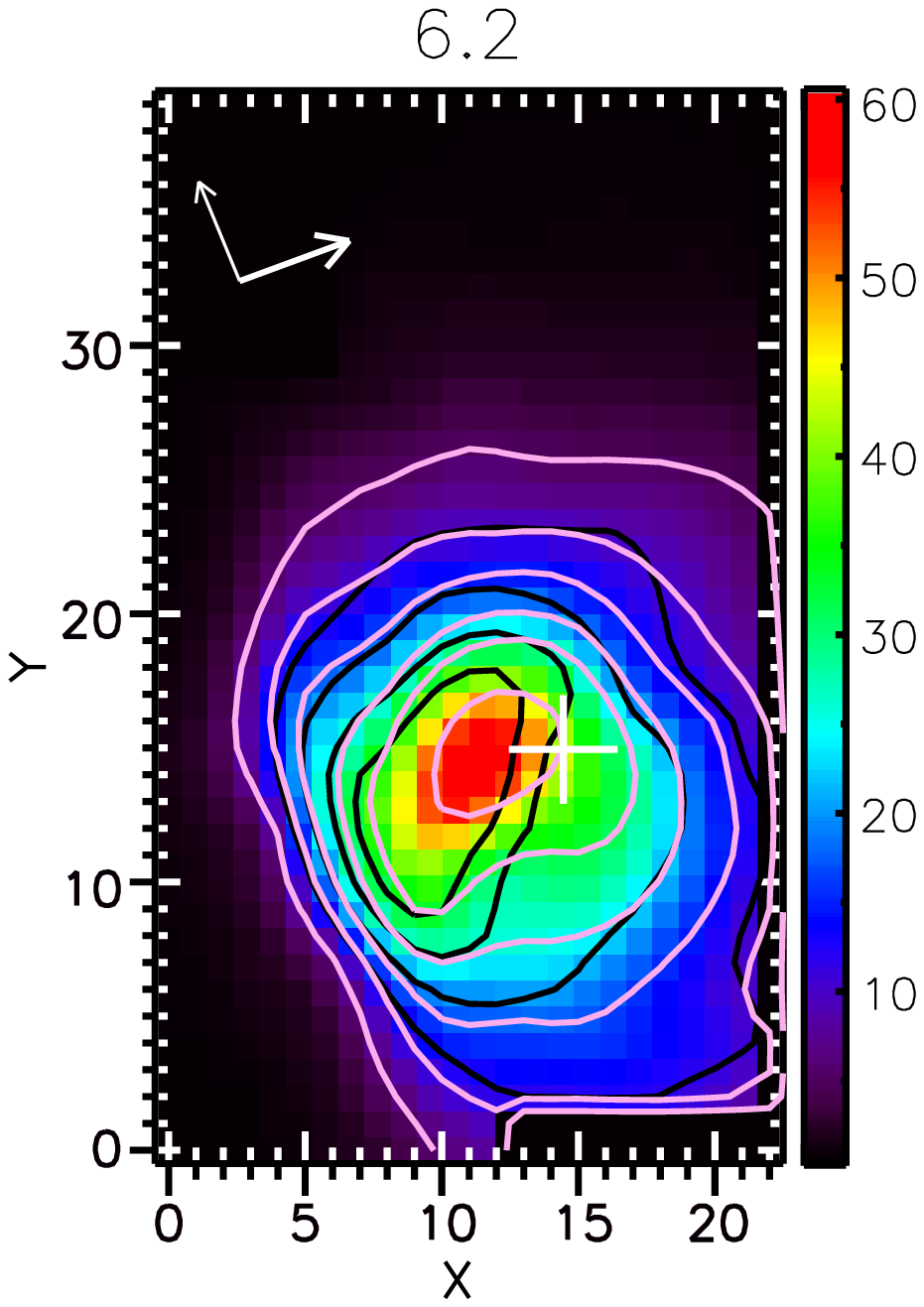}
\includegraphics[clip,trim =0.cm 2.5cm 0.8cm 2cm,width=0.20\textwidth]{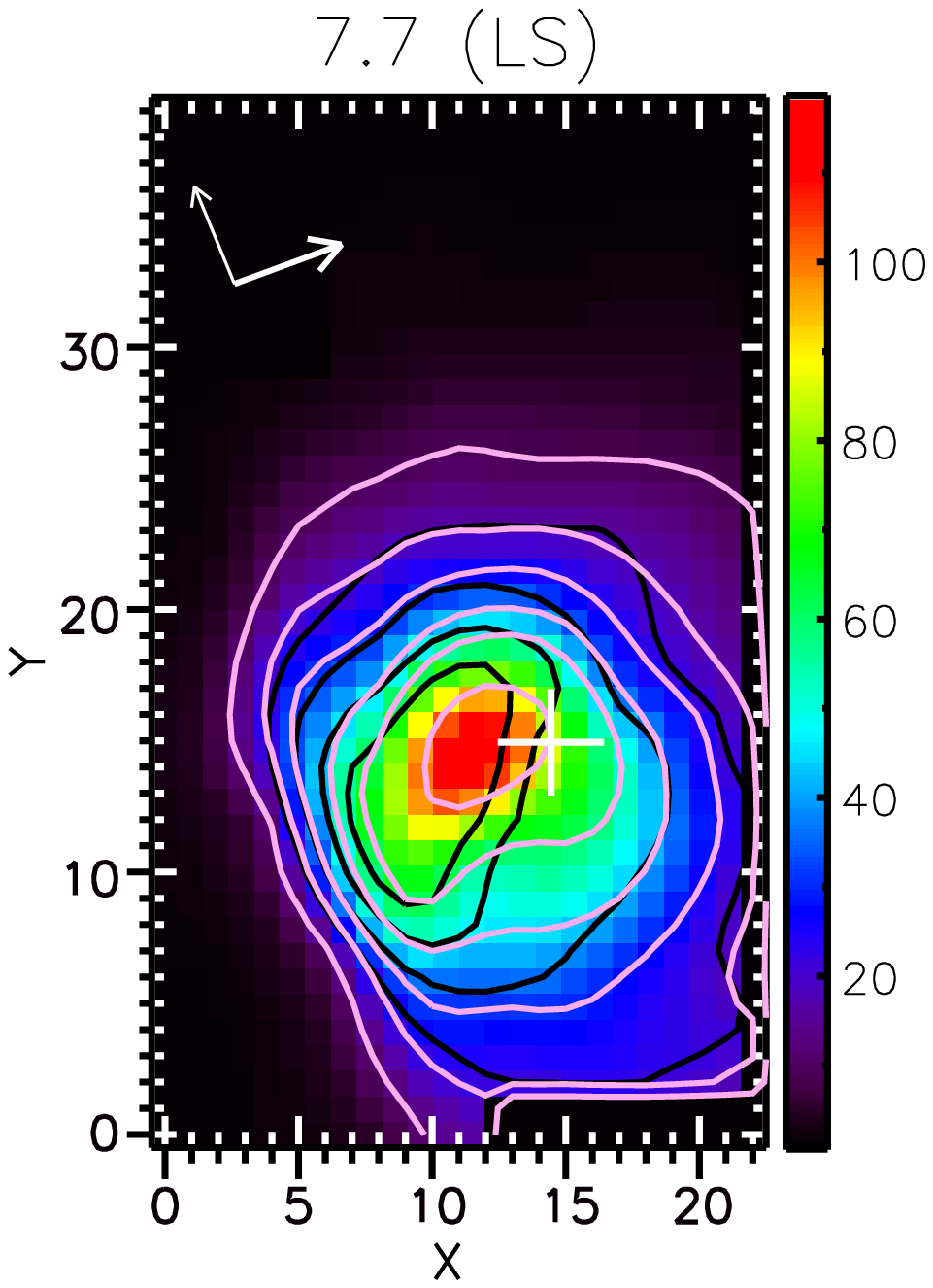}
\includegraphics[clip,trim =0.cm 2.5cm 0.8cm 2cm,width=0.20\textwidth]{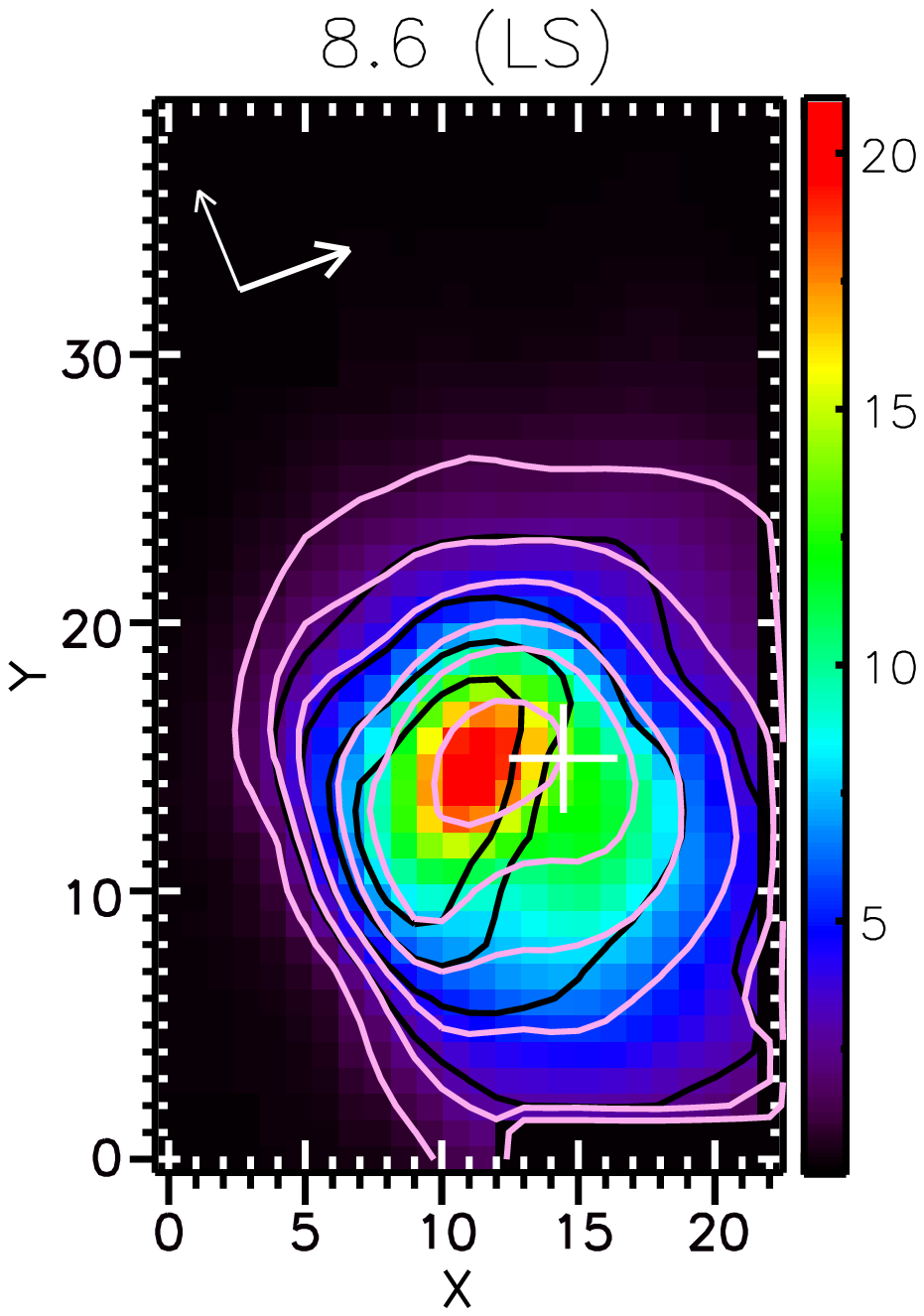}
\includegraphics[clip,trim =0.cm 2.5cm 0.8cm 2cm,width=0.20\textwidth]{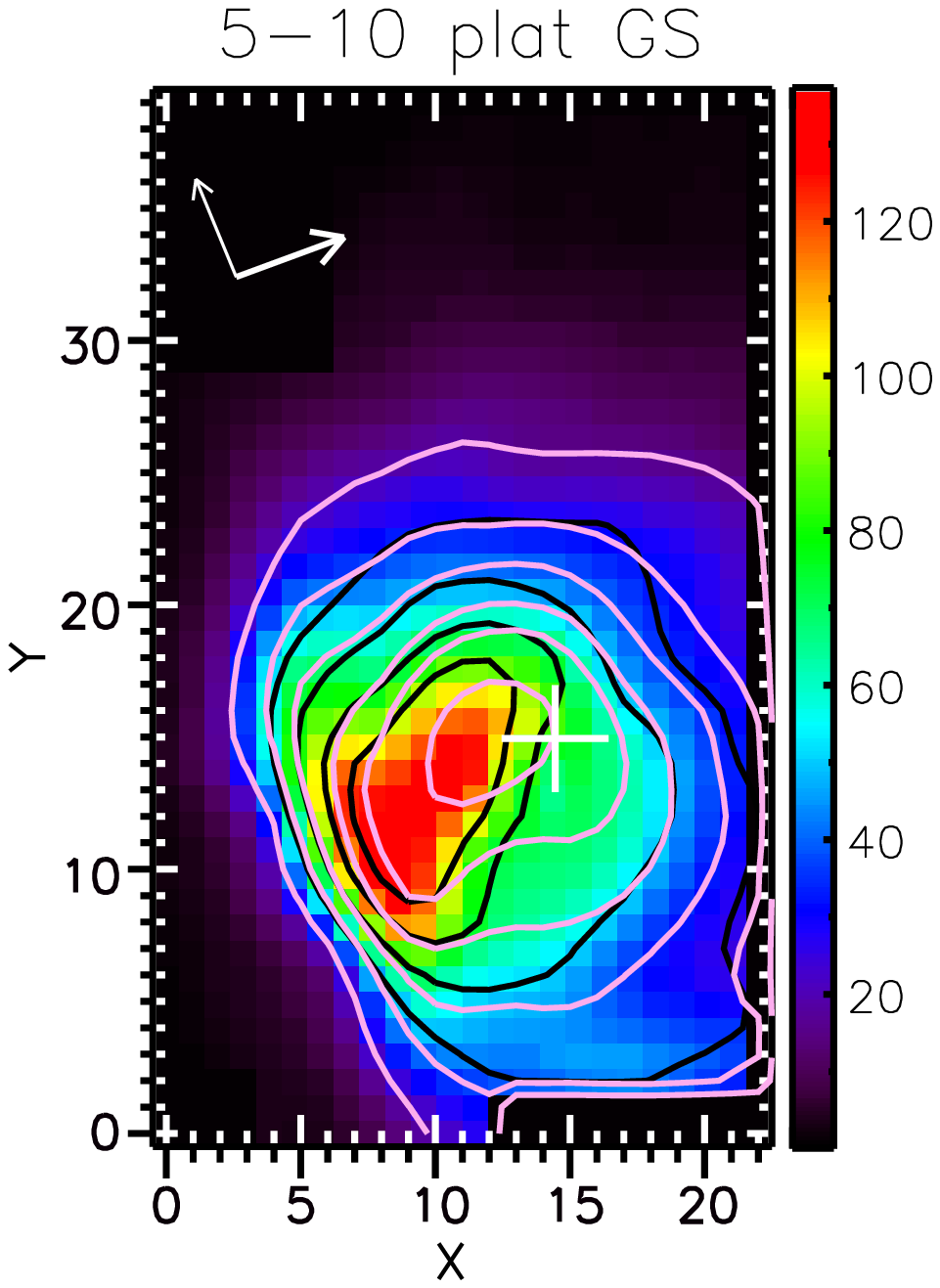}}
\resizebox{\hsize}{!}{%
\includegraphics[clip,trim =0.cm 2.5cm 0.8cm 2cm,width=0.20\textwidth ]{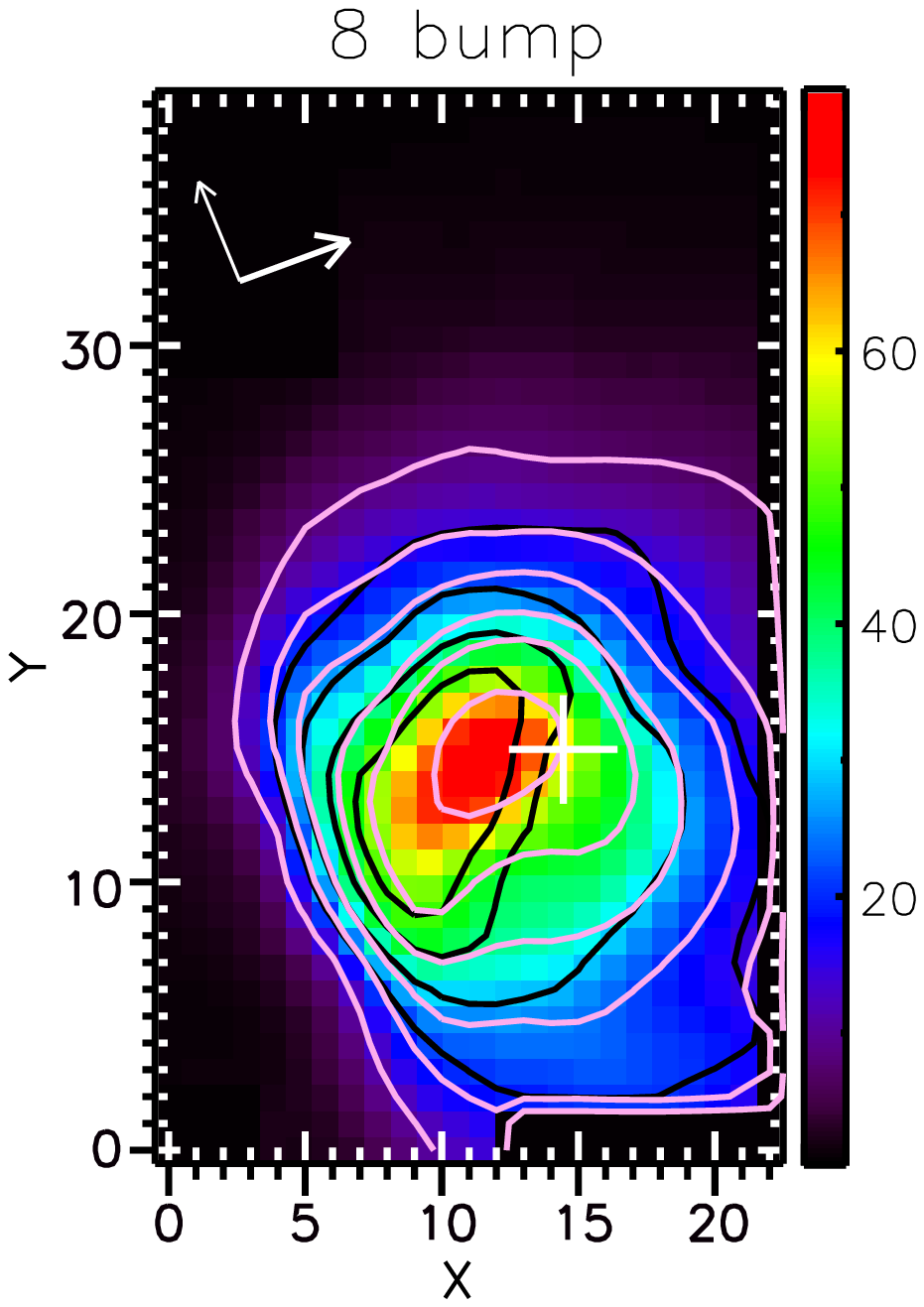}
\includegraphics[clip,trim =0.cm 2.5cm 0.8cm 2cm,width=0.20\textwidth]{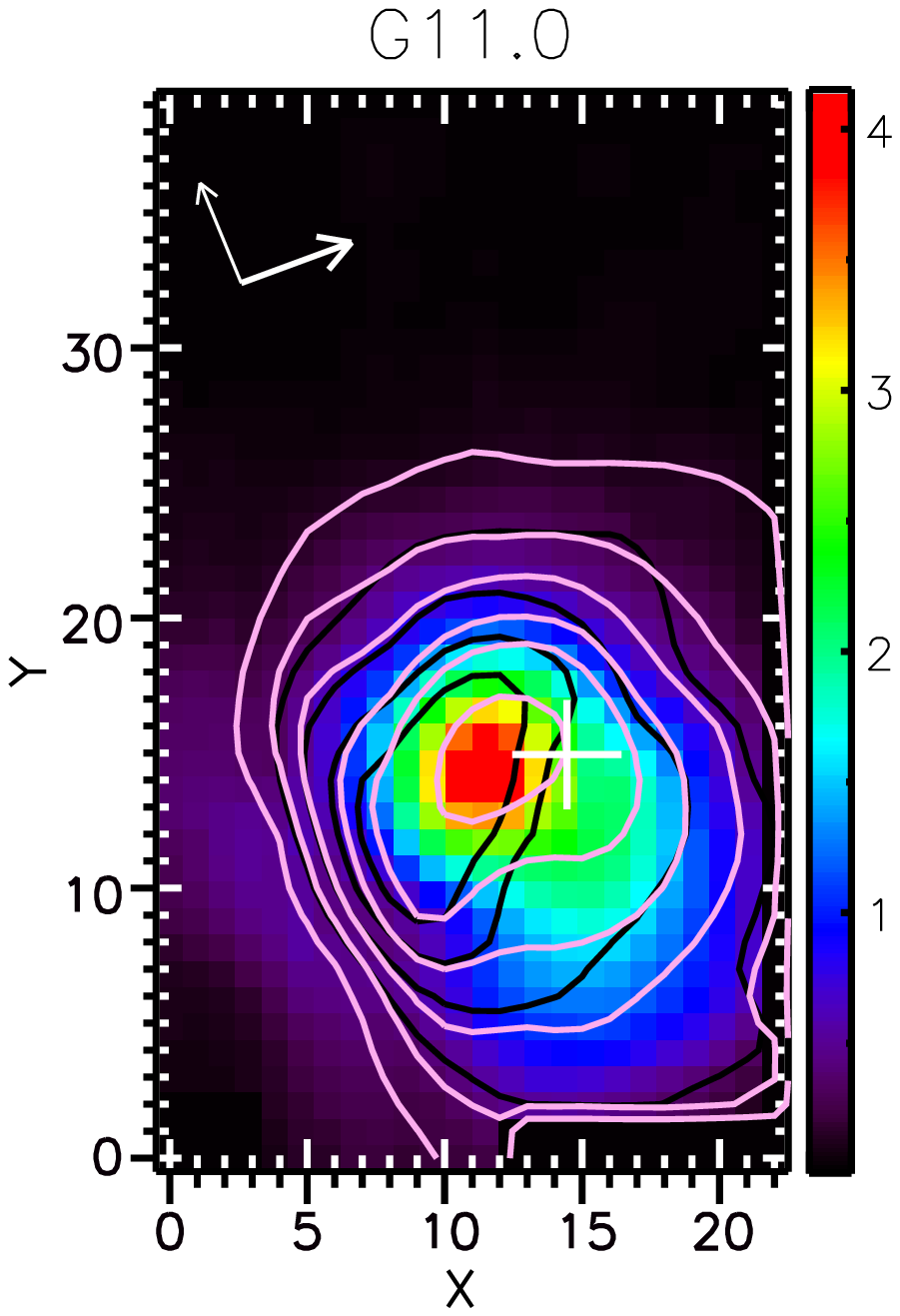}
\includegraphics[clip,trim =0.cm 2.5cm .8cm 2cm,width=0.20\textwidth]{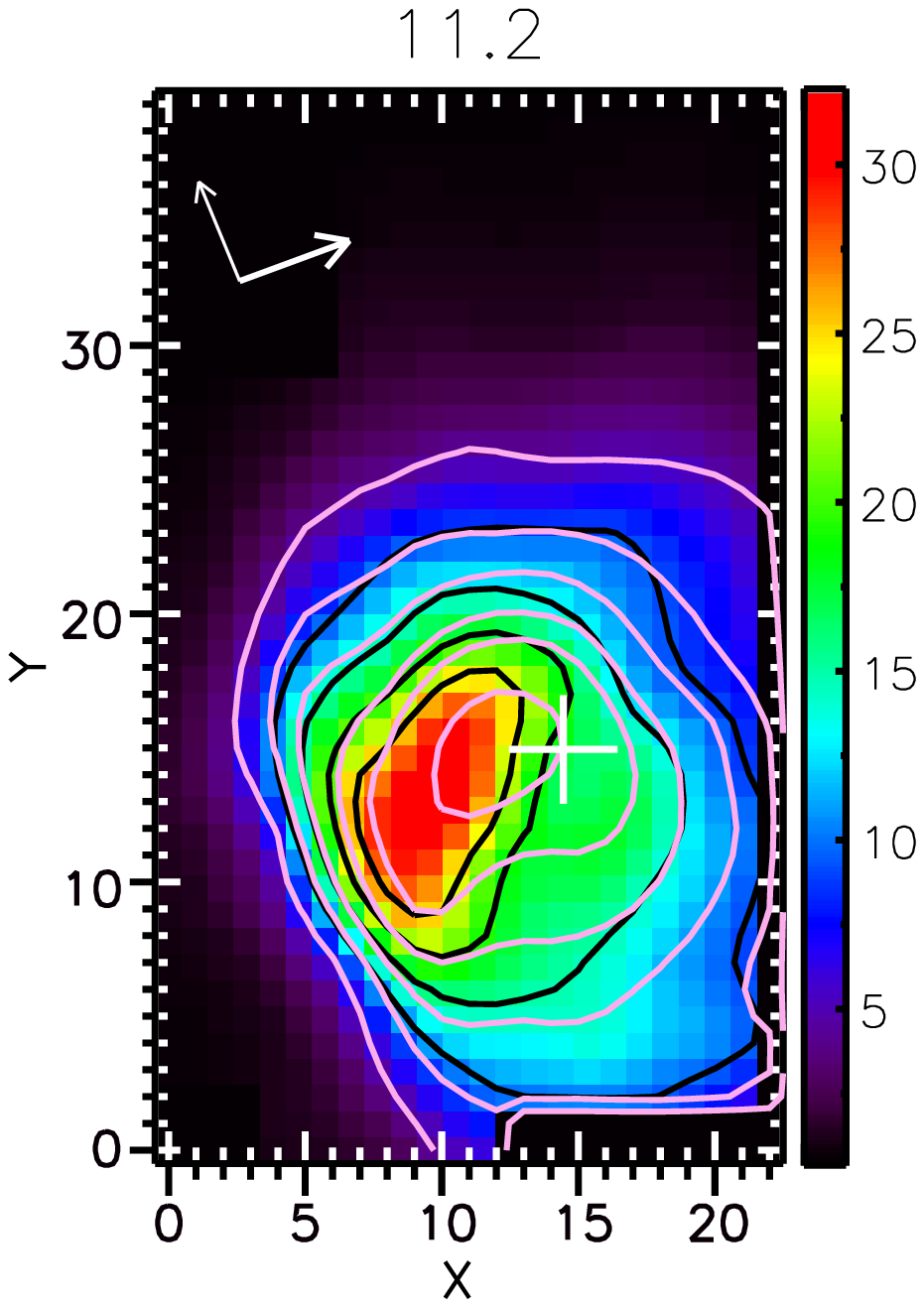}
\includegraphics[clip,trim =0.cm 2.5cm .8cm 2cm,width=0.20\textwidth]{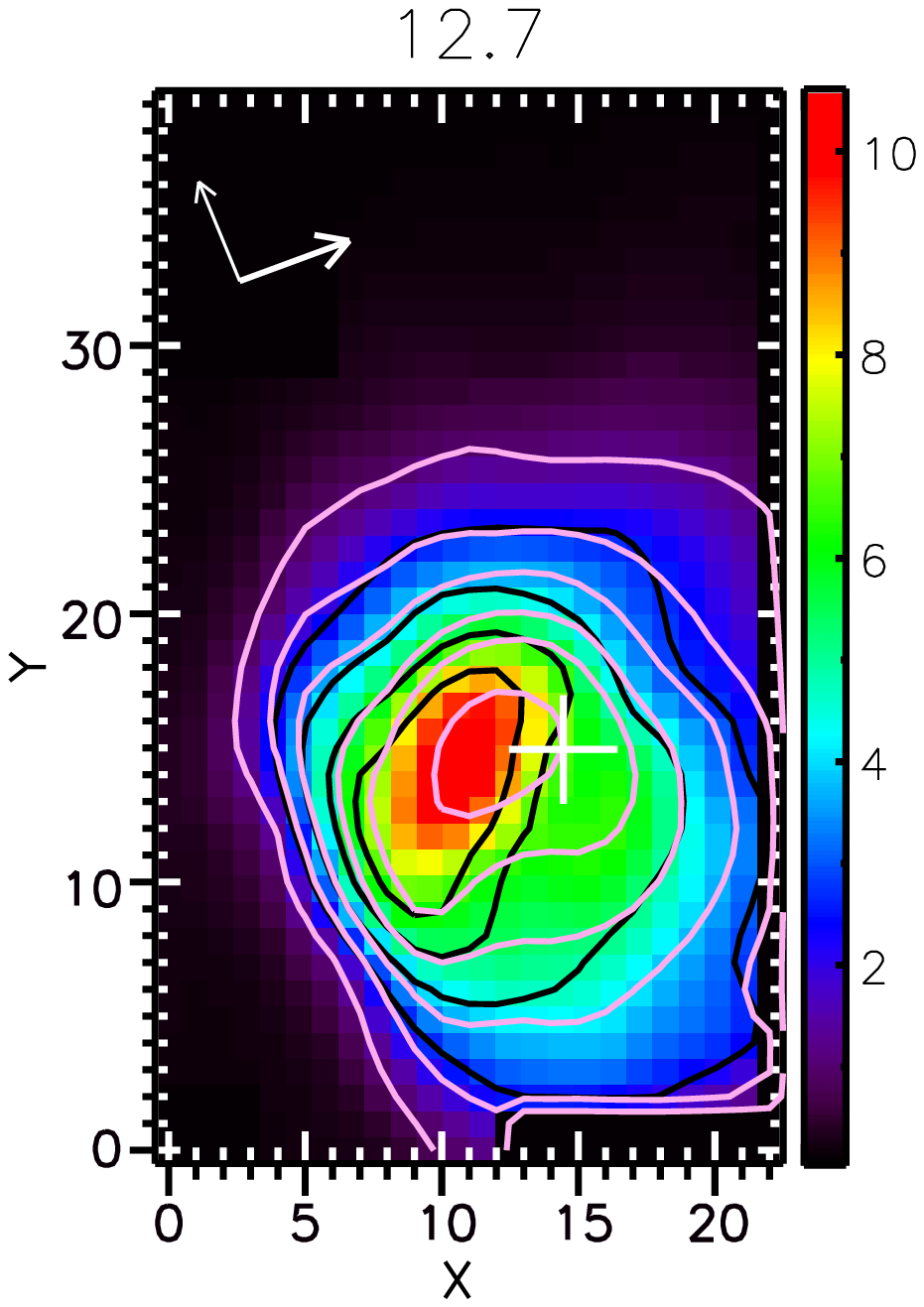}
\includegraphics[clip,trim =0.cm 2.5cm .8cm 2cm,width=0.20\textwidth]{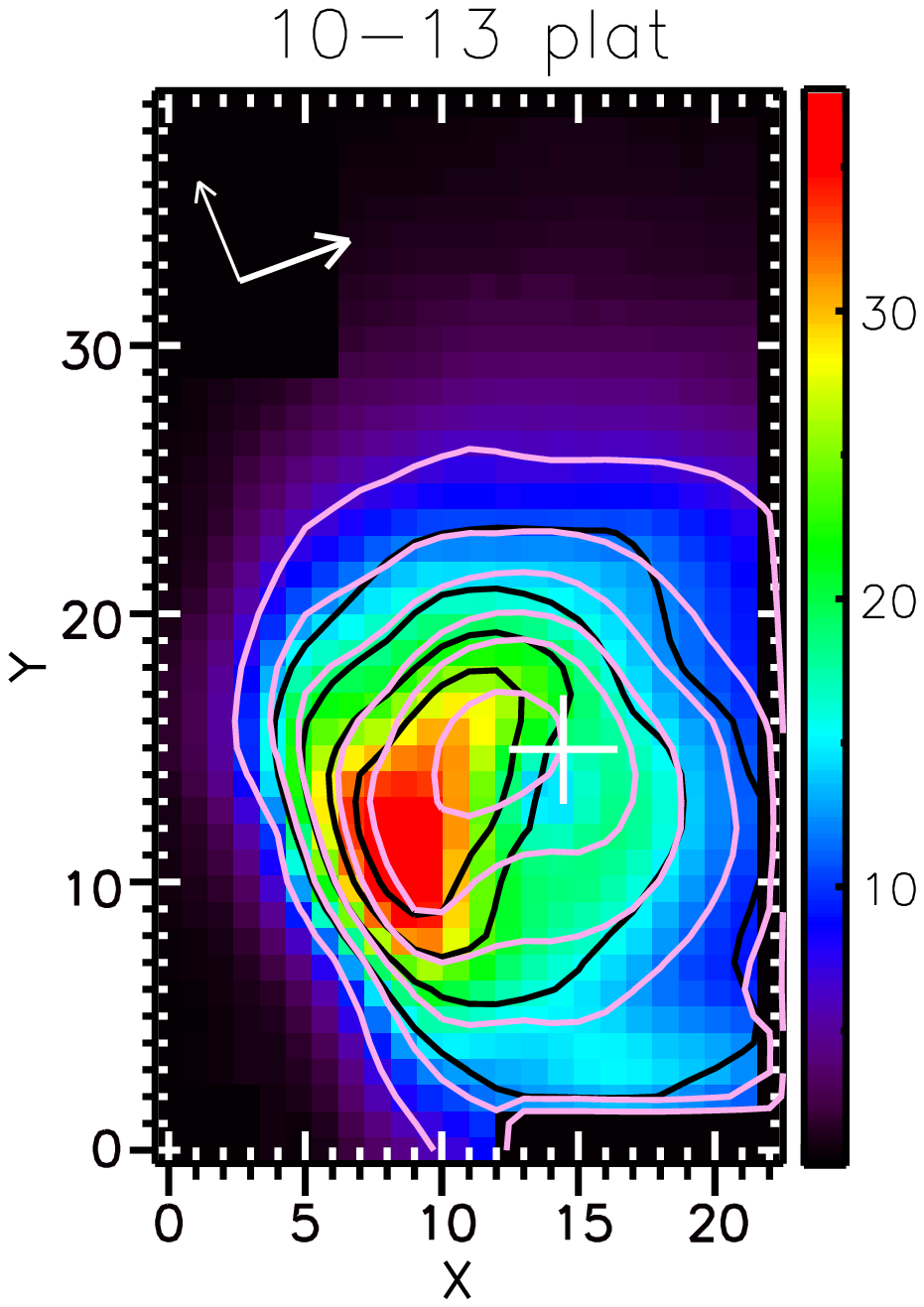}}
\resizebox{\hsize}{!}{%
\includegraphics[clip,trim =0.cm 2.5cm .8cm 2cm,width=0.20\textwidth]{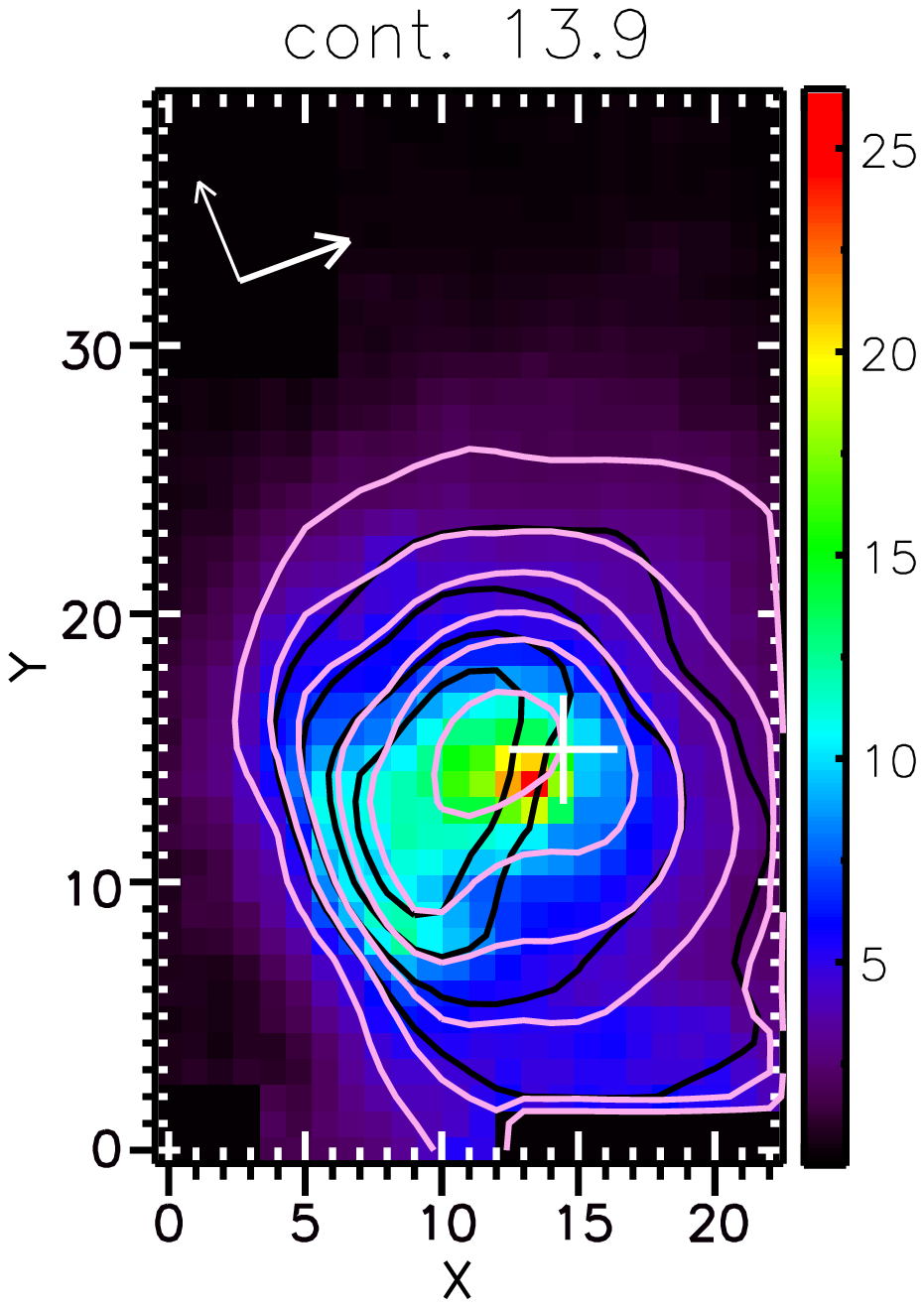}
\includegraphics[clip,trim =0.cm 2.5cm .8cm 2cm,width=0.20\textwidth]{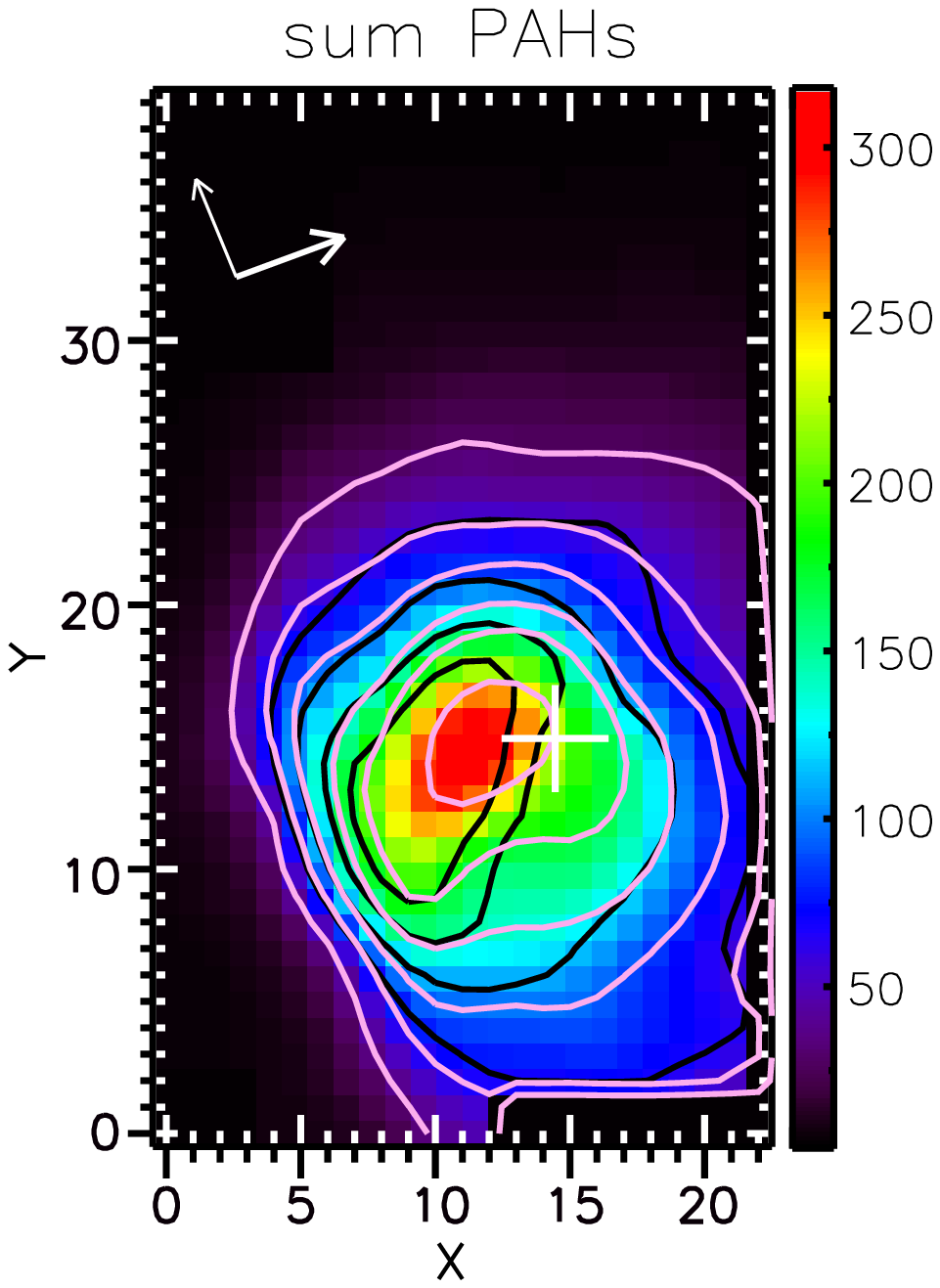}
\includegraphics[clip,trim =0.cm 2.5cm .8cm 2cm,width=0.20\textwidth]{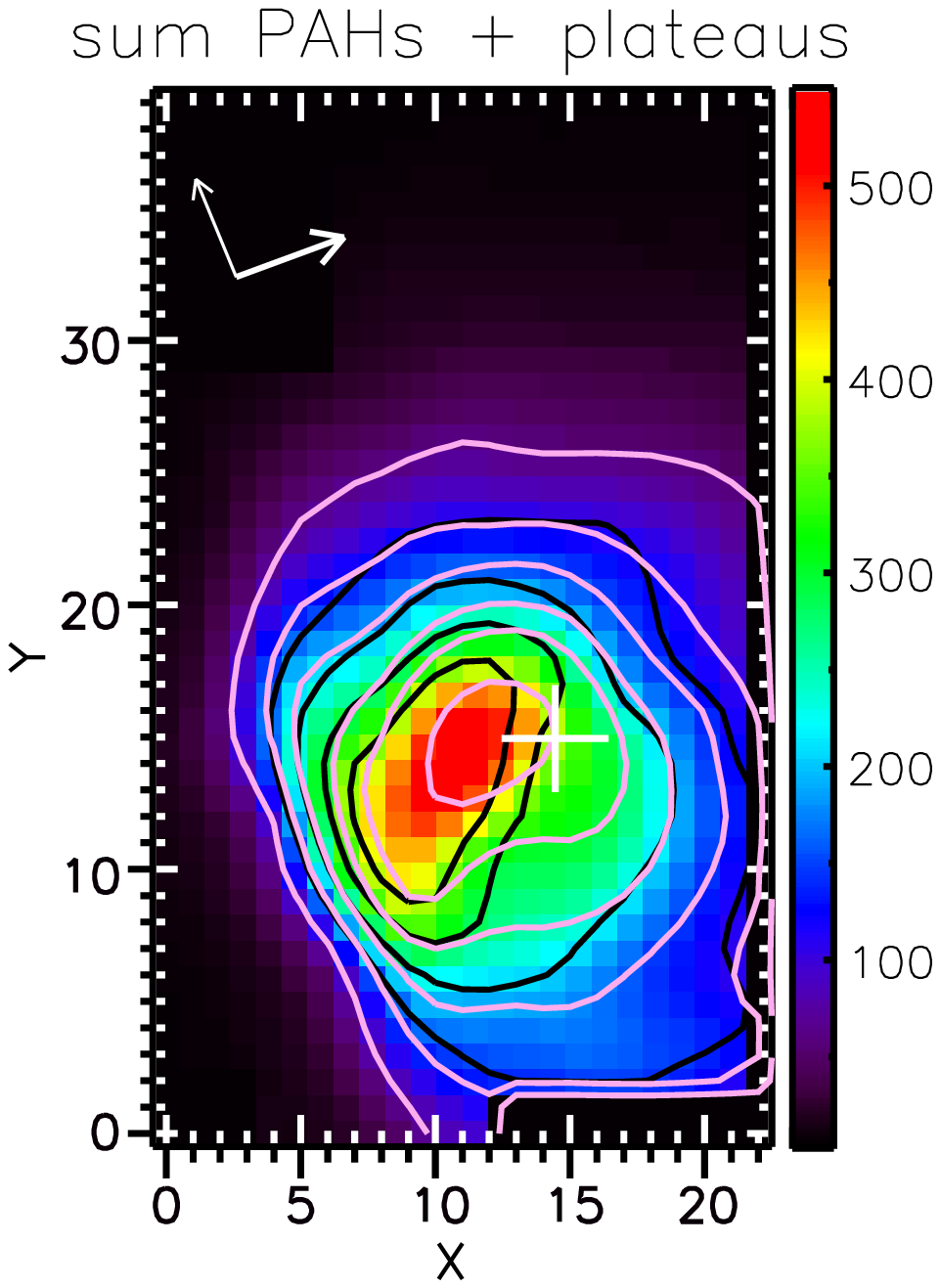}
\includegraphics[clip,trim =0.cm 2.5cm .8cm 2cm,width=0.20\textwidth]{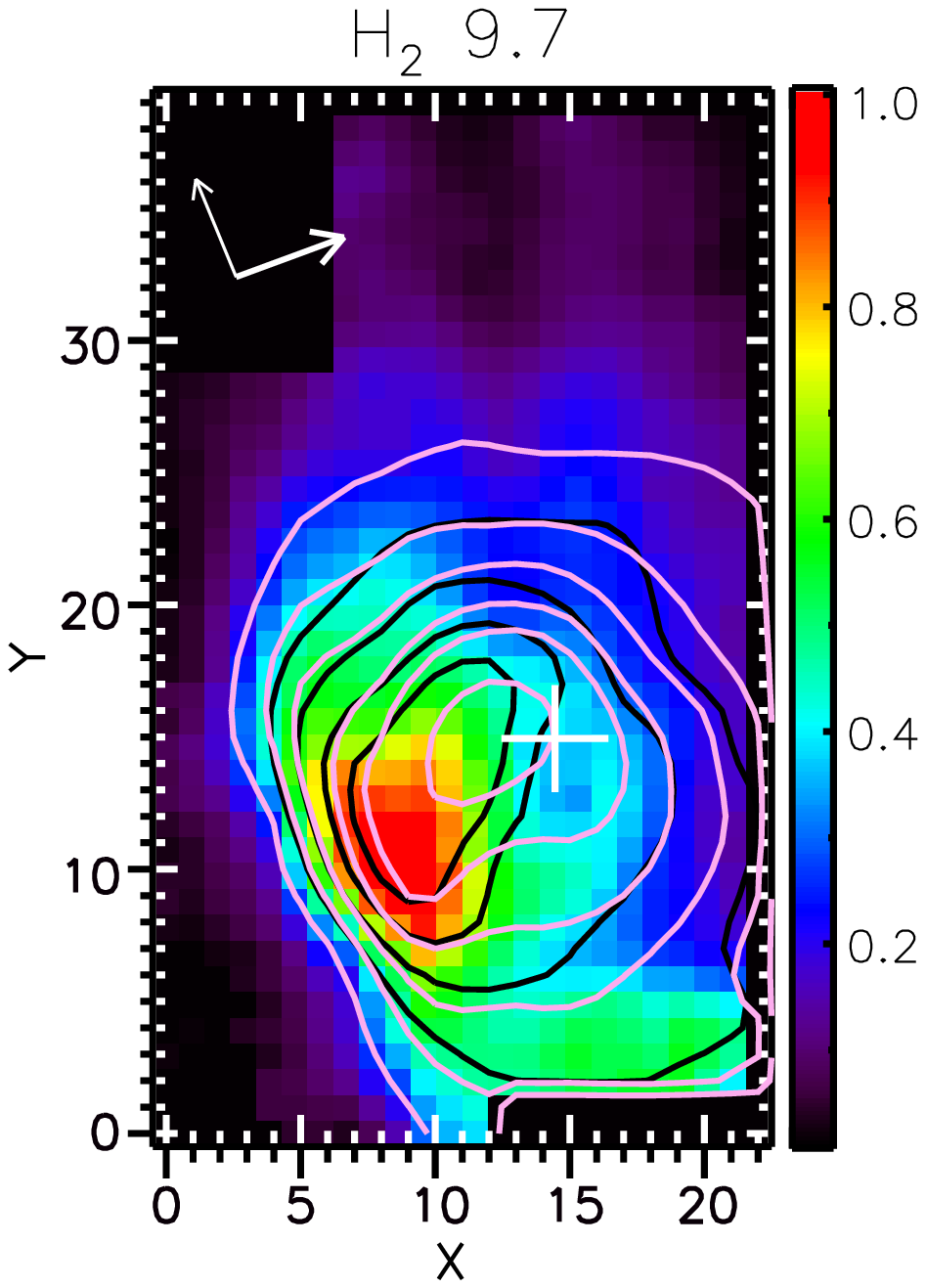}
\includegraphics[clip,trim =0.cm 2.5cm .8cm 2cm,width=0.20\textwidth]{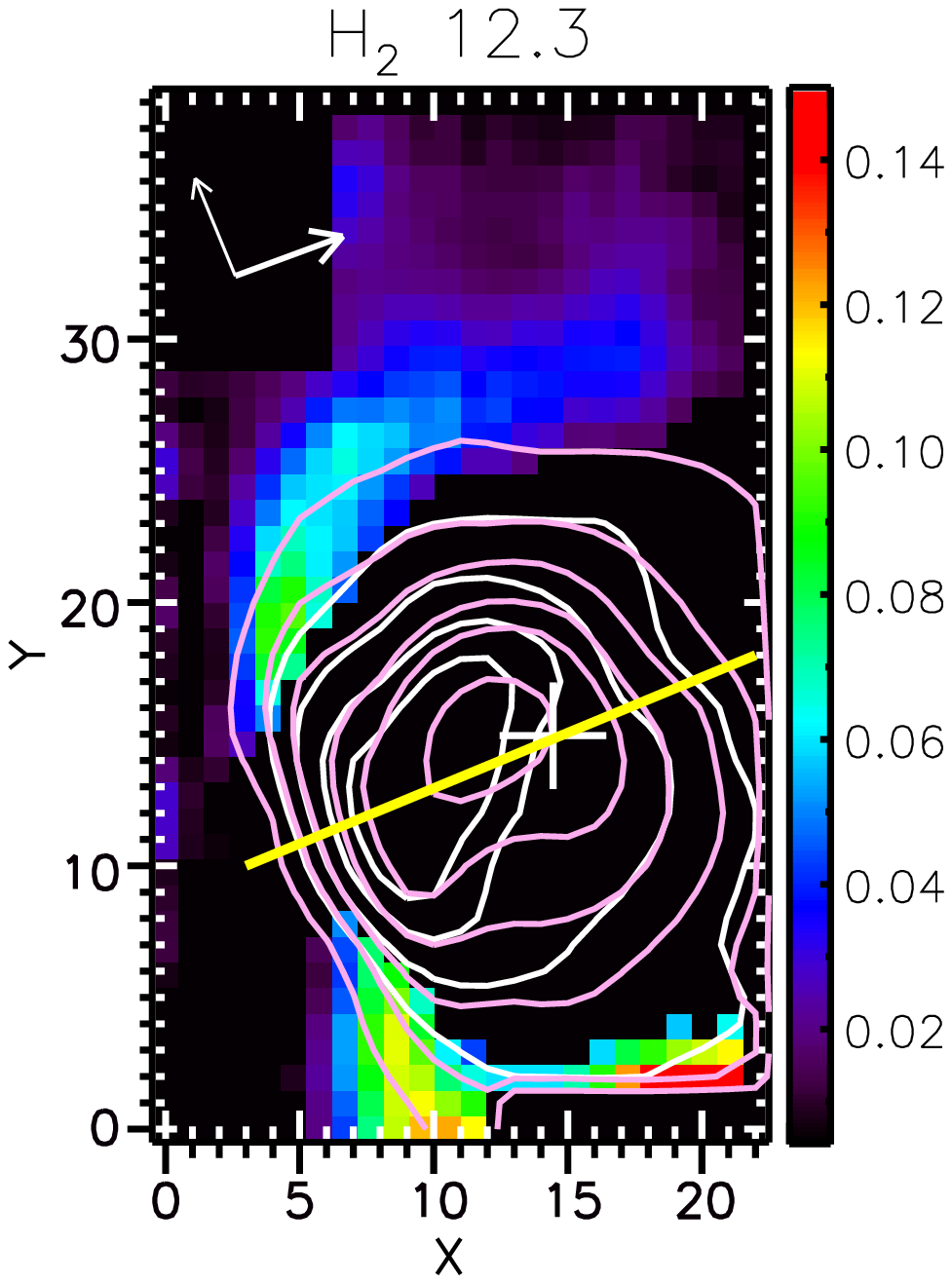}}
\end{center}
\caption{Spatial distribution of the emission features in the 5-- 15~$\mu$m IRS~SL data within the NGC~1333 SVS~3 region. PAH band and continuum intensities are given in units of 10$^{-6}$~W~m$^{-2}$ ~sr$^{-1}$ and 100~MJy~sr$^{-1}$ respectively. Contours of the 11.2 and 7.7~$\mu$m emission are shown respectively in black (1.0, 1.4, 1.8, and 2.2~$\times$~10$^{-5}$~W~m$^{-2}$~sr$^{-1}$) and pink (1.0, 2.0, 3.0, 4.5, 6.0 and 9.5~$\times$~10$^{-5}$~W~m$^{-2}$~sr$^{-1}$). North and East are indicated by the thick and thin white arrows respectively in the upper left corner of each map. The position of SVS~3 is indicated by a white cross and the cross cut used in Section \ref{irs lp} is shown as a yellow line in the H$_{2}$ S(2) 12.3~$\mu$m map (bottom right). Note that the 11.2~$\mu$m emission contours are shown in white in the H$_{2}$ S(2) 12.3~$\mu$m map for clarity. Axes are given in IRS~SL pixel units.}
\label{irs maps}
\end{figure*}

In Figure \ref{irs maps}, we show the spectral maps of the fluxes of the various spectral components within the IRS~SL spectral cube of NGC~1333 including the PAH features, plateau components, and H$_{2}$ lines. We set the range of the colorbar to the minimum and maximum intensities shown in each map. Note that the black areas correspond to the pixels that have been masked for various reasons as described in Section \ref{IRS red}. 

The major trends present within these maps are as follows. To first order, each of the major PAH features appears to share a similar concentric ovular morphology with intensities peaking just to the south of the stellar source, SVS~3, and gradually drop with distance from this source. All PAH bands show deviations in symmetry from this concentric ovular morphology.
In particular, the southern edge of the RN has a much steeper drop in emission relative to the other edges. This corresponds to regions with increased extinction. 

The morphologies of the 6.2, 7.7, 8.6, and 11.0~$\mu$m PAH bands all show a similar condensed shape in their peak emission (shown in red), in contrast to the much more elongated peak emission in the spatial distribution of the 11.2~$\mu$m PAH band, the 5--10~$\mu$m plateau, and the 10--15~$\mu$m plateau. The spatial distribution of the 8~$\mu$m bump and the 12.7~$\mu$m PAH band is a mixture of these condensed and elongated peak morphologies. We also note the presence of a diffuse emission `plateau' in the PAH emission maps moving westward from the center in a pixel range of approximately x~=~10--20 and y~=~3--5, which may be attributed to the two YSOs that are located at the western edge of these map. This emission plateau is much more readily apparent in the 11.2~$\mu$m band and the emission components that share a similar peak morphology. Continuum emission at 13.9~$\mu$m is much more spatially concentrated compared to the PAH emission and peaks almost on top of SVS~3. The H$_{2}$ S(3) 9.7~$\mu$m morphology deviates from the above trends: it is clearly concentrated to the southwest of SVS~3 with a peak co-spatial with the 10--15~$\mu$m plateau and one near the YSO at the western edge of the map. We do not have 3~$\sigma$ detection of H$_{2}$ S(2) 12.3~$\mu$m line over much of the RN inhibiting any conclusive trends.

\begin{figure*}
\begin{center}
\resizebox{\hsize}{!}{%
\includegraphics[clip,trim =0.cm 2.5cm .8cm 2cm,width=0.20\textwidth]{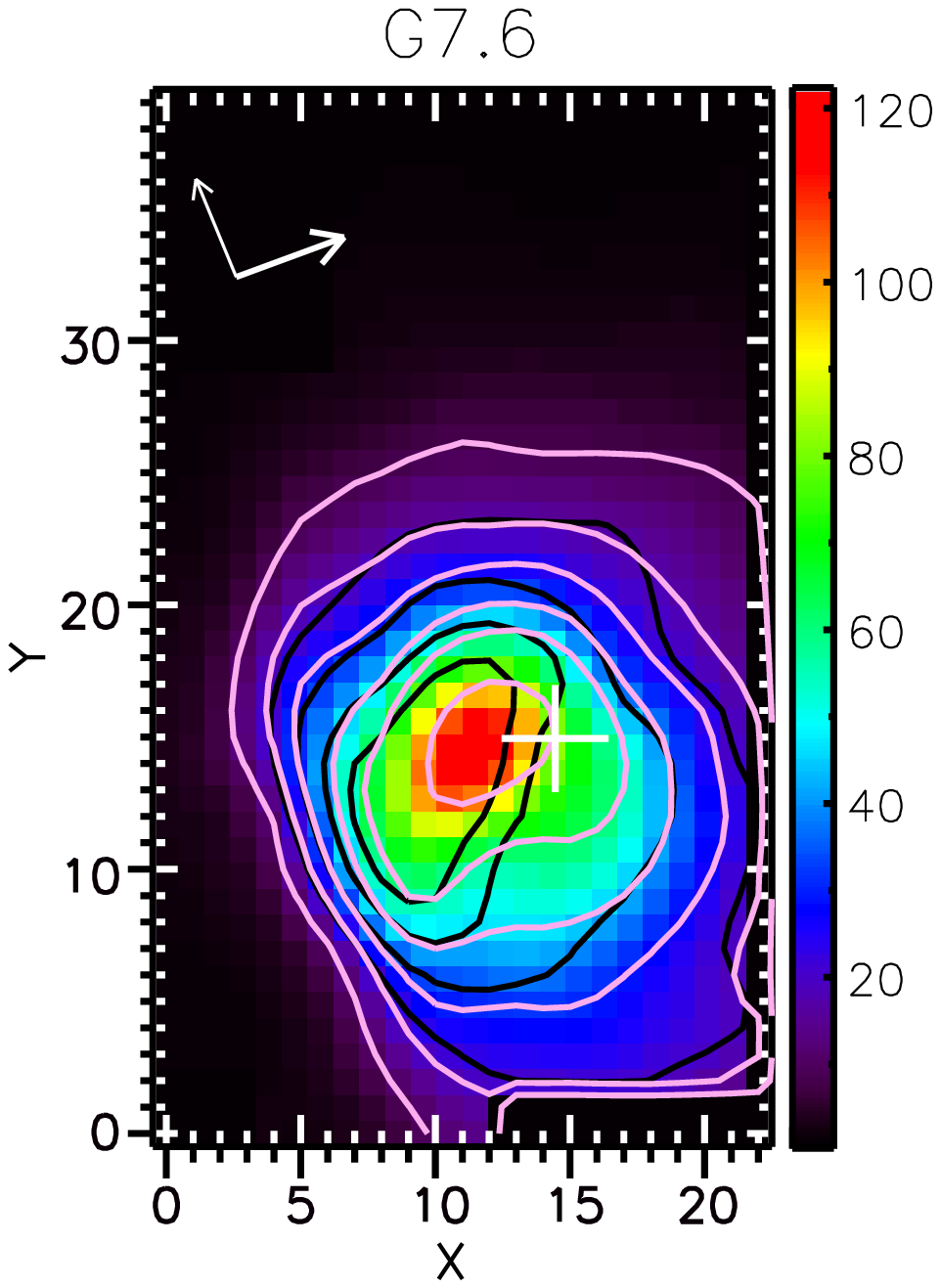}
\includegraphics[clip,trim =0.cm 2.5cm .8cm 2cm,width=0.20\textwidth]{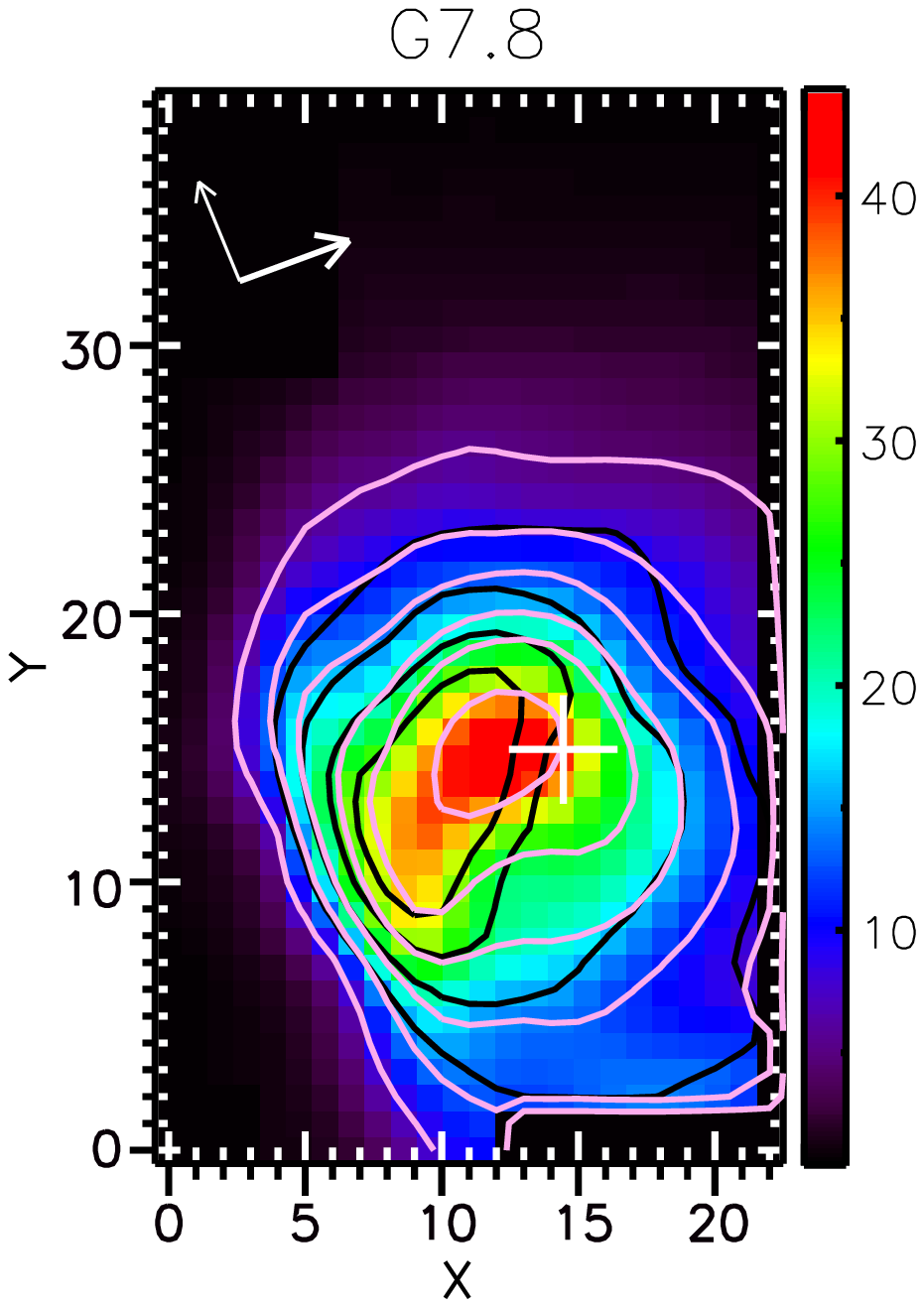}
\includegraphics[clip,trim =0.cm 2.5cm .8cm 2cm,width=0.20\textwidth]{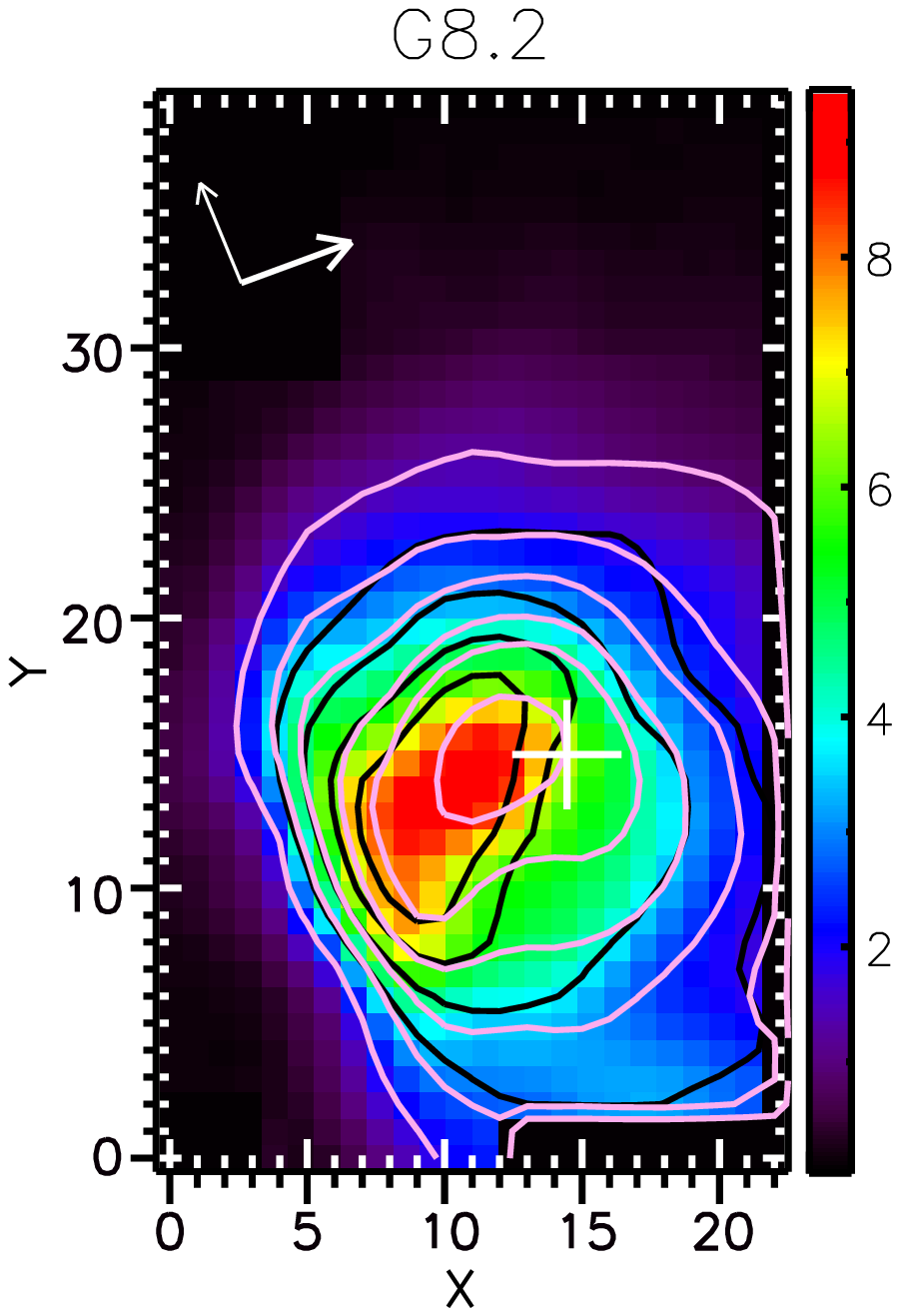}
\includegraphics[clip,trim =0.cm 2.5cm .8cm 2cm,width=0.20\textwidth]{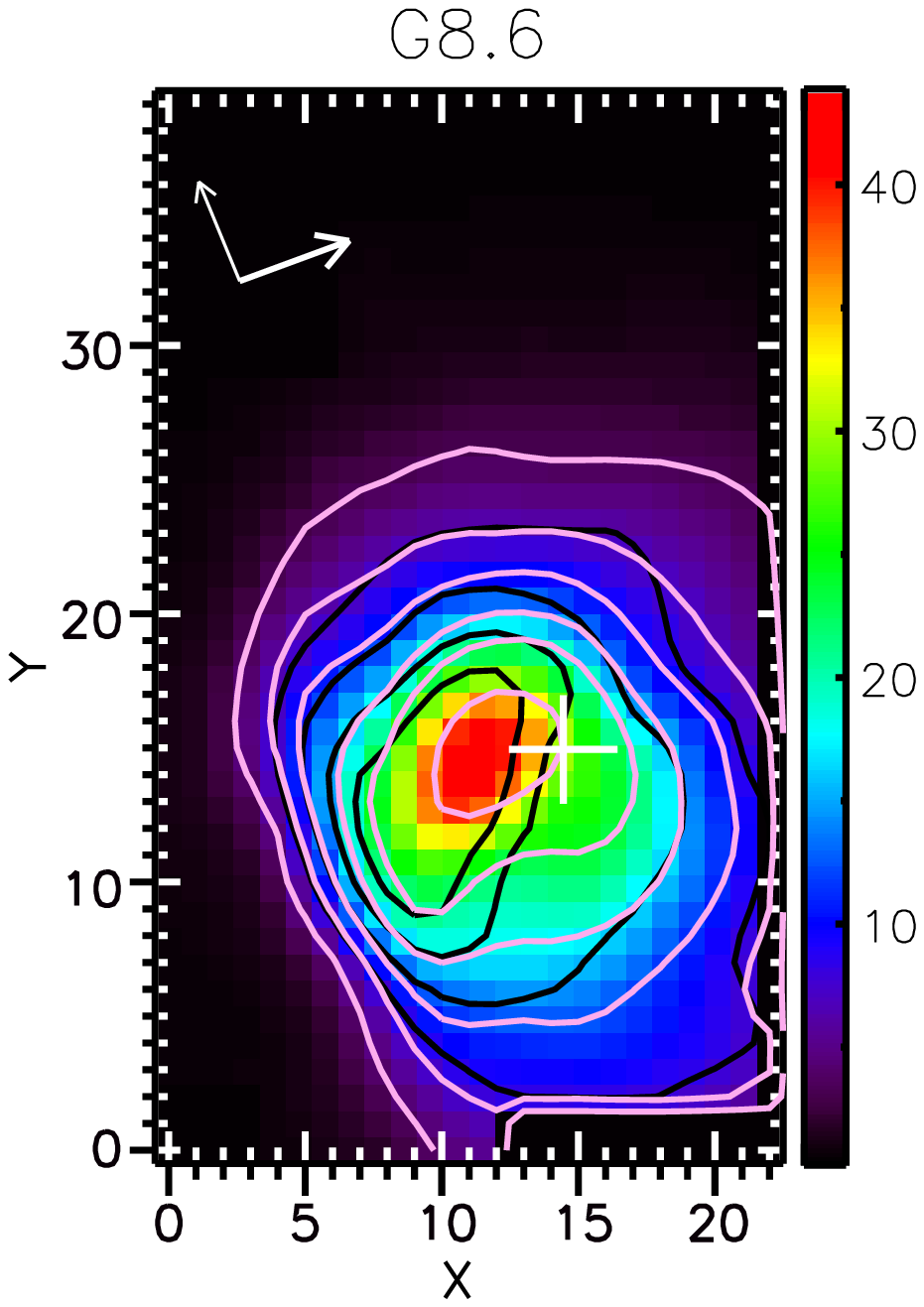}}
\end{center}
\caption{Spatial distribution of the Gaussian components in the 7--9~$\mu$m range of the IRS~SL data within the NGC~1333 SVS~3 region. Color bars of band intensities are given in units of 10$^{6}$~W~m$^{-2}$~sr$^{-1}$. Contours of the 11.2 and 7.7~$\mu$m emission are shown respectively in black and pink as in Figure~\ref{irs maps}. North and East are indicated by the thick and thin white arrows respectively in the upper left corner of each map. Axes are given in IRS~SL pixel units.}
\label{irsGS}
\end{figure*}

In Figure \ref{irsGS}, we present the spectral maps of the four Gaussian components in the 7--9~$\mu$m decomposition. Upon first inspection, it is evident that the G7.6 and G8.6~$\mu$m components show a remarkably similar morphology which is comparable to that of the major PAH bands in the 7--9~$\mu$m range. The spatial distribution of the G7.8 and G8.2~$\mu$m components are similar with an elongated shape in their peak emission as well as both having an extended emission `plateau' towards the western edge of the map. However, the peak intensity of the G7.8~$\mu$m component is more concentrated near the 7.7~$\mu$m emission peak while the G8.2~$\mu$m morphology better mimics that of the 11.2~$\mu$m band. 

\begin{figure*}
\begin{center}
\resizebox{\hsize}{!}{%
\includegraphics[clip,trim =0.cm 2.5cm .8cm 2cm,width=0.20\textwidth]{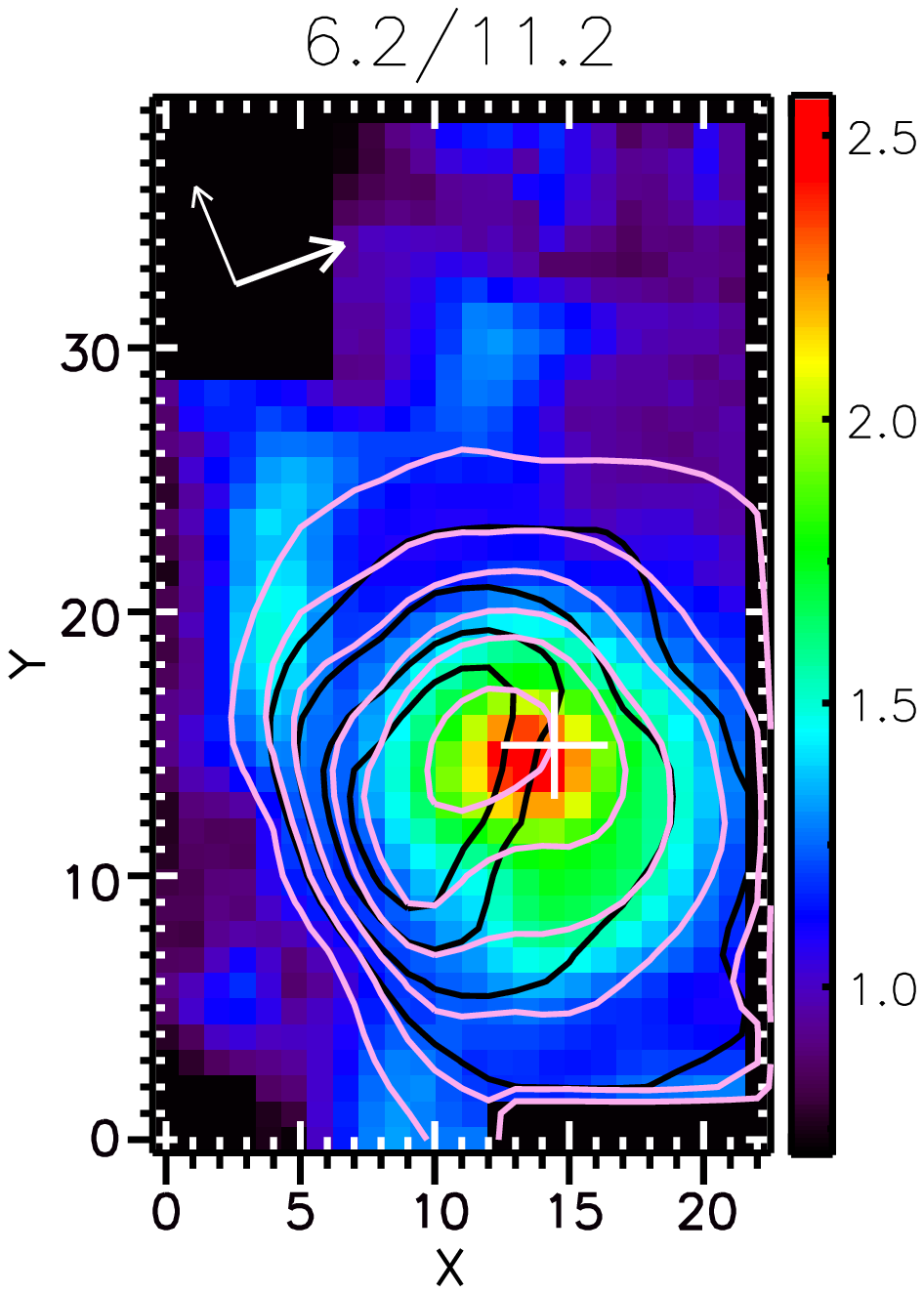}
\includegraphics[clip,trim =0.cm 2.5cm .8cm 2cm,width=0.20\textwidth]{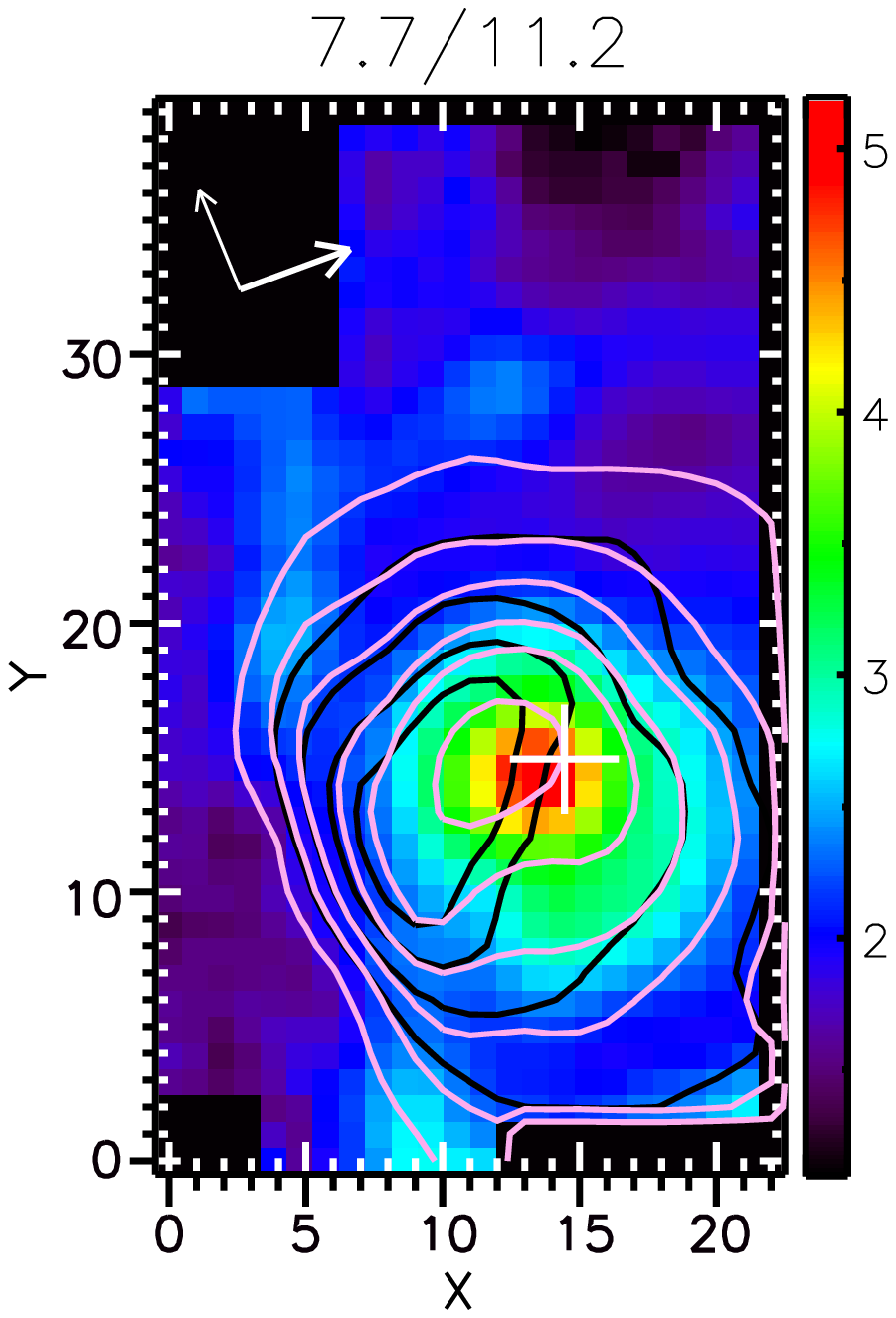}
\includegraphics[clip,trim =0.cm 2.5cm .8cm 2cm,width=0.20\textwidth]{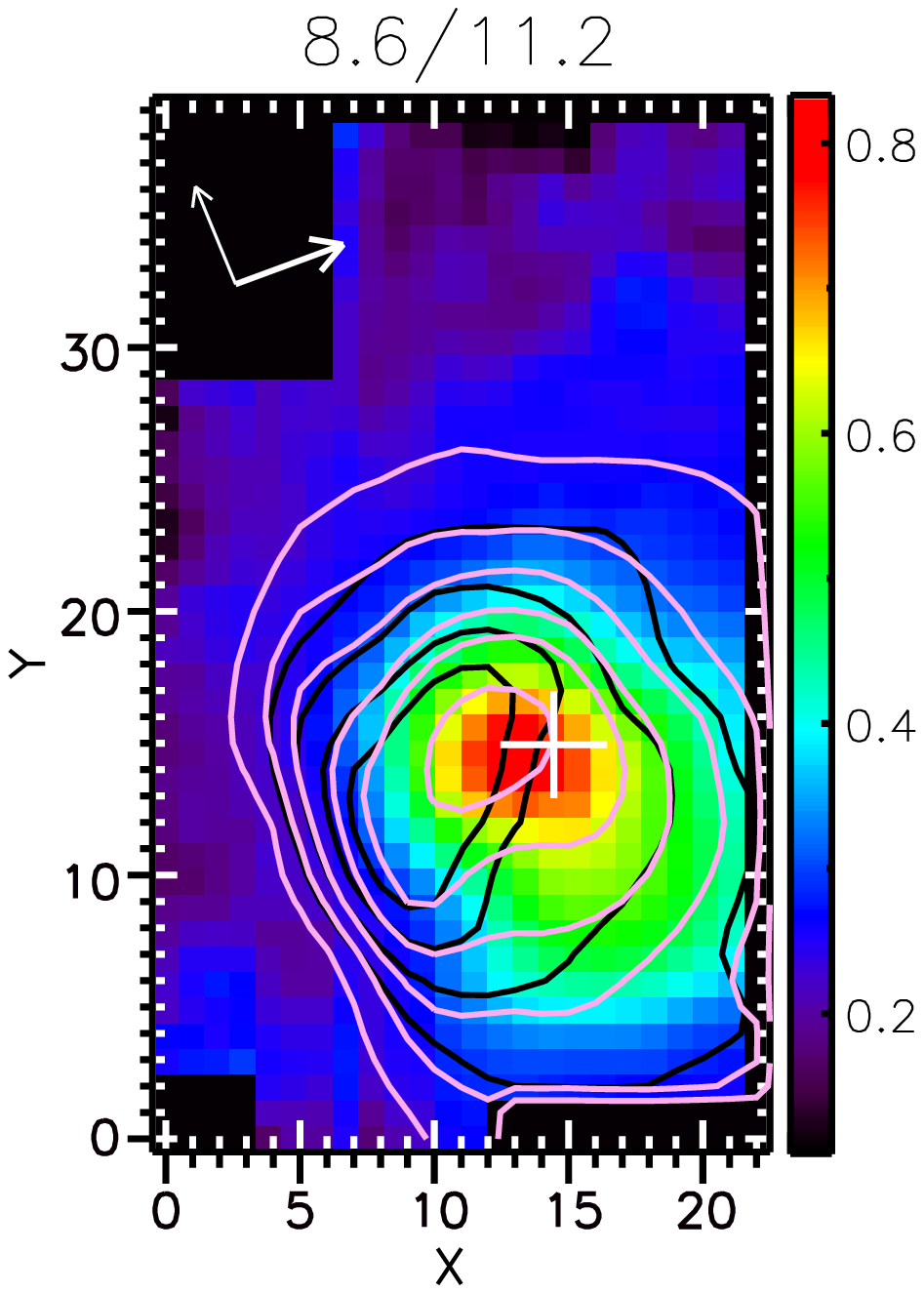}
\includegraphics[clip,trim =0.cm 2.5cm .8cm 2cm,width=0.20\textwidth]{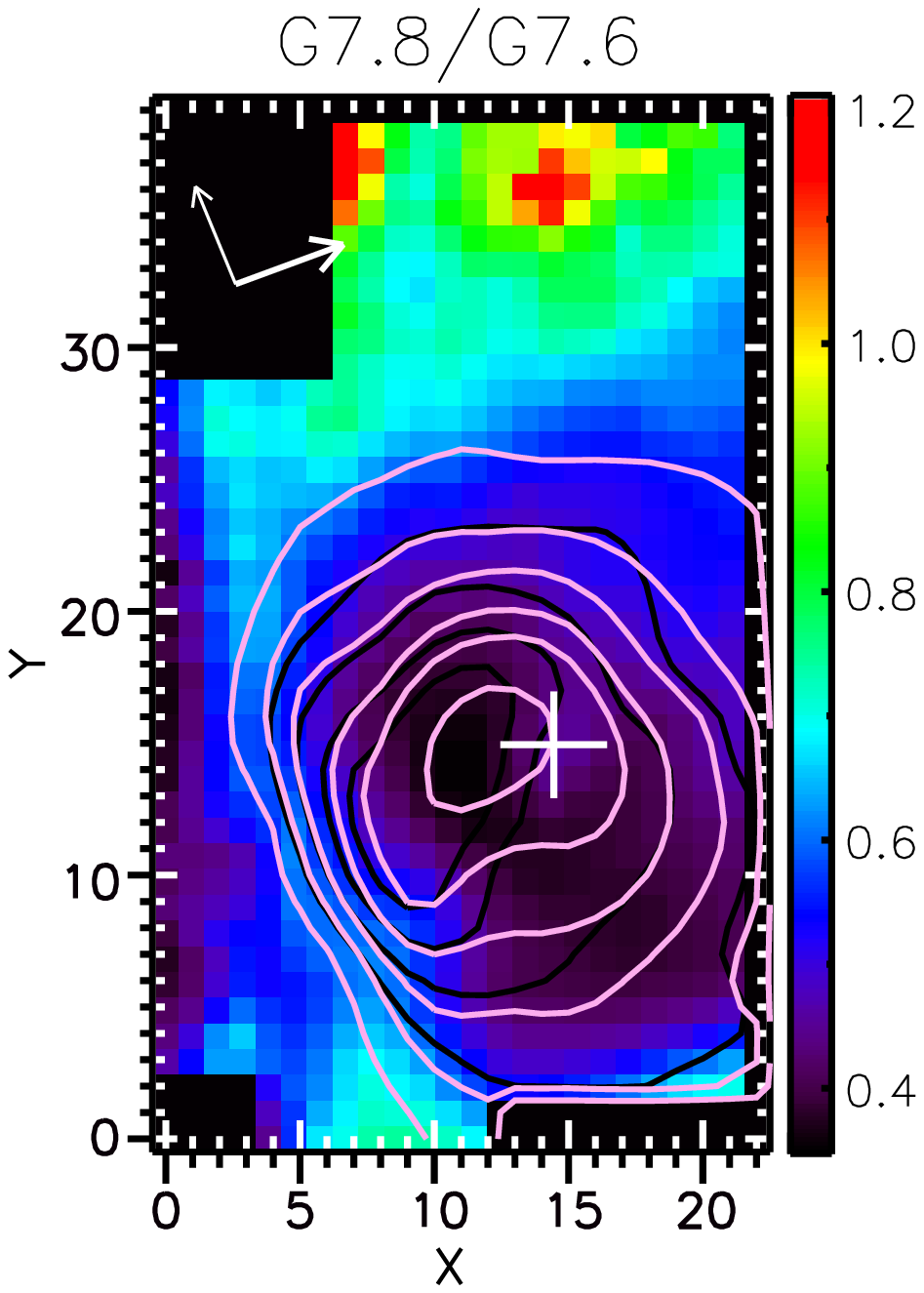}}
\end{center}
\caption{Spatial distribution of ratios of major PAH bands and Gaussian components in the 7--9~$\mu$m range of the IRS~SL data within the NGC~1333 SVS~3 region. Contours of the 11.2 and 7.7~$\mu$m emission are shown respectively in black and pink as in Figure~\ref{irs maps}. North and East are indicated by the thick and thin white arrows respectively in the upper left corner of each map. Axes are given in IRS~SL pixel units.}
\label{irs_ratios}
\end{figure*}

In Figure \ref{irs_ratios}, we compare the spatial distribution of PAH emission ratios that have previously been shown to be relevant in PDR studies \citep[e.g.][]{gal08,boe13,sto17}. First, we consider the ratio of each of the respective major PAH bands in the 6--9~$\mu$m to the 11.2~$\mu$m. In each of these maps, we see a circular peak very close to the position of the star which are more concentrated and symmetric compared to the 7.7~$\mu$m emission peak archetype seen above. 
In each case, the overall morphology deviates from circular symmetry with enhanced emission towards the west, i.e. north of the 11.2~$\mu$m peak and in the direction of the YSO located at the bottom of the map. Furthermore, the spatial distribution of the 8.6/11.2 emission is more extended compared to the 6.2/11.2 and 7.7/11.2 morphology. We note two long filaments (vertical and diagonally in pixel coordinates respectively) in the 6.2/11.2 and 7.7/11.2 ratios that extends into the eastern reaches of these maps.

The spatial distribution of the G7.8/G7.6 ratio exhibits a minimum throughout much of the nebula relative to the diffuse outskirts where an appreciable increase is found in this ratio. We note maxima located in the eastern edge of this map where PAH emission is essentially at its lowest\footnote{We note that the 7--9~$\mu$m decomposition had generally much poorer fits to the 7.7~$\mu$m peak in these eastern pixels. This is attributed to a weak bump at $\sim$~7.2~$\mu$m which the G7.6~$\mu$m component tries to account for in the Gaussian decomposition fit resulting in an underestimation of the peak in the 7.7~$\mu$m band. Consequently we find that these pixels correspond to  G7.8/G7.6~$>$~1.0, thus we consider these pixels to be unreliable in tracing this ratio.}.

\subsubsection{IRS Correlation Plots}
\label{slcorr}

\begin{figure*} 
\begin{center}
\resizebox{\hsize}{!}{%
\includegraphics[clip,trim =.5cm 0cm .5cm 1cm,width=0.14\textwidth]{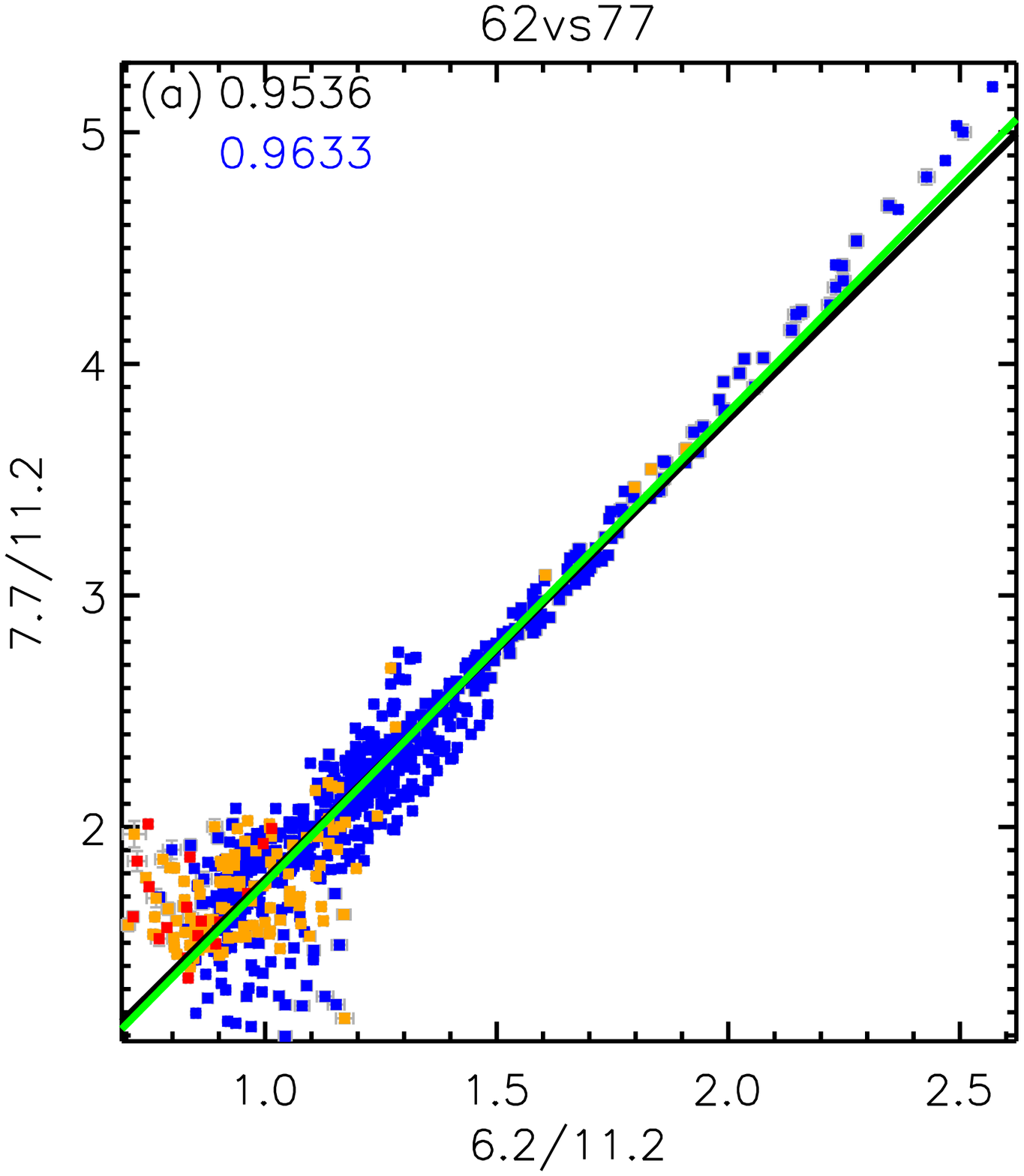}
\includegraphics[clip,trim =.5cm 0cm .5cm 1cm,width=0.14\textwidth]{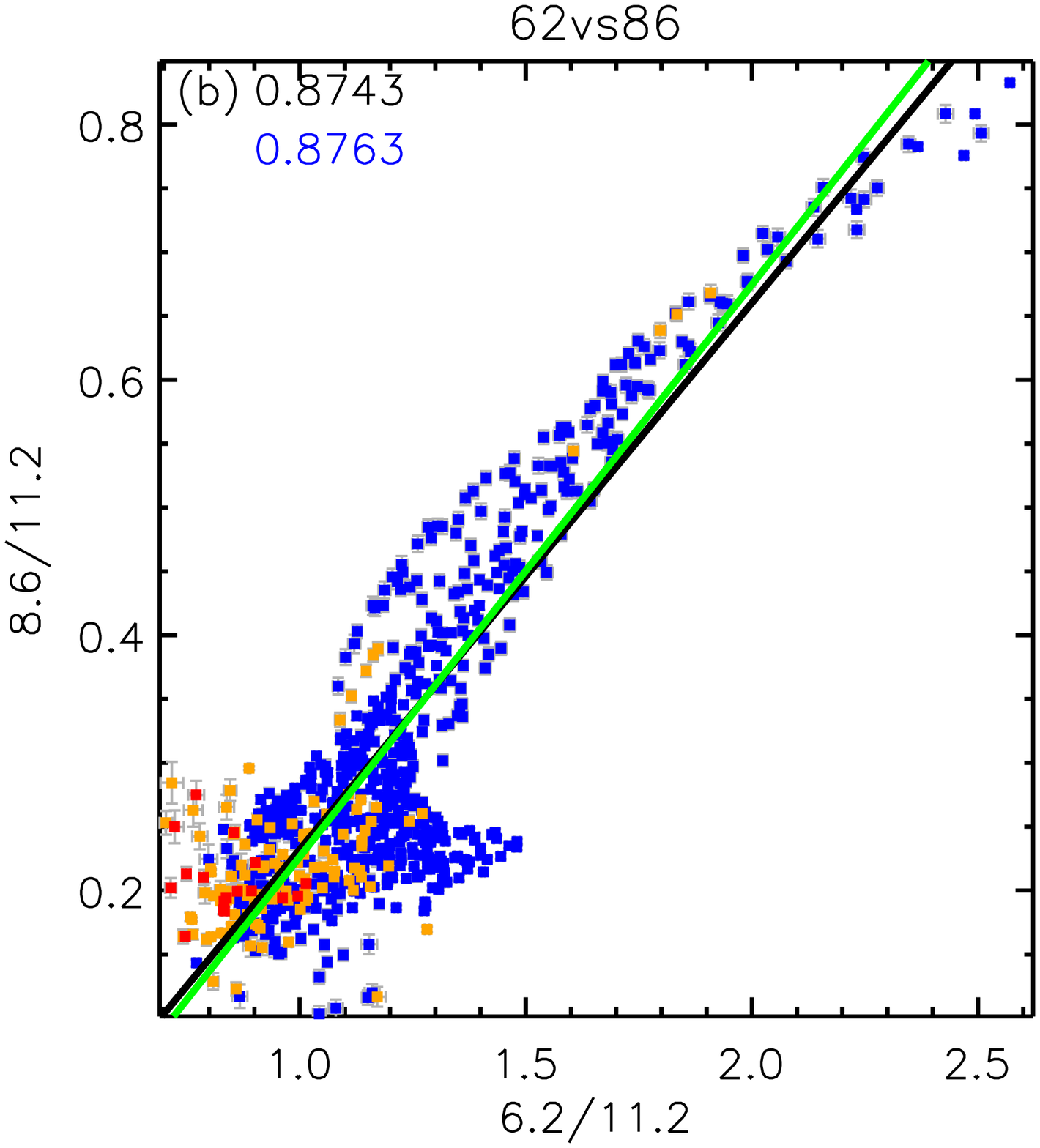}
\includegraphics[clip,trim =.5cm 0cm .5cm 1cm,width=0.14\textwidth]{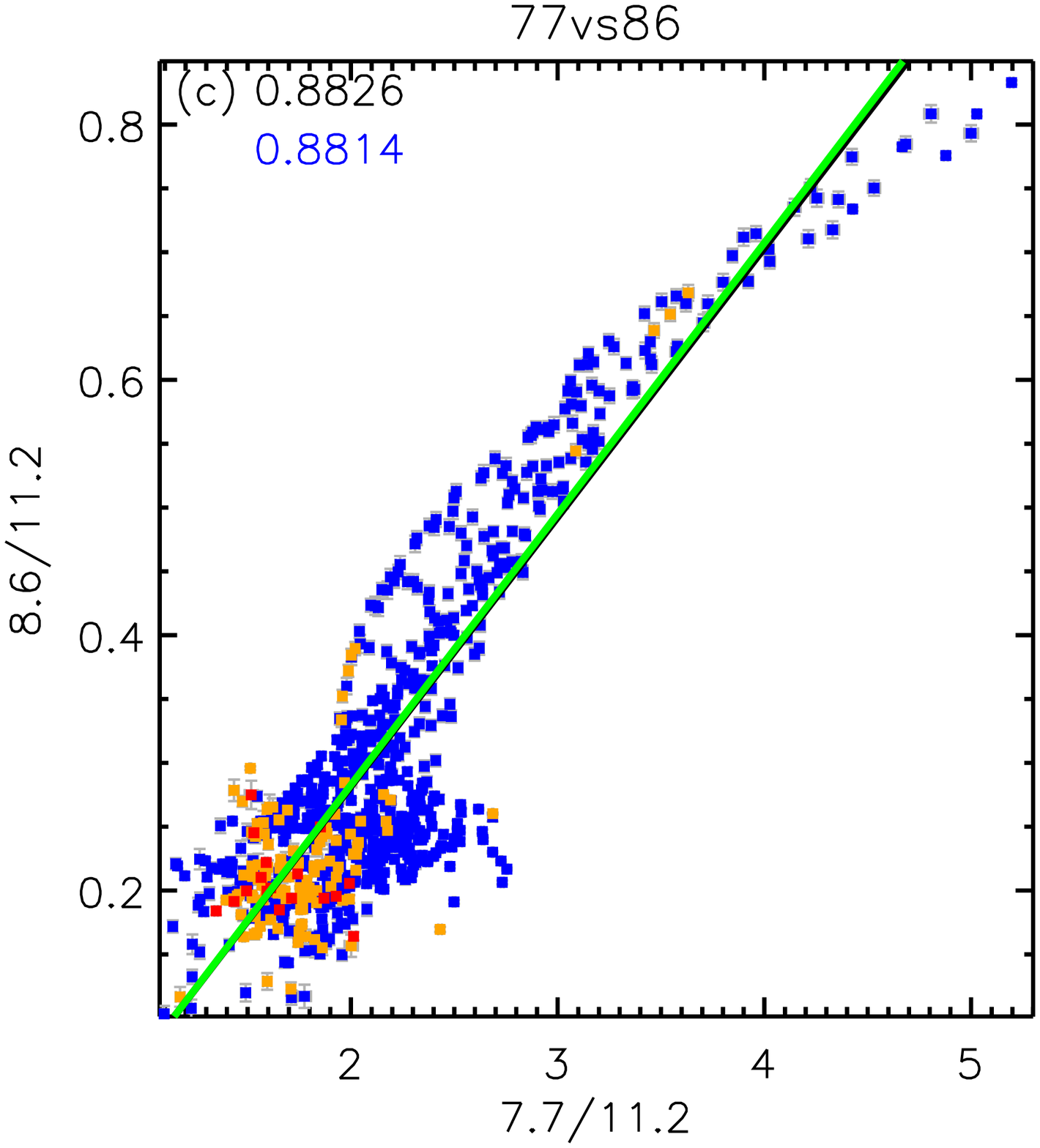}
\includegraphics[clip,trim =.5cm 0cm .5cm 1cm,width=0.14\textwidth]{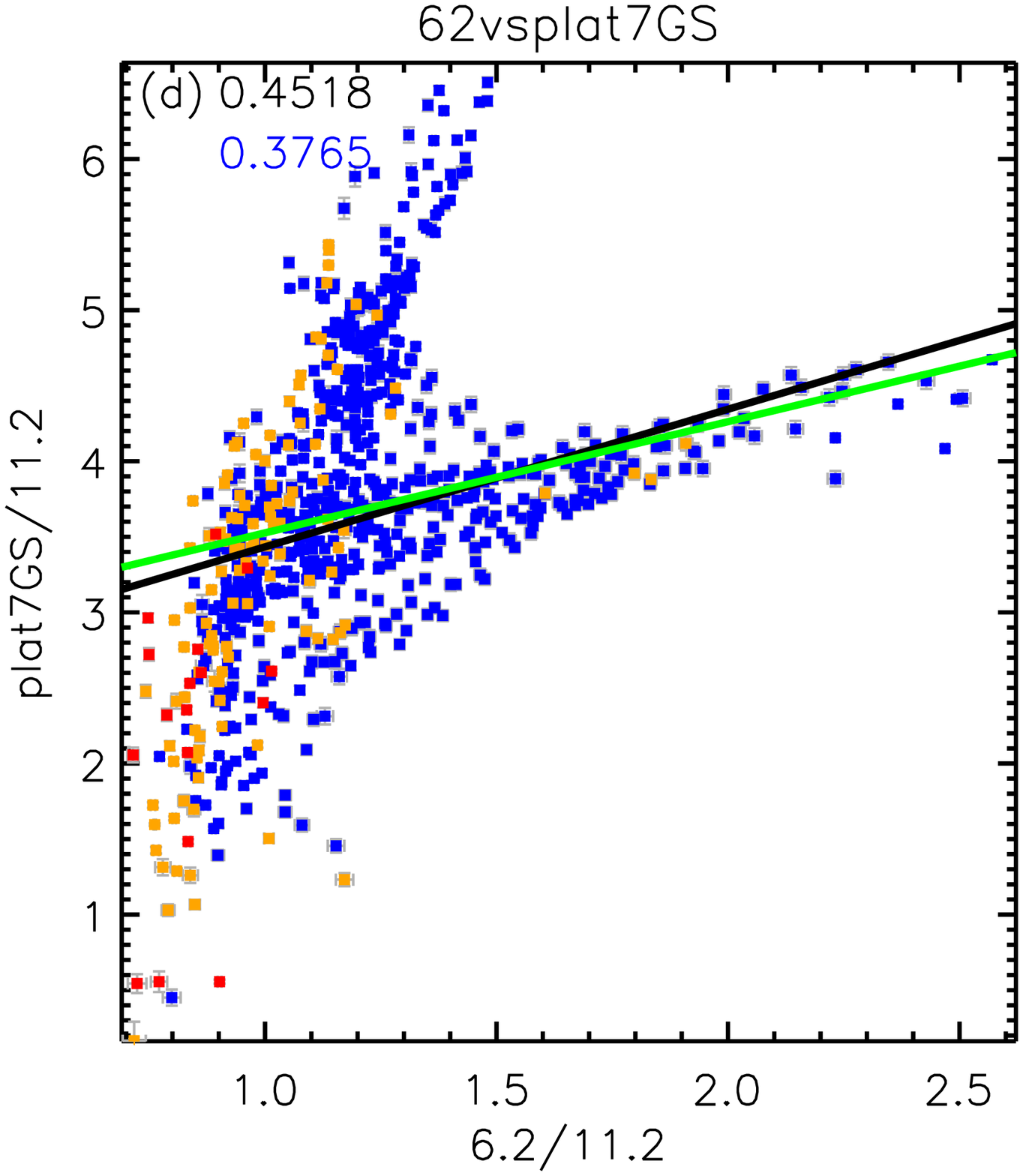}}
\resizebox{\hsize}{!}{%
\includegraphics[clip,trim =.5cm 0cm .5cm 1cm,width=0.14\textwidth]{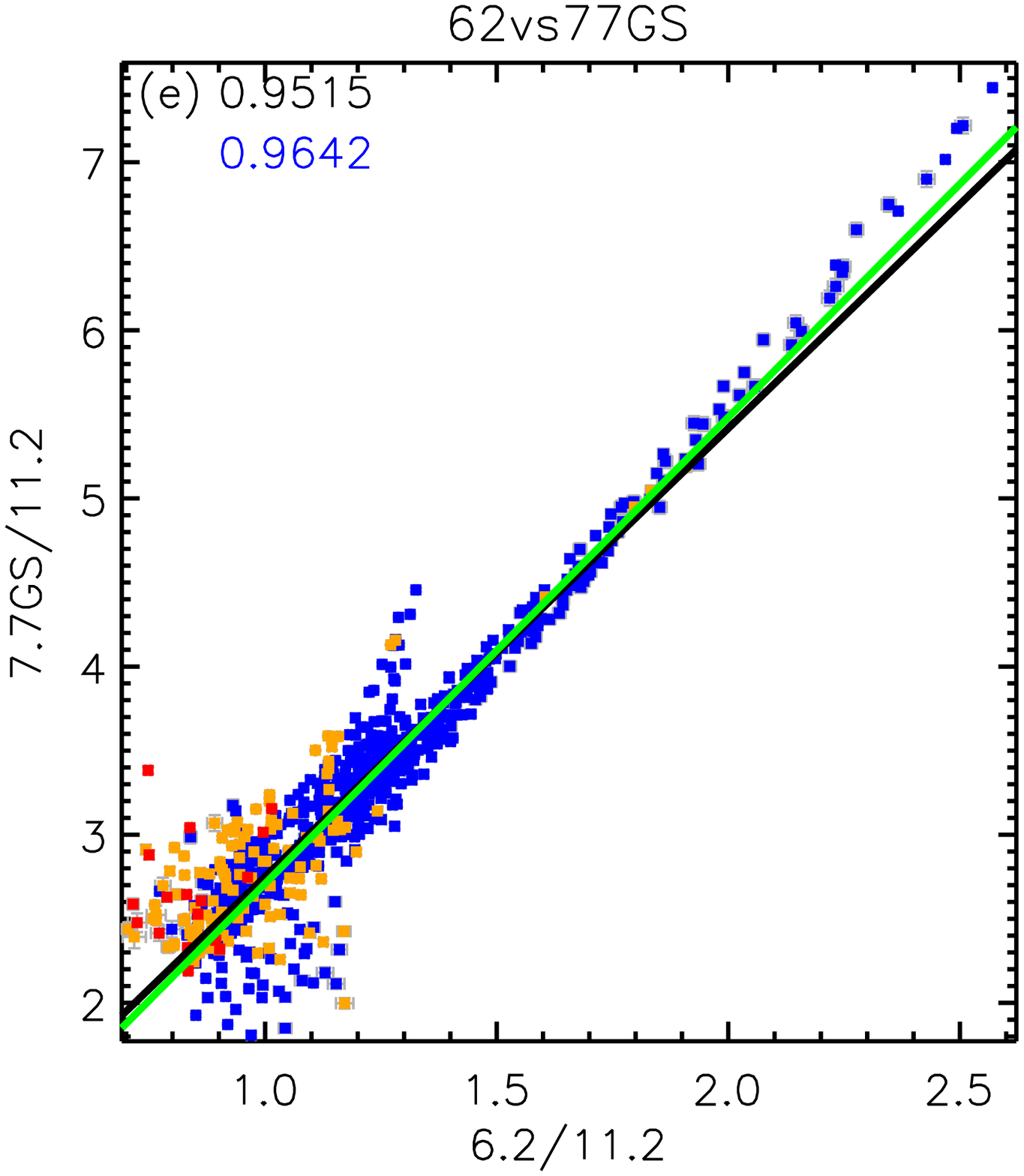}
\includegraphics[clip,trim =.5cm 0cm .5cm 1cm,width=0.14\textwidth]{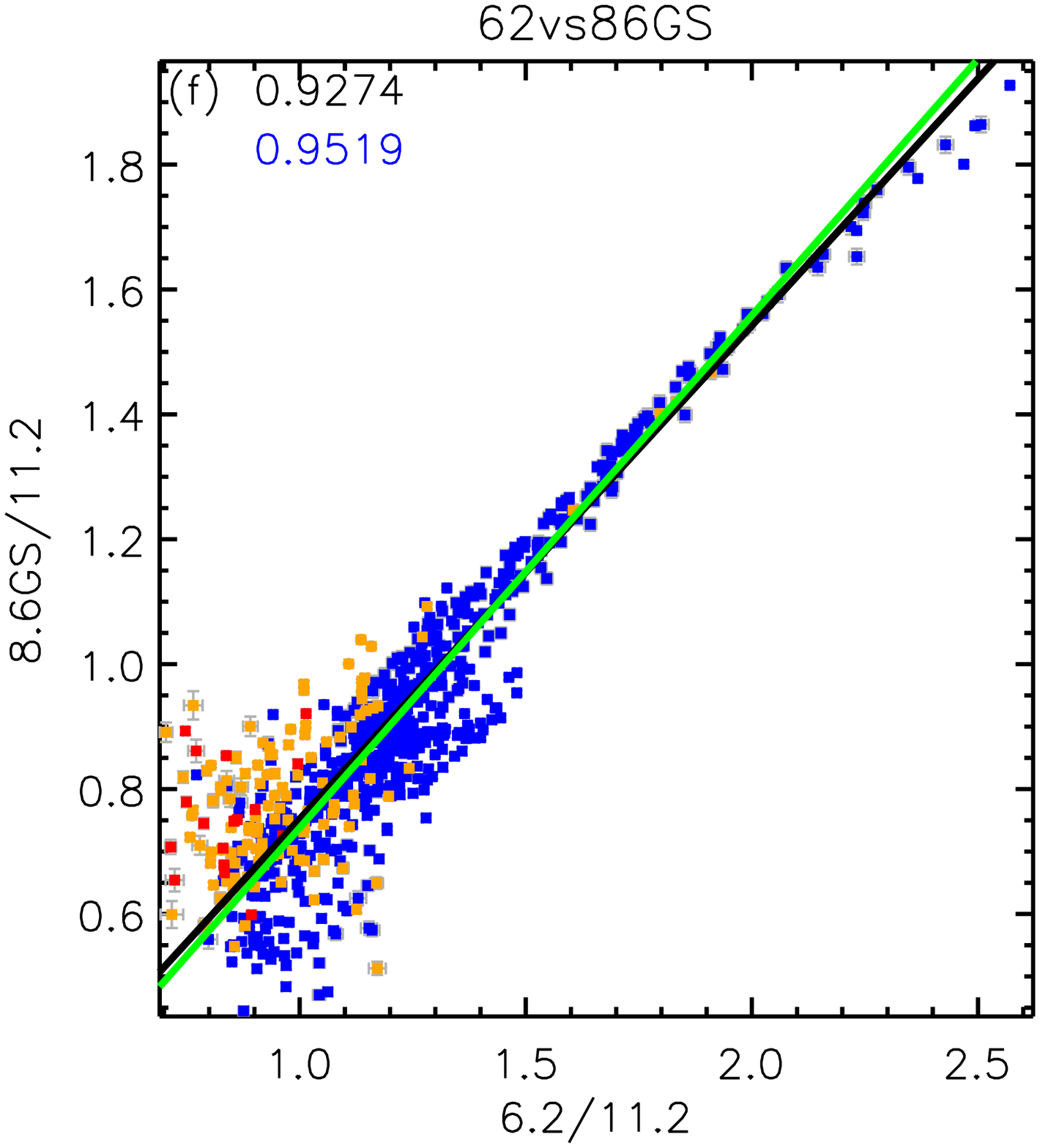}
\includegraphics[clip,trim =.5cm 0cm .5cm 1cm,width=0.14\textwidth]{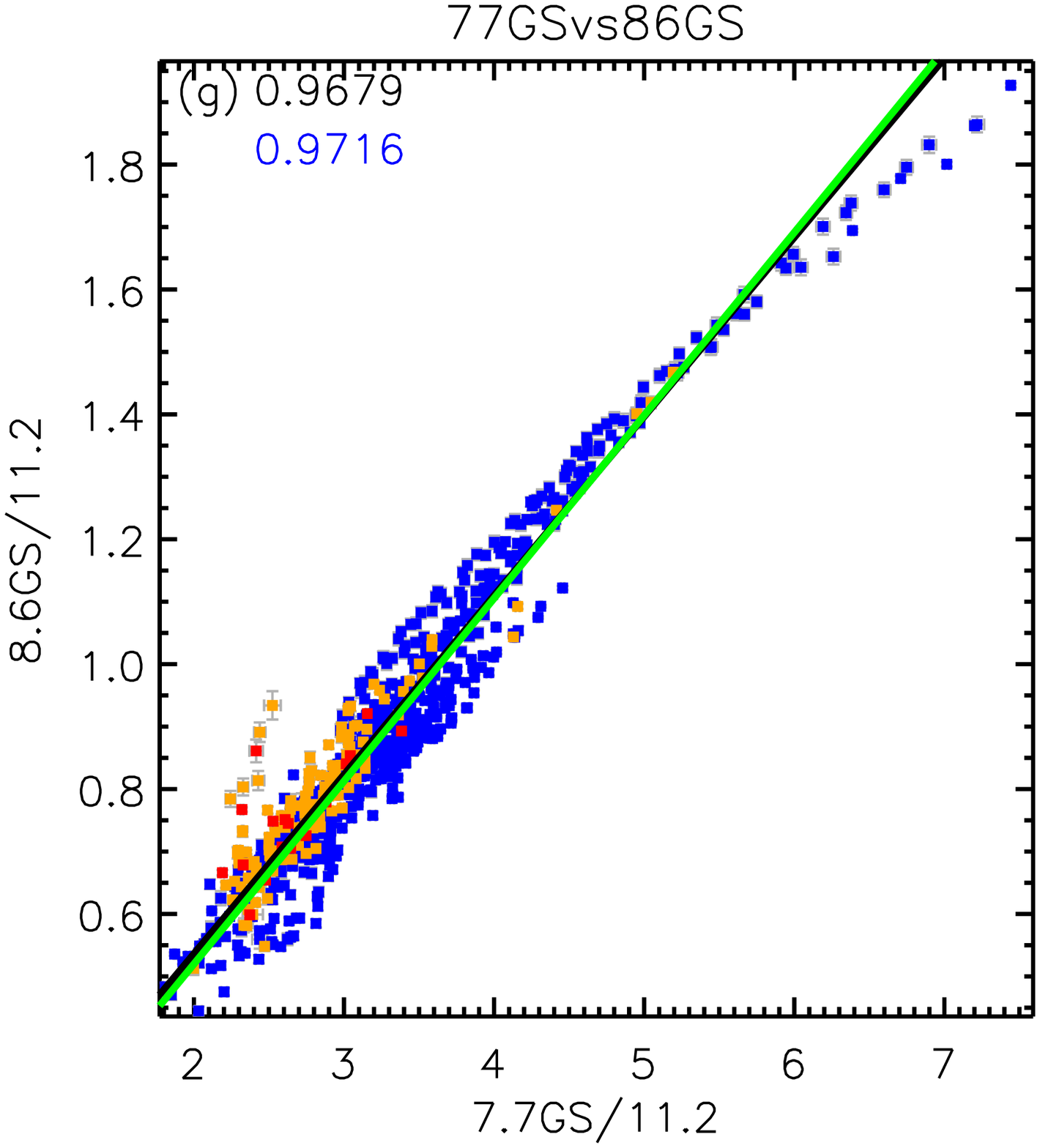}
\includegraphics[clip,trim =.5cm 0cm .5cm 1cm,width=0.14\textwidth]{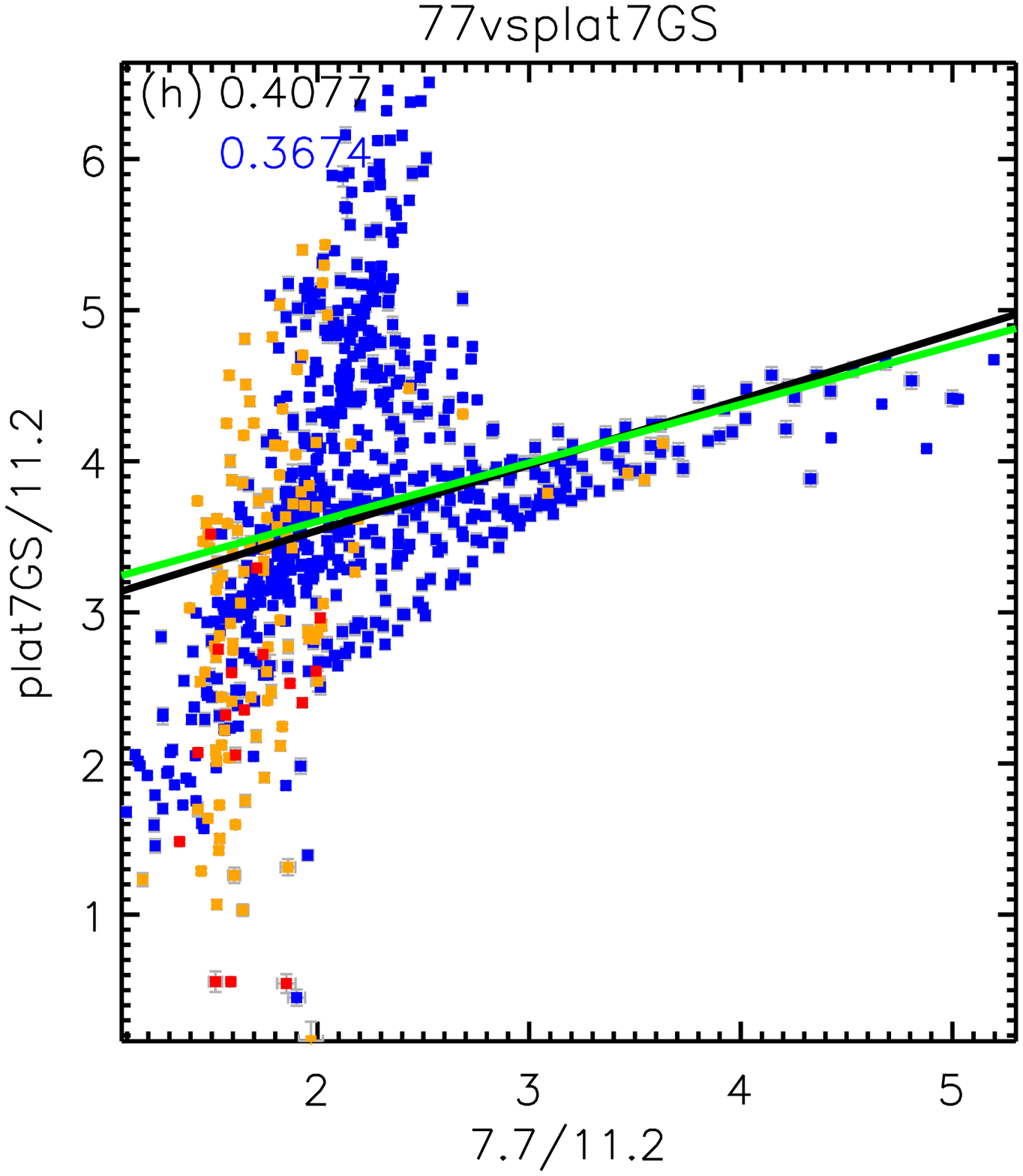}}
\resizebox{\hsize}{!}{%
\includegraphics[clip,trim =.5cm 0cm .5cm 1cm,width=0.14\textwidth]{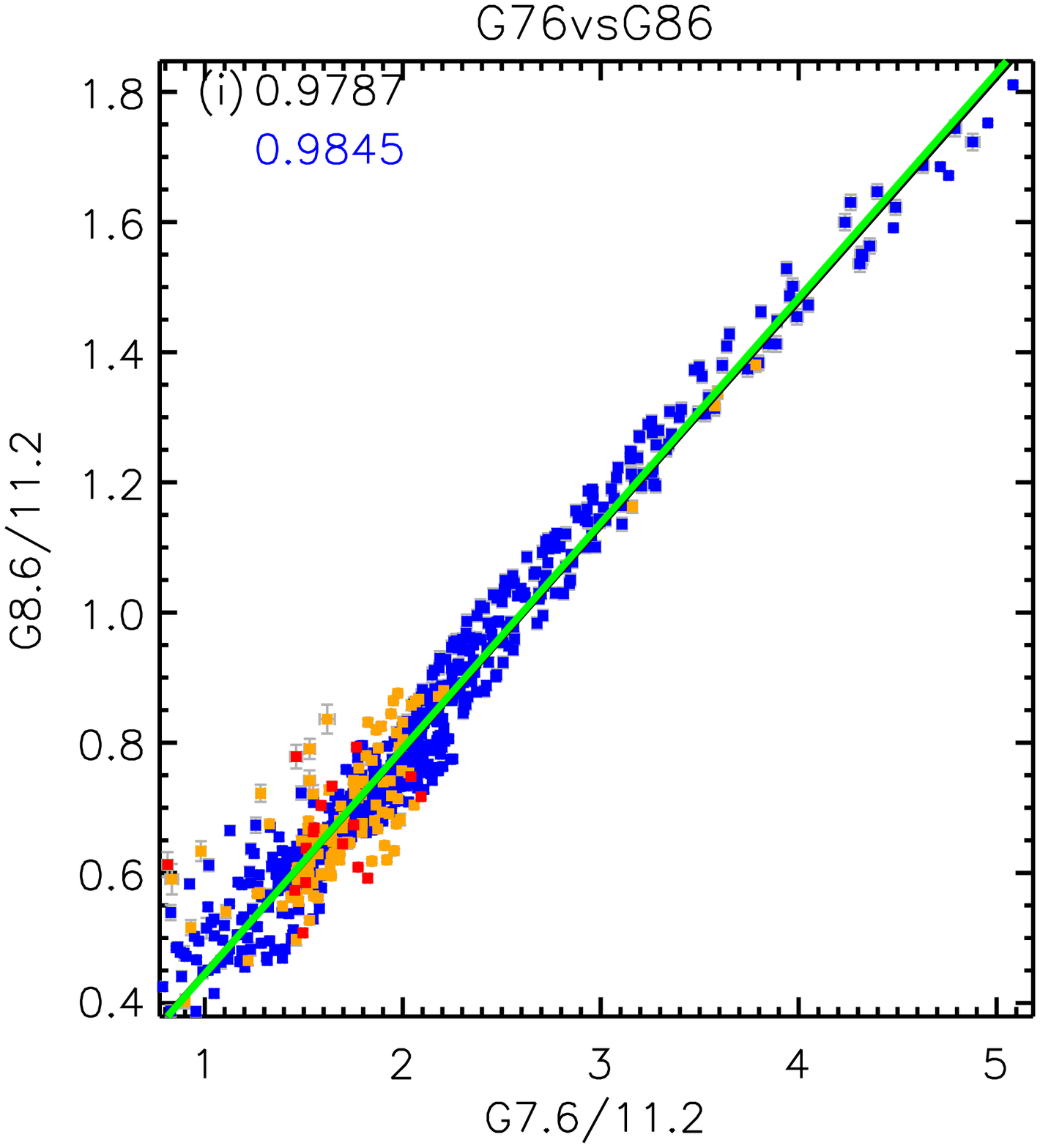}
\includegraphics[clip,trim =.5cm 0cm .5cm 1cm,width=0.14\textwidth]{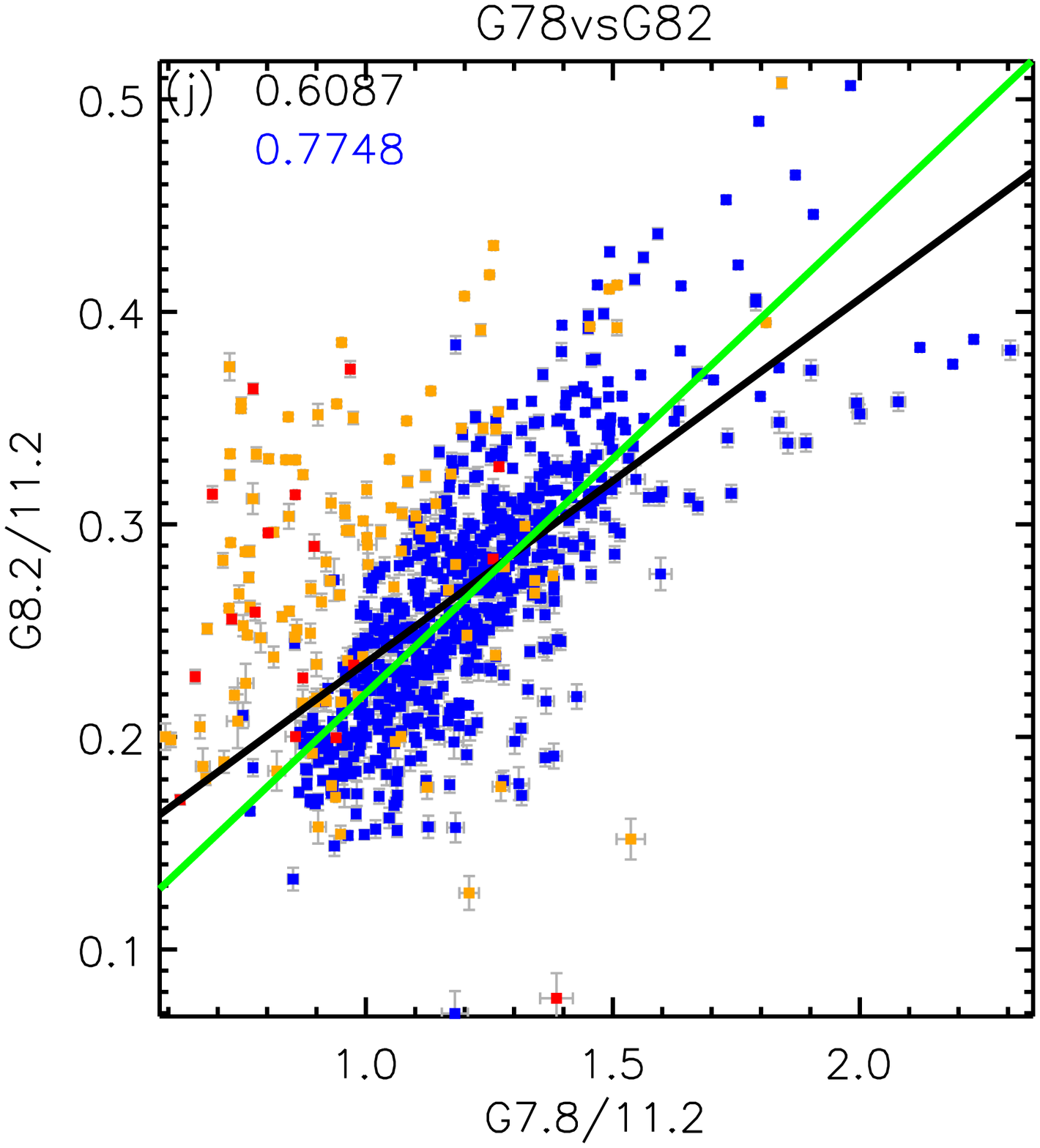}
\includegraphics[clip,trim =.5cm 0cm .5cm 1cm,width=0.14\textwidth]{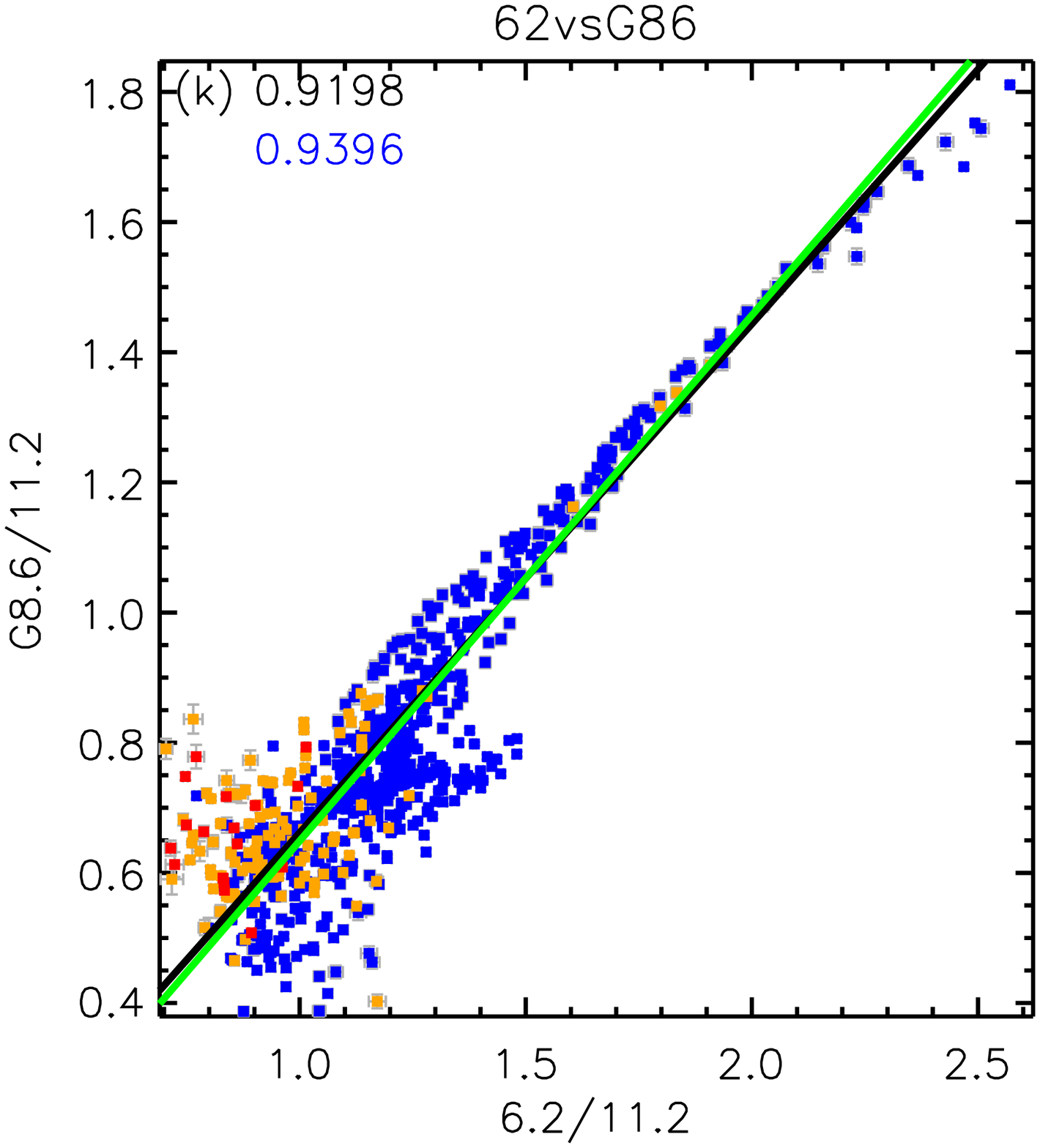}
\includegraphics[clip,trim =.5cm 0cm .5cm 1cm,width=0.14\textwidth]{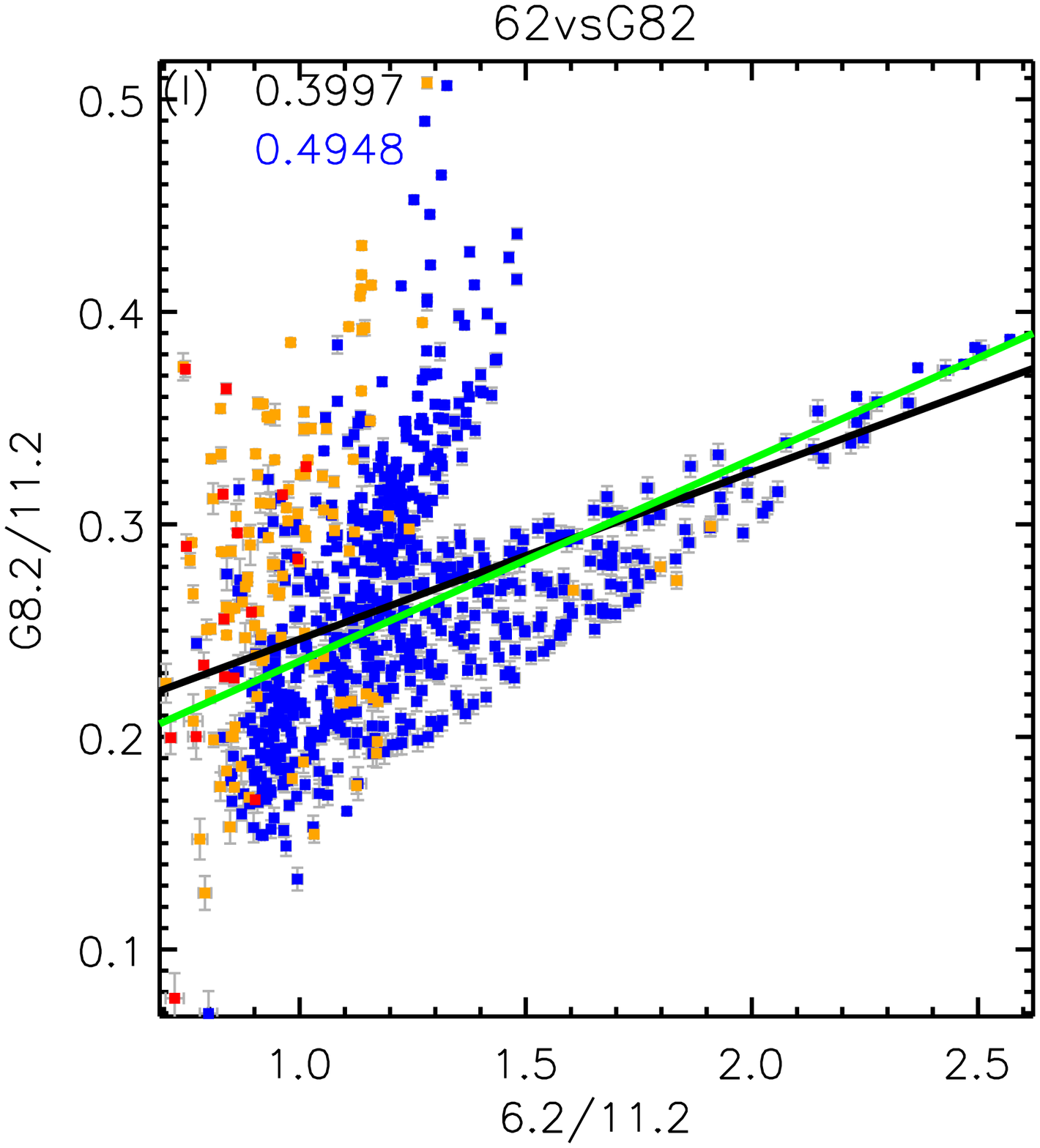}}
\resizebox{\hsize}{!}{%
\includegraphics[clip,trim =.5cm 0cm .5cm 1cm,width=0.14\textwidth]{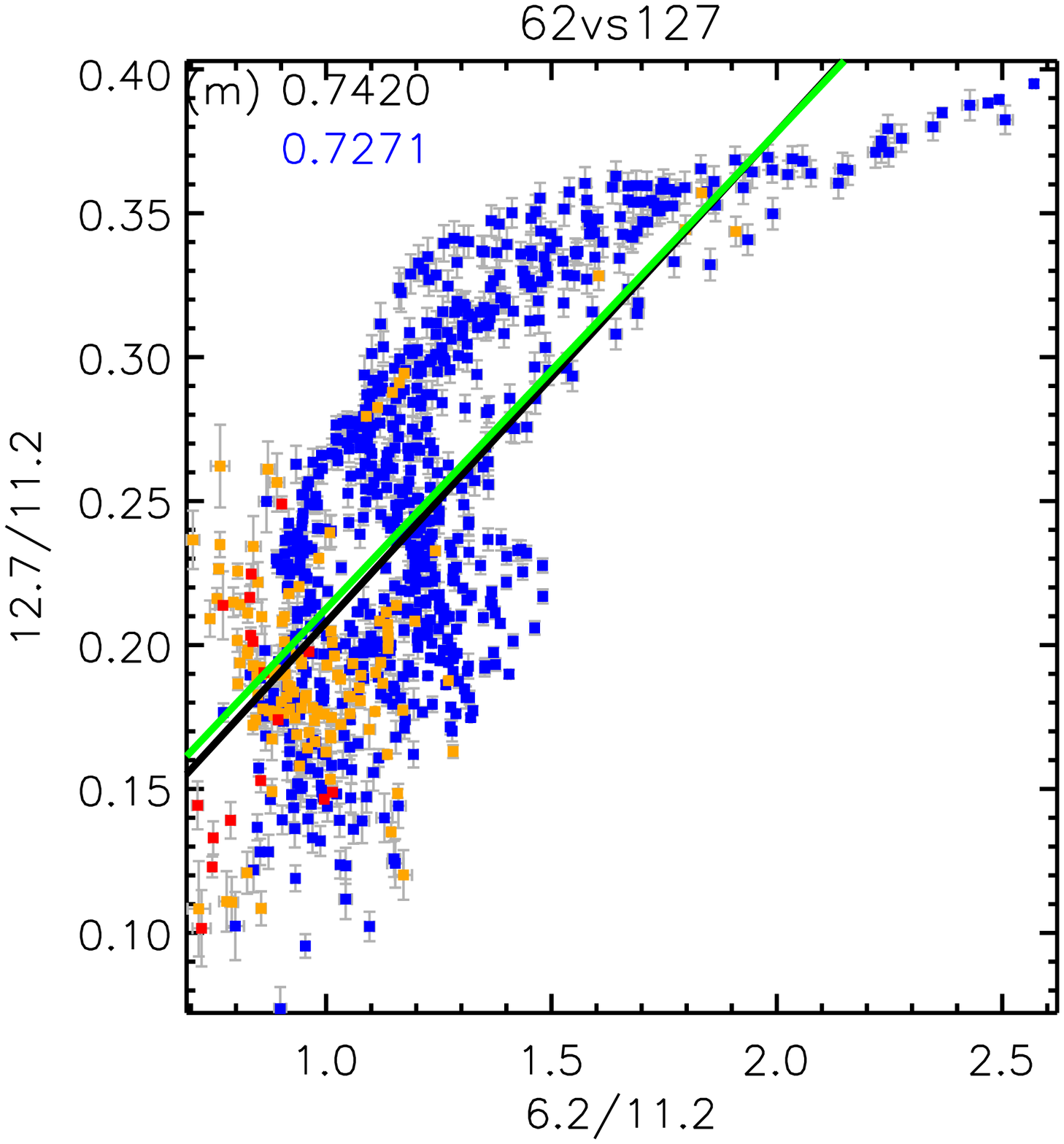}
\includegraphics[clip,trim =.5cm 0cm .5cm 1cm,width=0.14\textwidth]{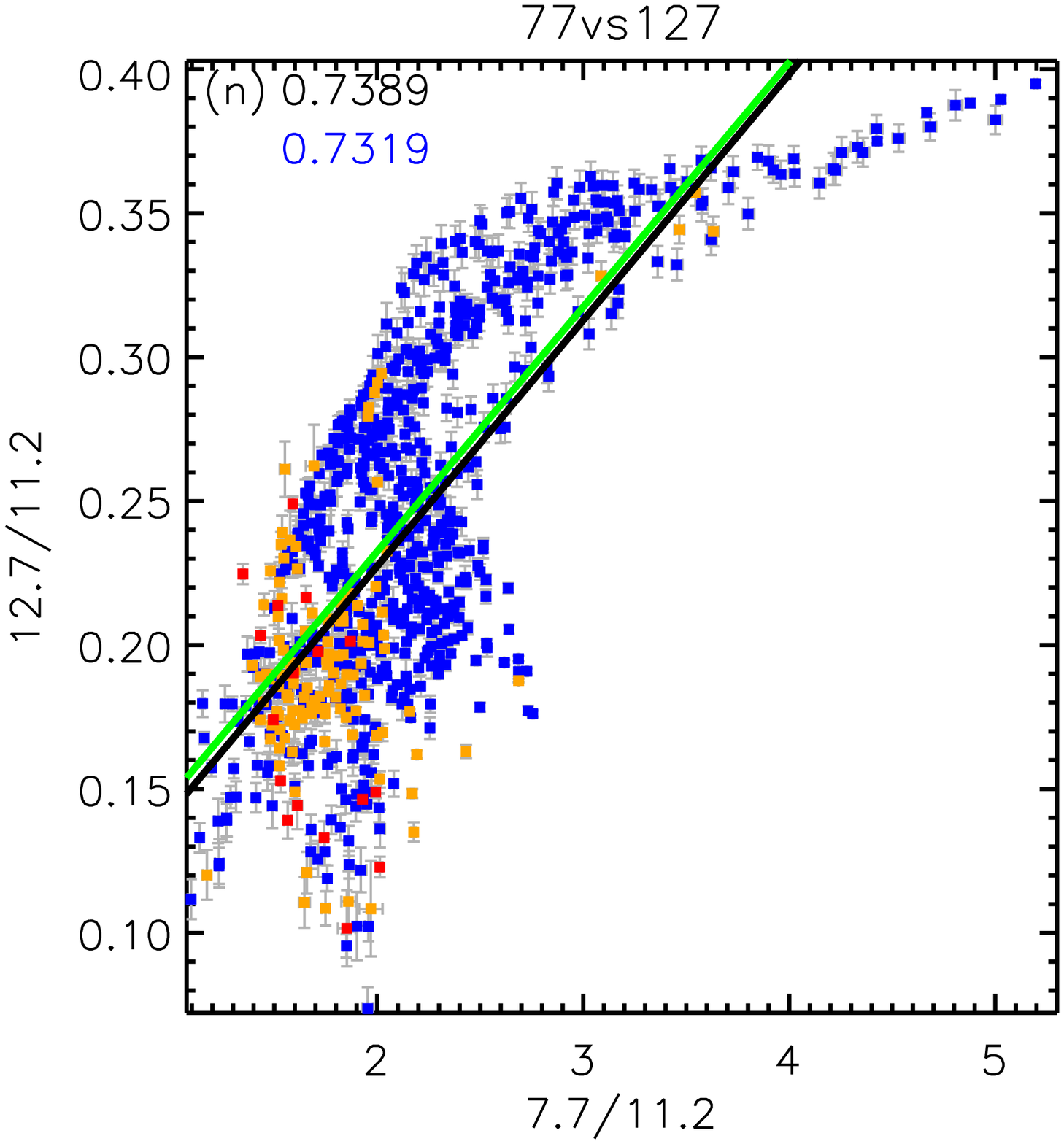}
\includegraphics[clip,trim =.5cm 0cm .5cm 1cm,width=0.14\textwidth]{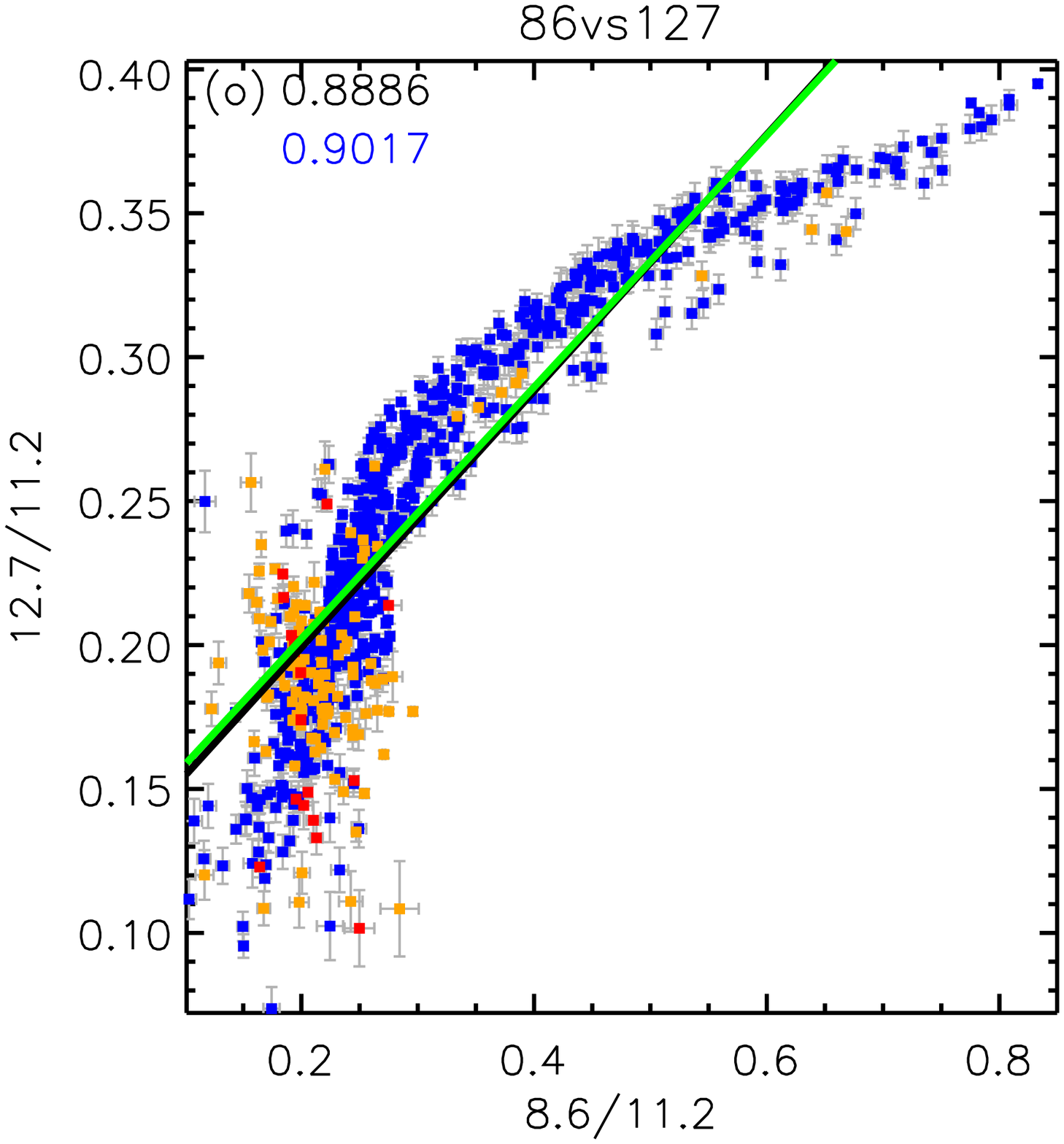}
\includegraphics[clip,trim =.5cm 0cm .5cm 1cm,width=0.14\textwidth]{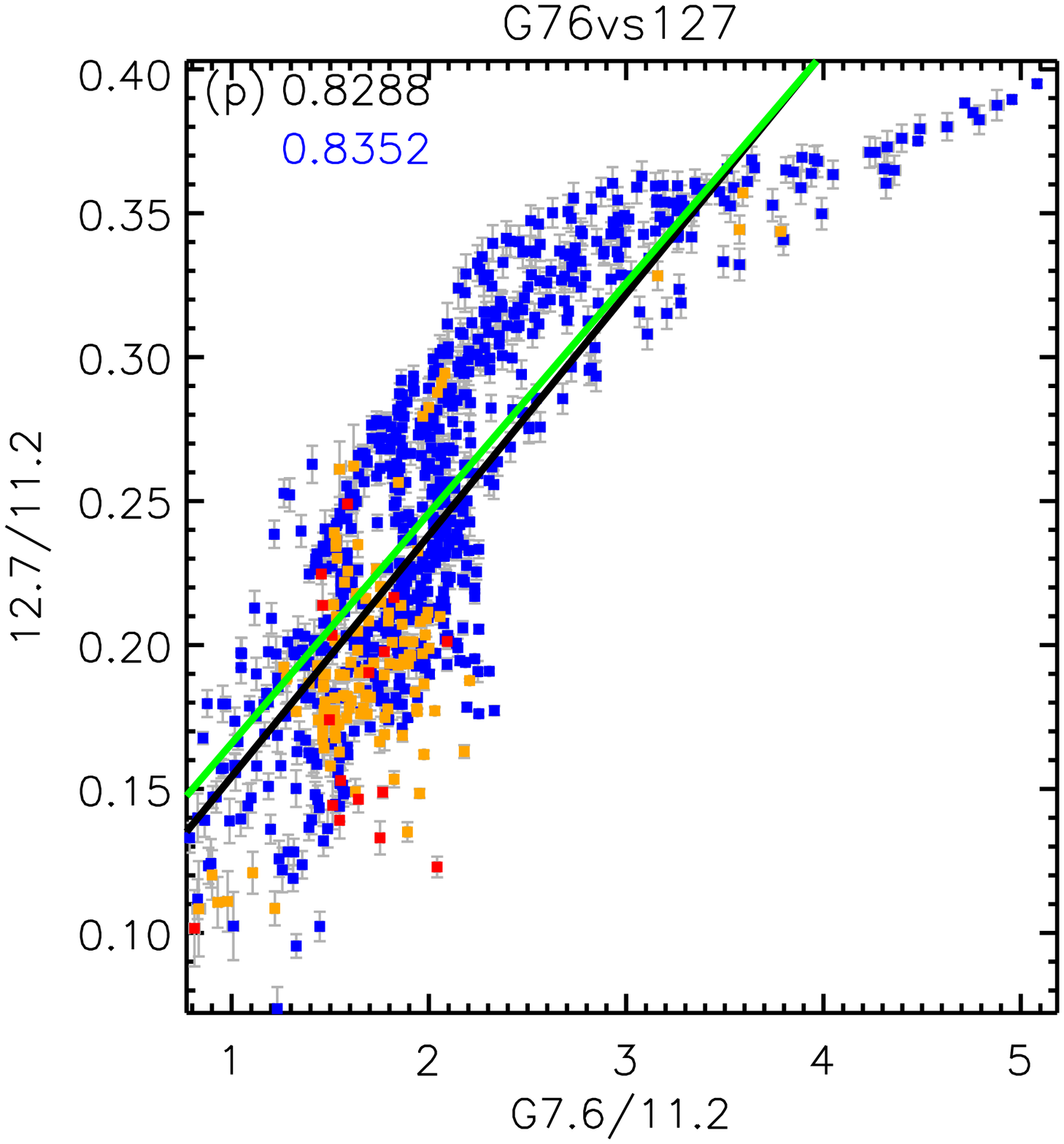}}
\resizebox{\hsize}{!}{%
\includegraphics[clip,trim =.5cm 0cm .5cm 1cm,width=0.14\textwidth]{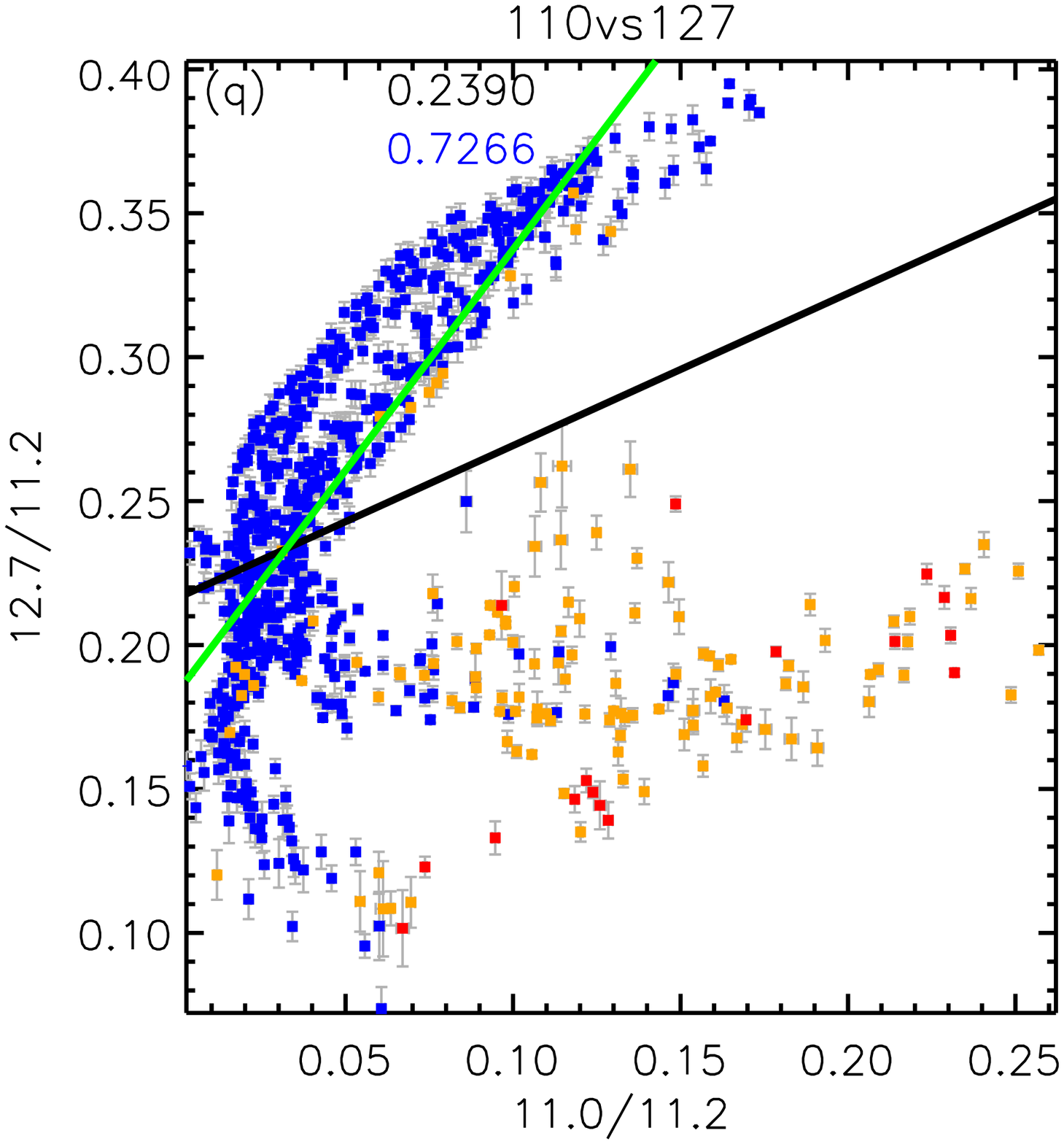}
\includegraphics[clip,trim =.5cm 0cm .5cm 1cm,width=0.14\textwidth]{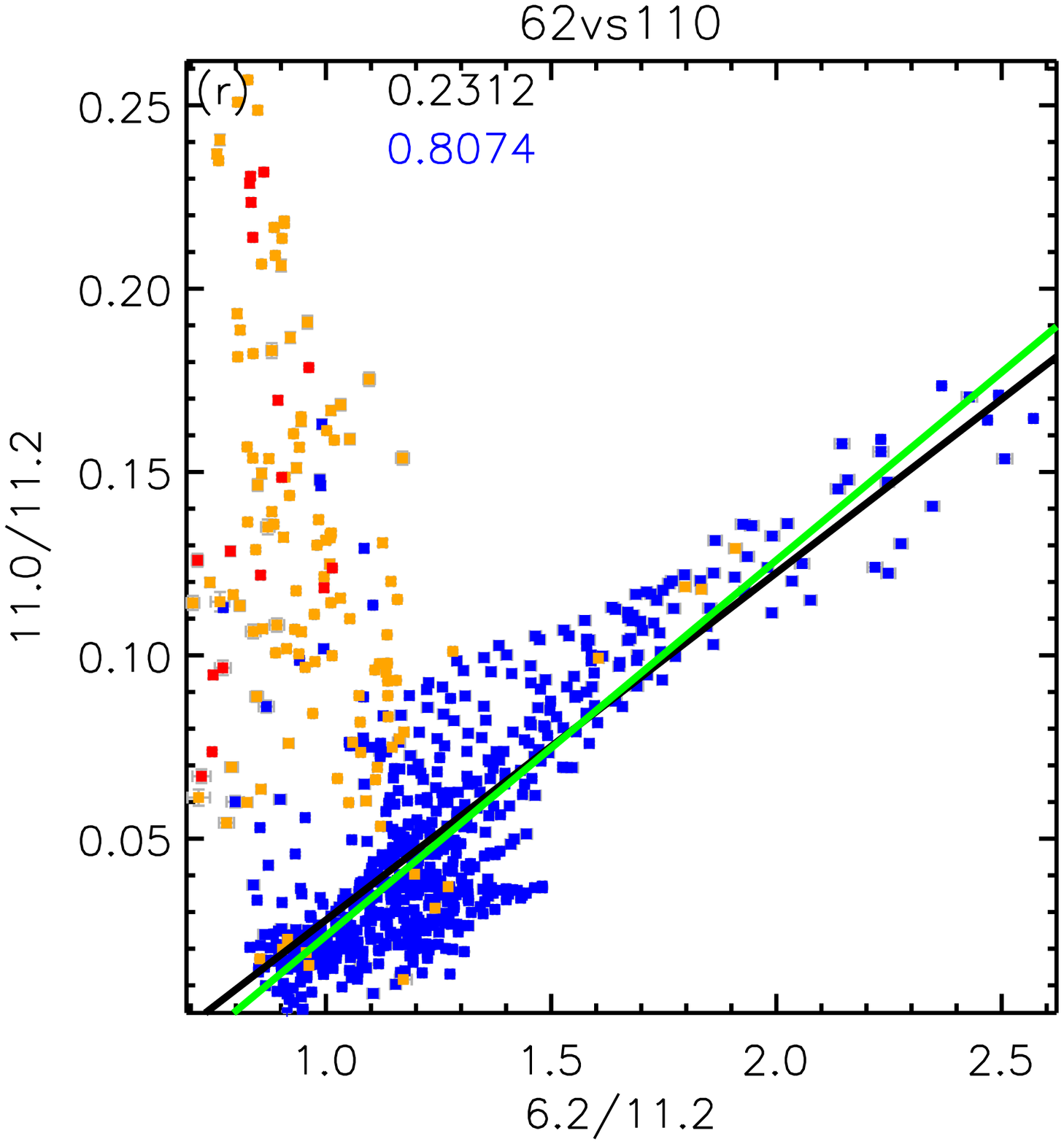}
\includegraphics[clip,trim =.5cm 0cm .5cm 1cm,width=0.14\textwidth]{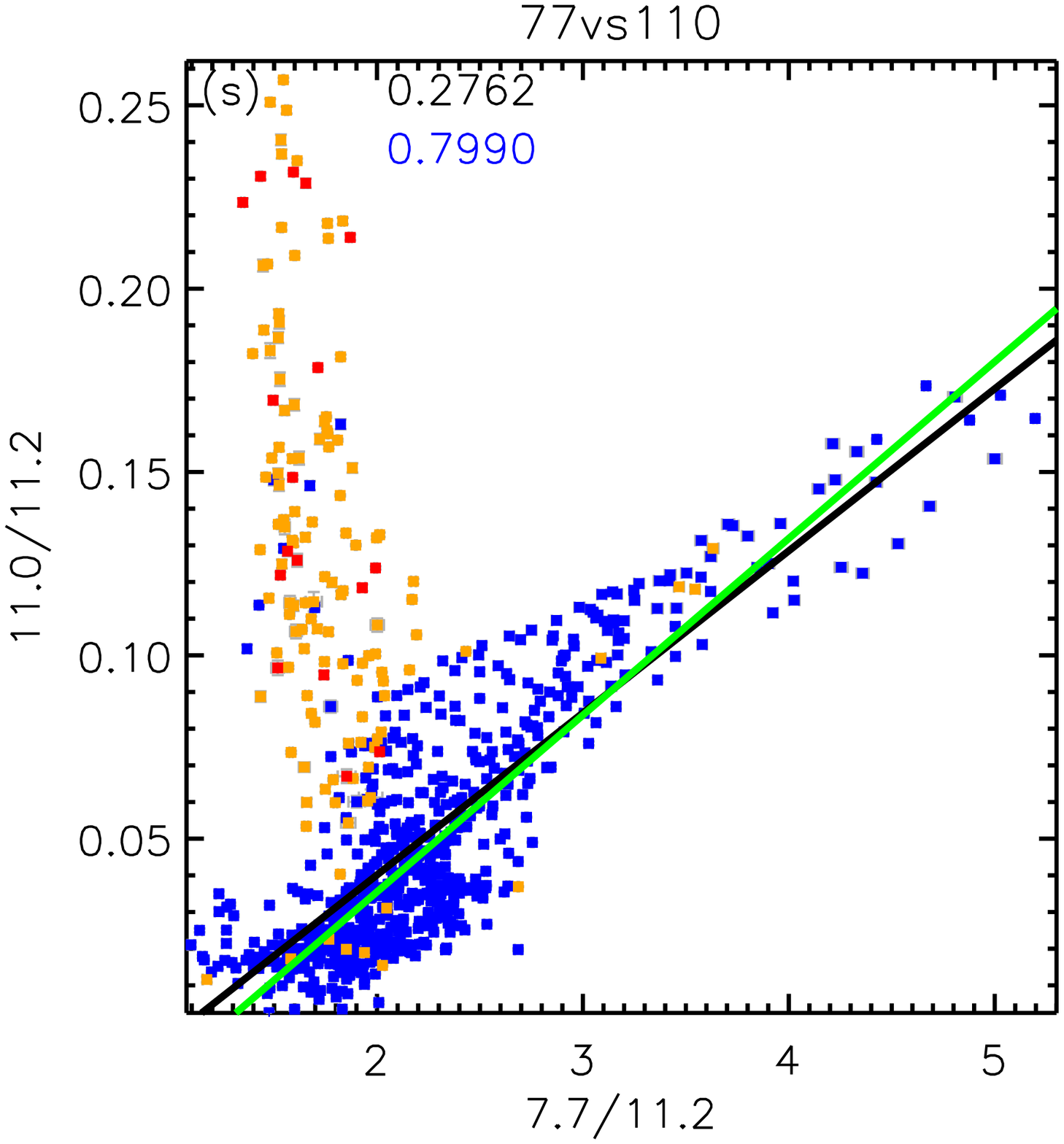}
\includegraphics[clip,trim =.5cm 0cm .5cm 1cm,width=0.14\textwidth]{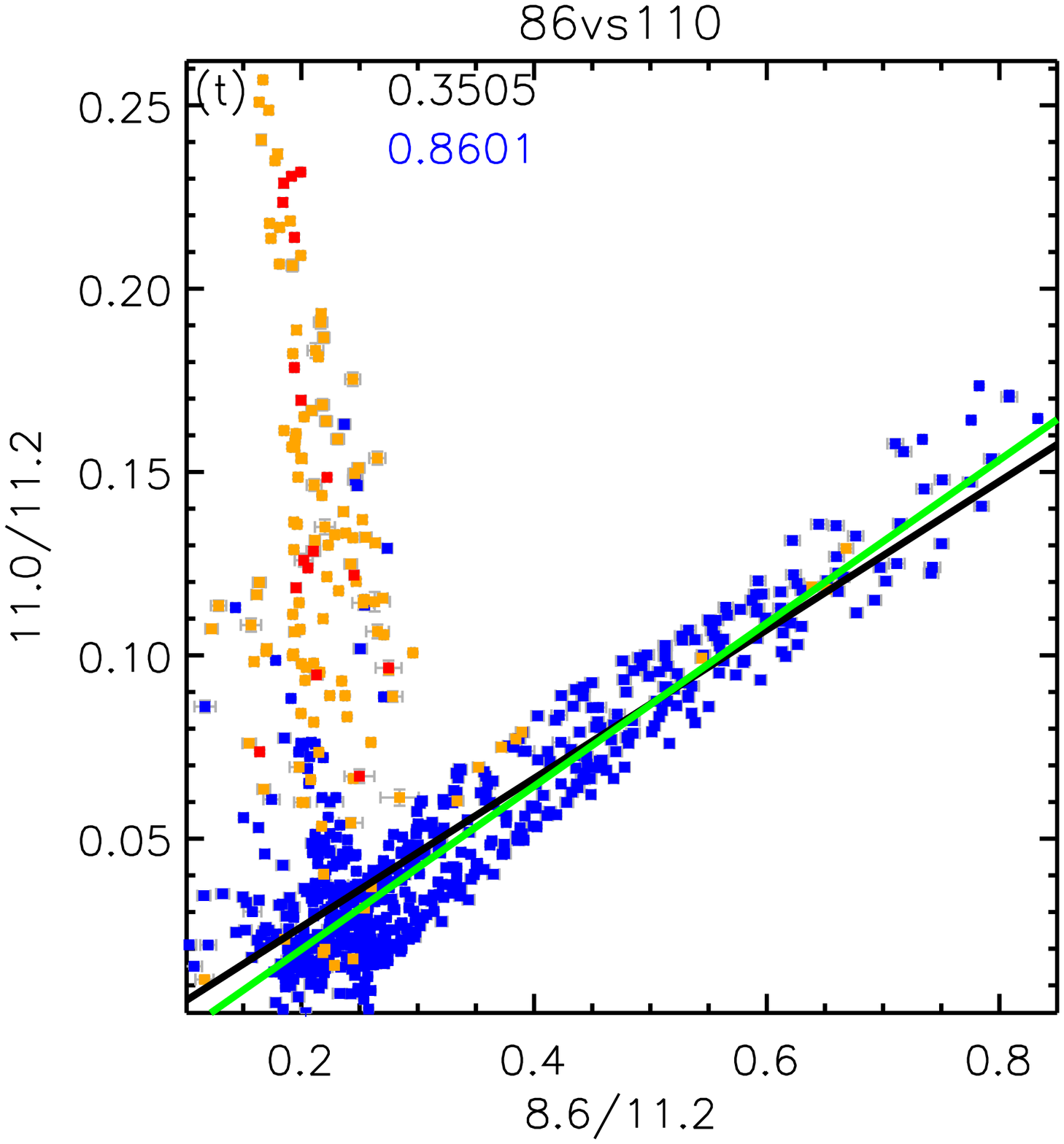}
}
\end{center}
\caption{Correlation plots between PAH features in NGC~1333. The data is sorted into three separate regimes based on the degree of extinction (Section \ref{IRS cont}): points with A$_{k}$~$<$~0.5 are given as blue squares, points with 0.5~$<$~A$_{k}$~$<$~1 are shown as orange squares, and points with A$_{k}$~$>$~1 are shown as red squares. Correlation coefficients are given in the top left of each panel for the entire map in black and for only the pixels with A$_{k}$~$<$~0.5 in blue. Linear fits are shown for all pixels in black and for only the pixels with A$_{k}$~$<$~0.5 in green.}
\label{irs_corr}
\end{figure*}

Figure \ref{irs_corr} presents a selection of observed intensity correlations found between PAH related emission components within the IRS~SL data. To investigate the effects of extinction in our data, we group this data based into three regimes of extinction: low A$_{K}$~$<$~0.5; intermediate 0.5~$<$~A$_{k}$~$<$~1; and high A$_{k}$~$>$~1. In general, all correlations show an improvement when we only account for the low extinction pixels, which fortunately coincides with the bulk of the RN. 

We find very tight correlations between the 6.2 and 7.7~$\mu$m LS continuum subtracted (hereafter LS) PAH bands (Figure \ref{irs_corr} (a)). The 8.6~LS~$\mu$m band also shows a strong correlation with both of these bands with some appreciable deviations (Figure \ref{irs_corr} (b) and (c)). Notably, we find that the 6.2, 7.7~GS and 8.6~$\mu$m GS continuum subtracted (hereafter GS) bands show even tighter correlations with each other than the LS counterparts (Figure \ref{irs_corr} (e), (f) and (g)). This is most evident for the correlations involving the 8.6~GS~$\mu$m band. In addition, at high values, the 7.7 and 8.6 (both LS and GS) level off instead of linearly increase.

The G7.6 and G8.6~$\mu$m components have the best correlation of all features consider here (Figure \ref{irs_corr} (i)), while the 6.2~$\mu$m band and the G8.6~$\mu$m also show very tight correlations (Figure \ref{irs_corr} (k)). In contrast, the G7.8 and G8.2~$\mu$m only correlate modestly (Figure \ref{irs_corr} (j)).

The 12.7~$\mu$m band correlates strongest with the 8.6 LS $\mu$m band (Figure \ref{irs_corr} (o)). However, instead of exhibiting a linear correlation, there are two regimes present. At low 8.6/11.2 values, a steep increase in the 12.7/11.2 is seen while at intermediate and high 8.6/11.2 values, the corresponding increase in 12.7/11.2 is more moderate. A similar behavior is seen in the relation of the 12.7~$\mu$m band with the 6.2, 7.7, and G7.6~$\mu$m bands overall though they exhibit enhanced ``scatter'' which seem to occur in a similar pattern (Figure \ref{irs_corr} (m), (n), and (p)). The 11.0~$\mu$m band shows a bifurcated pattern for all correlations considered. One of the branches is entirely composed of spectra with moderate to high extinction while the other branch is dominated by regions with low extinction. Given the uncertainty related to extinction measurements, we restrict ourselves to low extinction areas. The 11.0~$\mu$m band exhibit moderate correlations with the 6.2, 7.7, and 8.6~$\mu$m bands (Figure \ref{irs_corr} (r), (s) and (t)). Its behaviour with the 12.7~$\mu$m band shows a similar pattern as that of the 12.7 versus 8.6~$\mu$m band: for the lower range of 12.7/11.2 values, little to no variation is seen in the 11.0/11.2 and only for the upper range of 12.7/11.2 values, a clear correlation is observed (Figure \ref{irs_corr} (q)).

The 5--10~$\mu$m plateau does not correlate well with the 6.2 and 7.7~$\mu$m bands, primarily because of a bi--linear trend that seems to stem from this plateau. This bi--linear trend is also evident between the 6.2 and G8.2~$\mu$m correlation with both branches containing pixels with low extinction.

\begin{figure}
\begin{center}

\includegraphics[clip,trim =0.cm 1.2cm 0.cm 2cm,width=8.8cm]{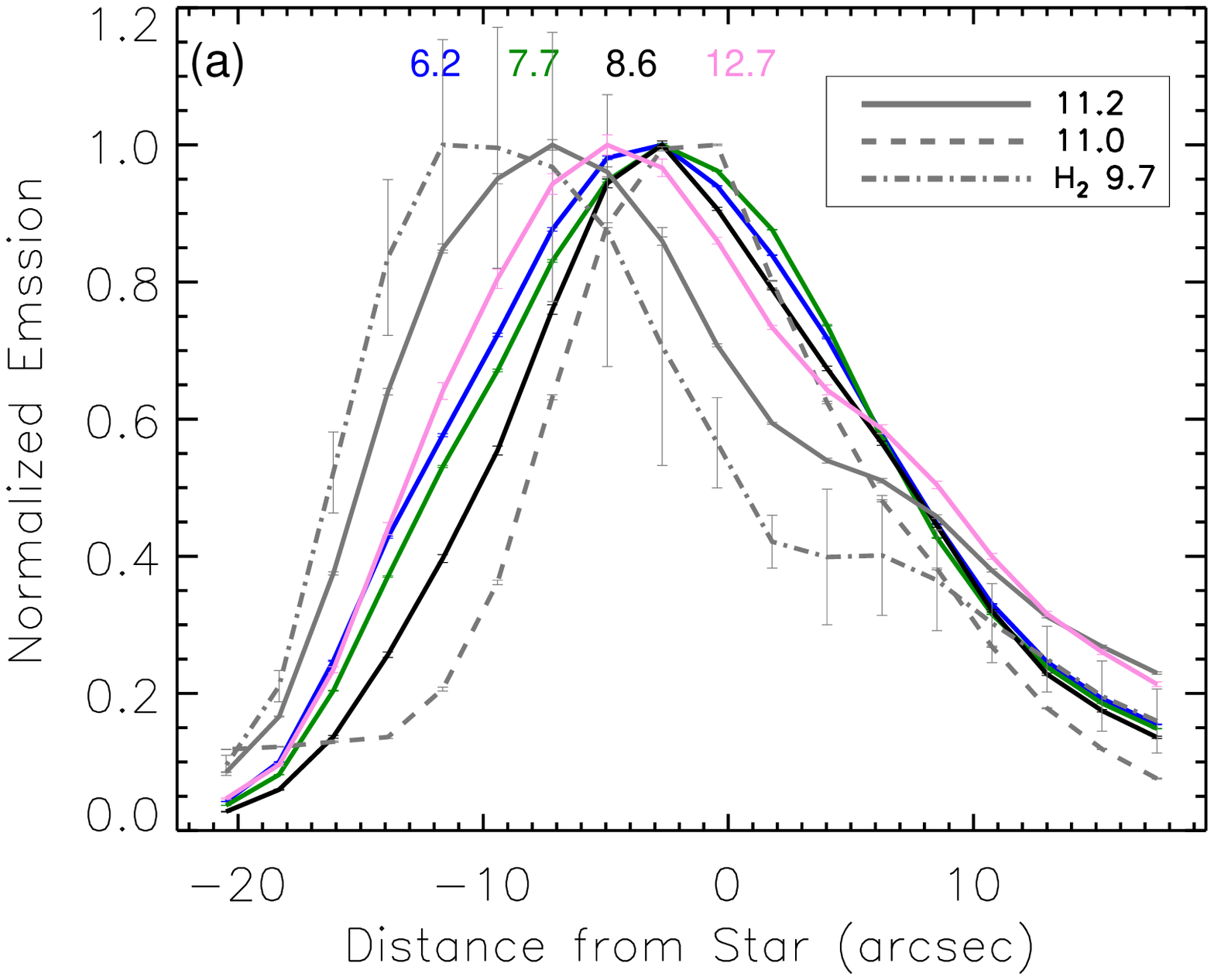}
\includegraphics[clip,trim =0.cm 1.2cm 0.cm 2cm,width=8.8cm]{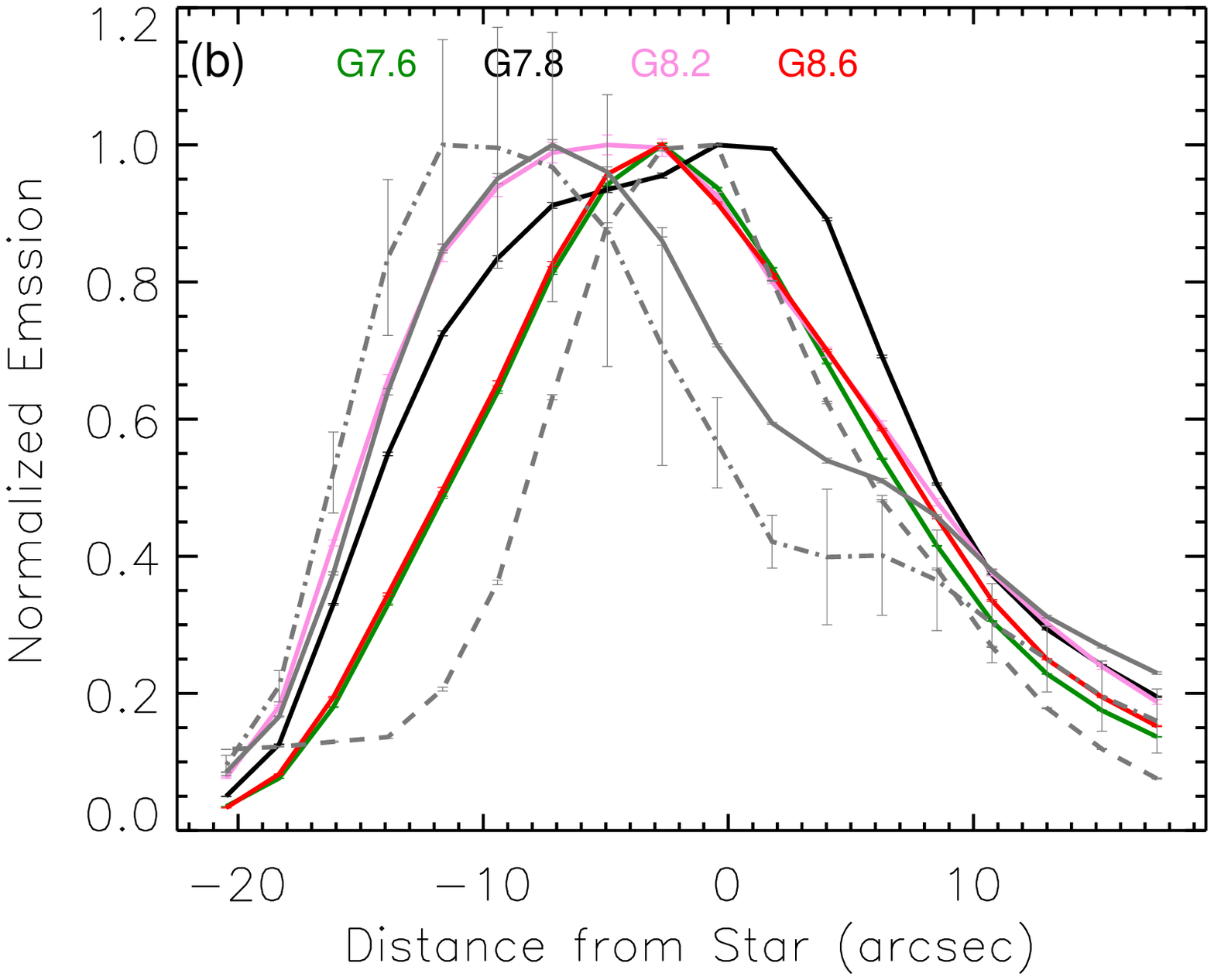}
\includegraphics[clip,trim =0.cm 0cm 0.cm 2cm,width=8.8cm]{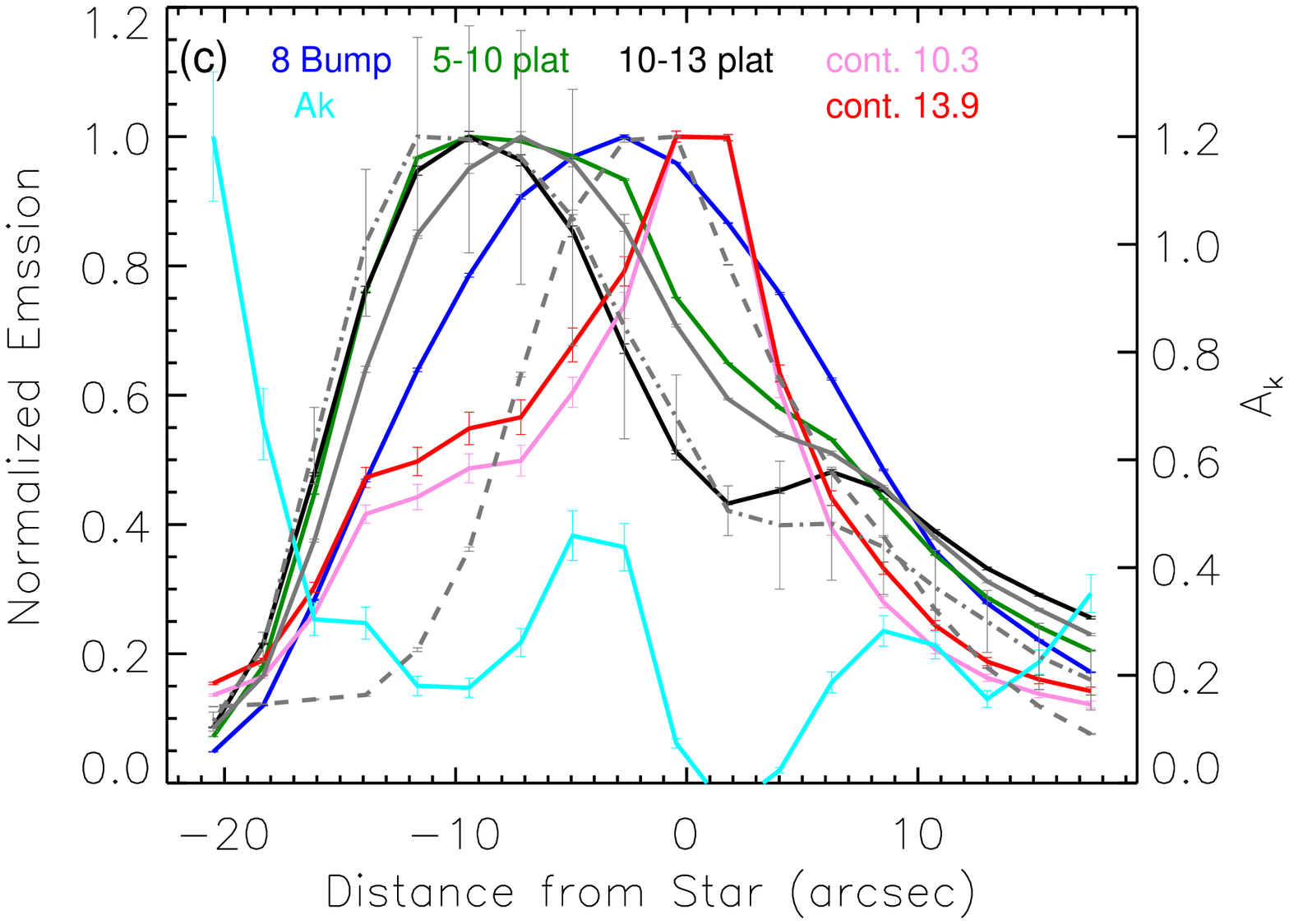}

\end{center}

\caption{NGC~1333 cross cuts normalized to the peak values for each emission feature across SVS~3 from S (negative distance) to N (positive distance) along a cross cut which is shown in Figure \ref{irs maps}. The nomenclature is given in Table \ref{table:1}. 
Cross cuts of the 11.0 and 11.2 $\mu$m bands as well as the H$_{2}$ 9.7 $\mu$m line are included in each panel for reference. Error bars for each emission feature are given in the same color as the associated cross cut. The magnitude of the extinction A$_{k}$ is given on the right y--axis (bottom panel). }
\label{ngc1333_nsline1}
\end{figure} 

\subsubsection{IRS Cross Cuts}
\label{irs lp}

\begin{figure*}
\begin{center}
\resizebox{.9\hsize}{!}{%
\includegraphics[clip,trim =0.cm 1.2cm 2.cm 2cm,width=8.8cm]{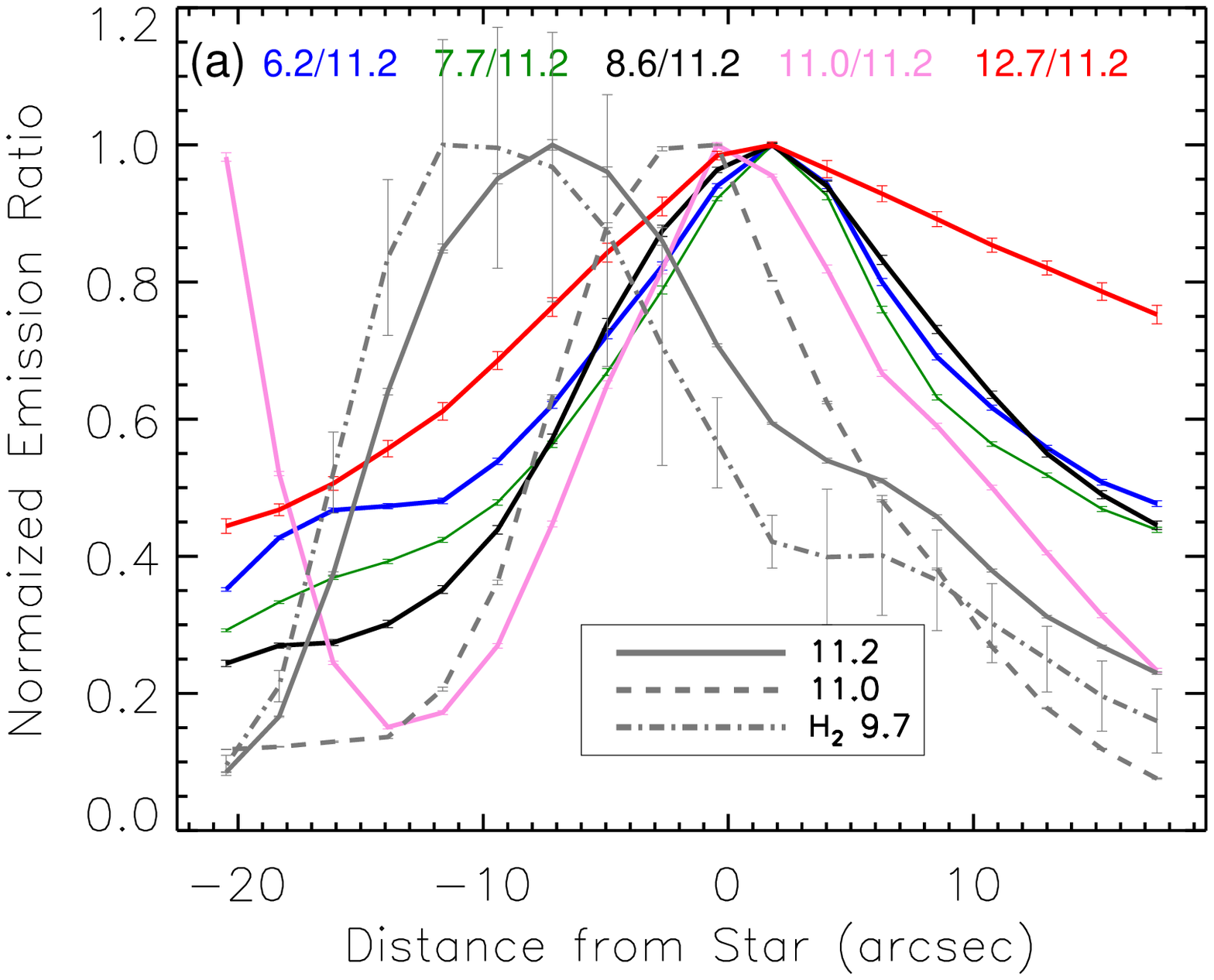}
\includegraphics[clip,trim =0.cm 1.2cm 2.cm 2cm,width=8.8cm]{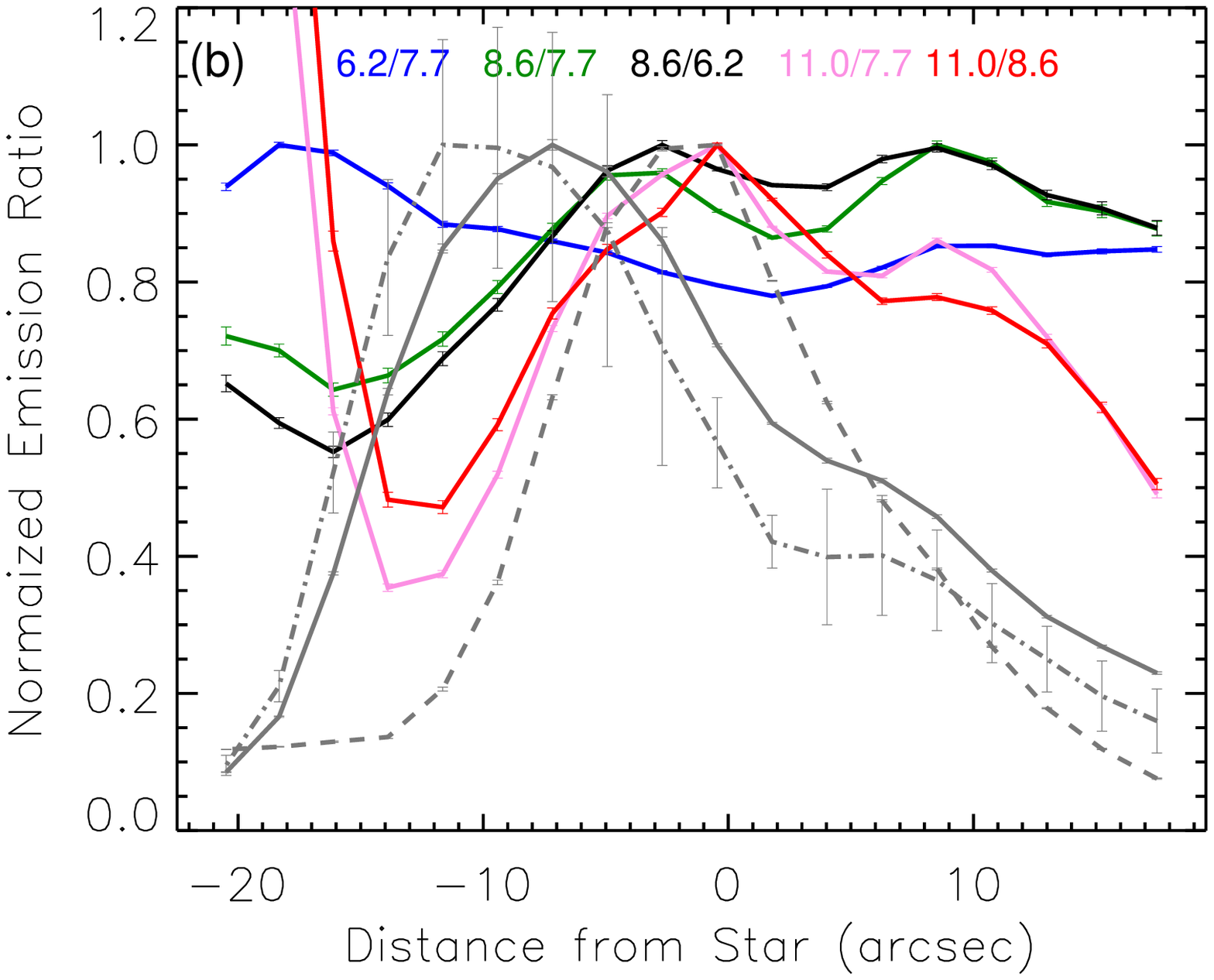}}
\resizebox{.9\hsize}{!}{%
\includegraphics[clip,trim =0.cm 0cm 2.cm 2cm,width=8.8cm]{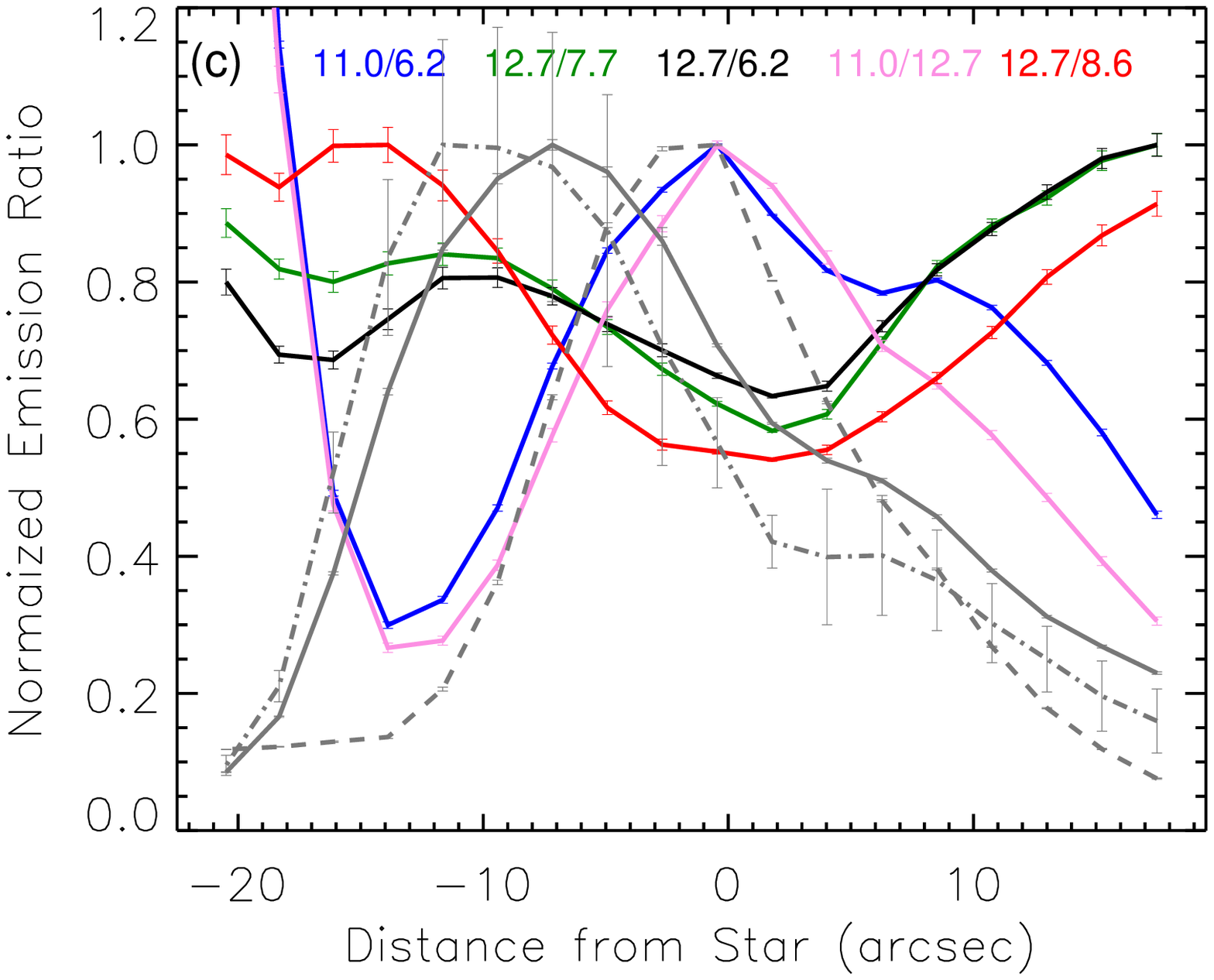}
\includegraphics[clip,trim =0.cm 0cm 2.cm 2cm,width=8.8cm]{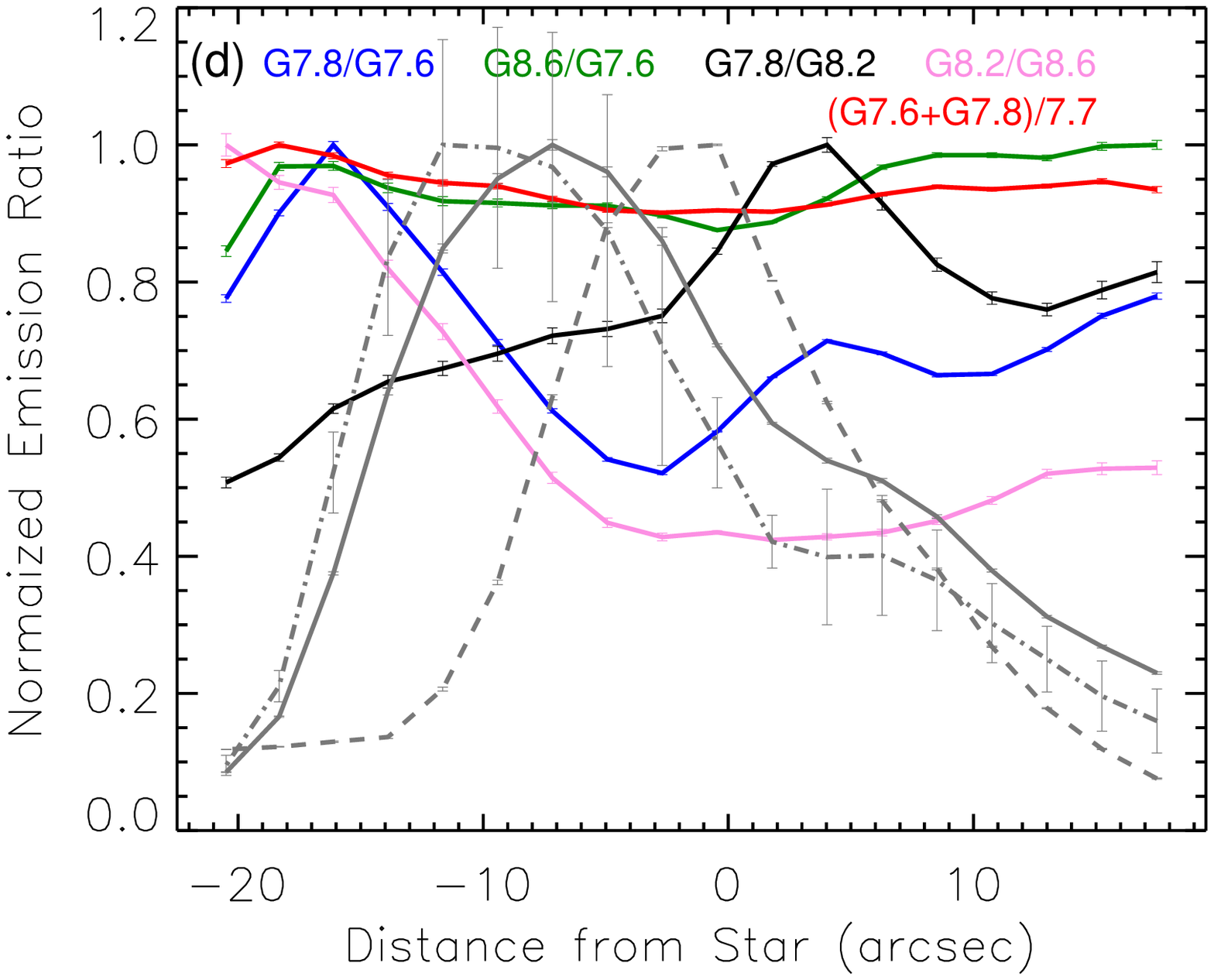}}
\end{center}
\caption{NGC~1333 emission ratios cross cuts normalized to the peak values for each emission ratio across SVS~3 from S (negative distance) to N (positive distance) along a cross cut which is shown in Figure \ref{irs maps}. For emission ratios involving the 11.0 $\mu$m band, we normalize to the ratio value at the 11.0 $\mu$m peak. Cross cuts of the 11.0 and 11.2 $\mu$m bands as well as the H$_{2}$ 9.7 $\mu$m line are included in each panel for reference. Error bars for each emission ratio are given in the same color as the associated cross cut. }
\label{ngc1333_nsline2}
\end{figure*}

Figures \ref{ngc1333_nsline1} and \ref{ngc1333_nsline2} show radial cuts of relevant emission features and their ratios\footnote{We give the nomenclature and normalization factors in Appendix \ref{ngc1333 irs components}.}. The radial cut is centered on SVS~3 in a South to North direction as illustrated in Figure \ref{irs maps}. A clear stratification in the emission peaks is present. Consistent with the results in Section \ref{slmaps}, the 11.0~$\mu$m peaks closest to SVS~3. The 6.2, 7.7, and 8.6~$\mu$m bands peak at approximately the same distance southwards of SVS~3, followed by the 12.7~$\mu$m band and then the 11.2~$\mu$m band which is furthest southwards of SVS~3 towards the H$_{2}$~9.7~$\mu$m peak (Figure~\ref{ngc1333_nsline1}~(a)). We note that 6.2 and 7.7~$\mu$m bands show increased emission south of the star compared to 8.6~$\mu$m band and the H$_{2}$~9.7~$\mu$m, 11.2~$\mu$m, and to a lesser extent the 12.7~$\mu$m emission exhibit a second, local maximum about 8$^{\prime\prime}$ north of SVS~3. The Gaussian components G7.6 and G8.6 have a radial profile similar to that of the 8.6~$\mu$m bands while the G8.2~$\mu$m band shows a broad peak encompassing both the 6--9~$\mu$m peak and the 11.2 $\mu$m peak, and the G7.8~$\mu$m band shows a broad distribution peaking slightly to the north of SVS~3 (Figure~\ref{ngc1333_nsline1}~(b)). The 10--13~$\mu$m plateau shows a similar radial profile as the H$_2$ emission with slight increased emission between the main peak and the second local maximum (i.e. towards the star) while the 5--10~$\mu$m plateau behaves similar to the 11.2~$\mu$m emission (Figure~\ref{ngc1333_nsline1}~(c)). Both plateaus are thus significantly displaced from the individual bands perched on top of these plateaus. The 8~$\mu$m~bump shows a broad radial profile peaking close to the peak of the 6.2, 7.7, and 8.6~$\mu$m bands. The underlying dust continuum measured at 10.3 and 13.9~$\mu$m show similar radial profiles which are very different from those of the PAH bands and plateaus (Figure~\ref{ngc1333_nsline1}~(c)). They have a similar 2~pixel wide flat peak on the star, which also corresponds to the peak of the G7.8~$\mu$m emission, and have a weaker shoulder from the peak of the 11.2~$\mu$m band to the H$_2$ emission peak. We refer to peak position of the dust continuum emission as the `dust peak' for the remainder of this section. Finally, the extinction A$_k$ is highest in the southern edge of the cross cut and shows a minimum around SVS~3 (Figure~\ref{ngc1333_nsline1}~(c)).                

The 6.2, 7.7, 8.6, and 12.7~$\mu$m emission, normalized to the 11.2~$\mu$m emission, all peak 2$^{\prime\prime}$ north of SVS~3, within the dust peak, but have variable width (Figure~\ref{ngc1333_nsline2}~(a)). The peak of the 11.0~$\mu$m emission, normalized to the 11.2~$\mu$m emission, nearly coincides with SVS~3 (i.e. it is one pixel closer to SVS~3 compared to the other bands) and is the sharpest. The sharp rise in the ratios involving the 11.0 $\mu$m band towards the southern edge of this cut is attributable to the strong effects of silicate extinction on the 11.0 $\mu$m band as mentioned in Section \ref{slcorr}. The 6.2/7.7 emission ratio is quite constant over this cross cut, with a slight rise towards the southern edge, and is thus distinct from the 8.6/6.2 and 8.6/7.7 ratios cross cut. The latter show very similar cuts that show slight variations north  of the 11.2~$\mu$m peak and drop off south of the 11.2~$\mu$m peak (Figure~\ref{ngc1333_nsline2}~(b)). Cross cuts of ratios involving the 11.0~$\mu$m emission all peak at the same location (Figure~\ref{ngc1333_nsline2}~(a),~(b),~(c)).  Cross cuts of ratios involving the 12.7~$\mu$m emission also peak at the same location and exhibit broad minima/maxima (Figure~\ref{ngc1333_nsline2}~(a),~(c)). Finally, we consider cross cuts involving the 7--9~$\mu$m Gaussian sub components (Figure~\ref{ngc1333_nsline2}~(d)). We note that the G8.6/G7.6 ratio remains approximately constant over the entire cut, consistent with these features showing the best correlation (Section~\ref{slcorr}). The cross cuts of the G7.8/G7.6 and G8.2/G8.6 ratios are also somewhat comparable with peaks towards the southern edge. Their differences largely originates in the variable G7.8/G8.2 ratios which shows a strong peak 2--4$^{\prime\prime}$ north of SVS~3. 
        
\begin{figure}[t]
\begin{center}
\includegraphics[clip,trim =1.cm 0cm 0.3cm 0cm,width=8.4cm]{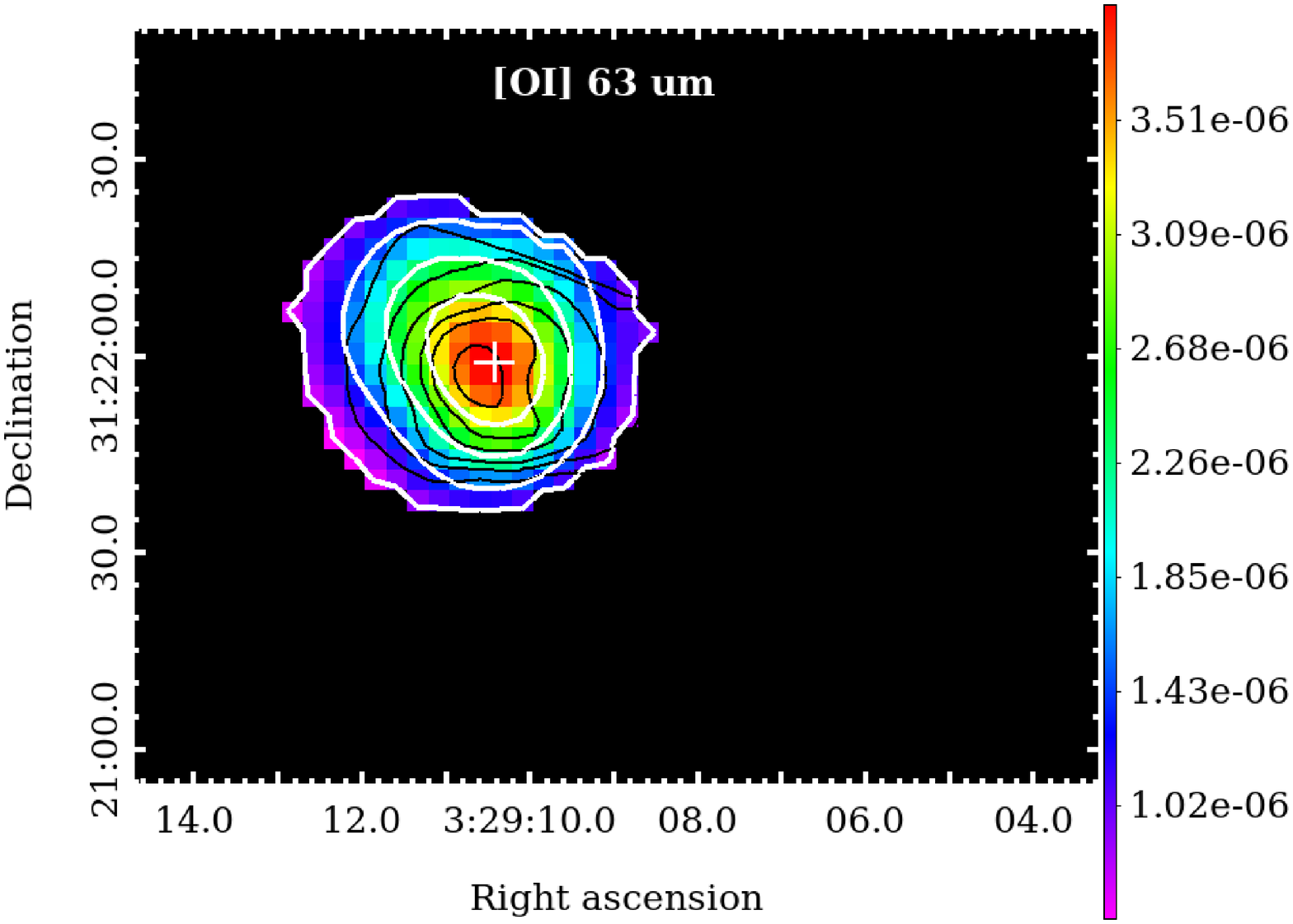}
\includegraphics[clip,trim =1.cm 0cm 0.3cm 0cm,width=8.4cm]{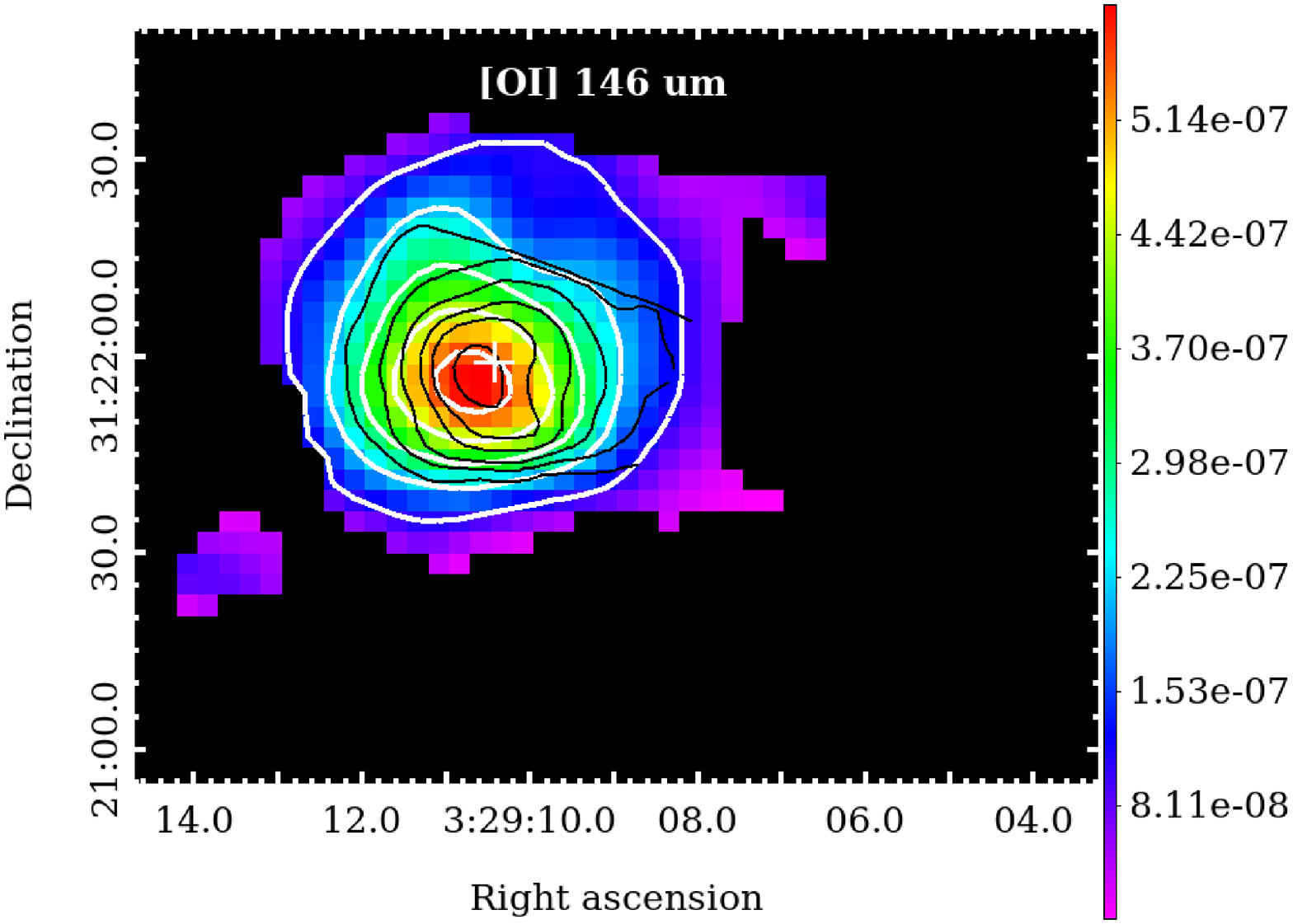}
\includegraphics[clip,trim =1.cm 0cm 0.3cm 0cm,width=8.4cm]{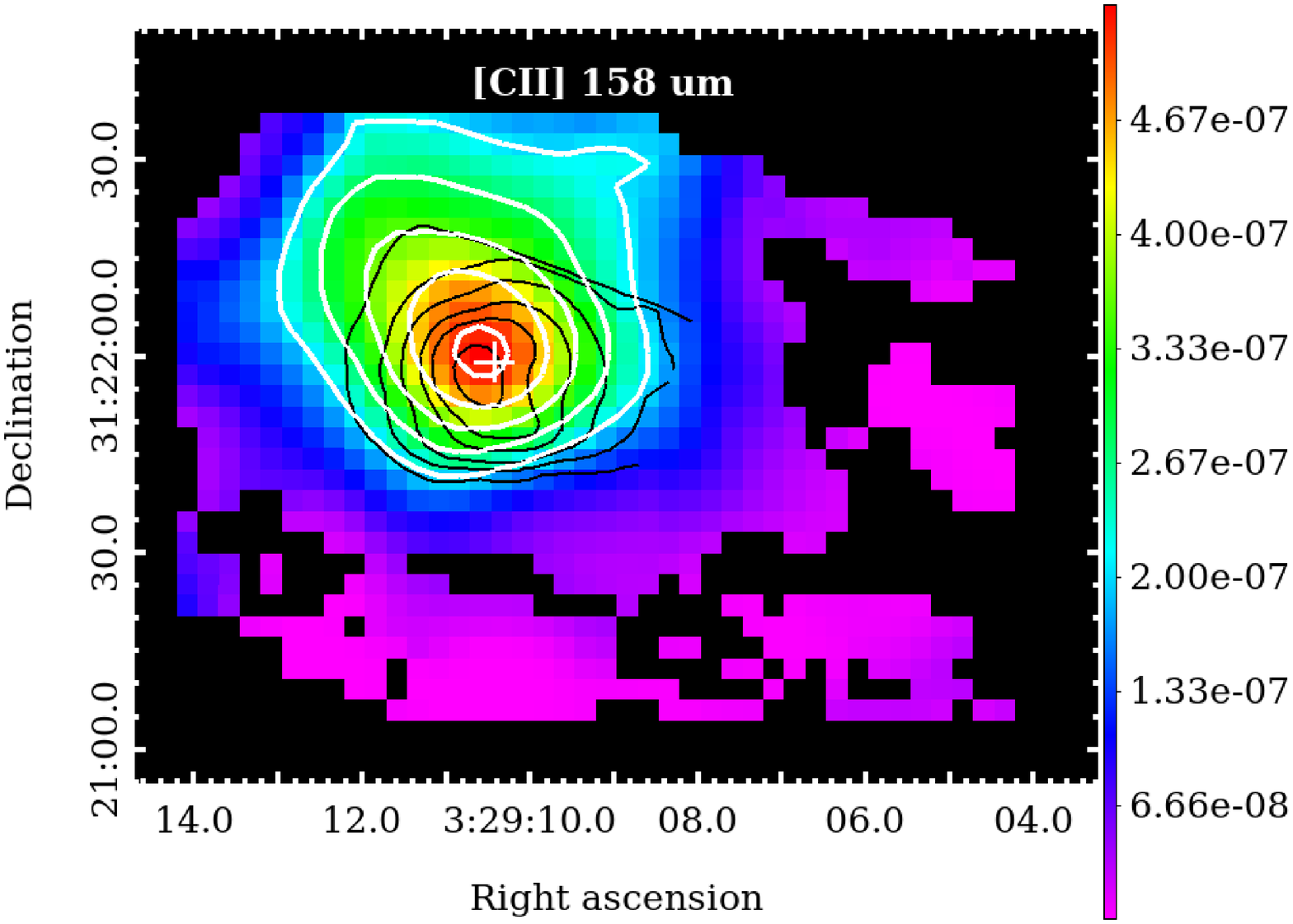}
\end{center}
\caption{FIR Cooling line fluxes. {\it Top} The [\OI] 63~$\mu$m line fluxes with white contours of 0.60, 1.43, 2.26, 3.09, 3.92~$\times$~10$^{6}$~W~m$^{-2}$~sr$^{-1}$. {\it Middle} The [\OI] 146~$\mu$m line fluxes with white contours of 1.06, 2.17, 3.27, 4.38, 5.48~$\times$~0$^{7}$~W~m$^{-2}$~sr$^{-1}$. {\it Bottom} The [\CII] 158~$\mu$m line fluxes with white contours of 2.05, 2.81, 3.59, 4.35, 5.12~$\times$~10$^{7}$~W~m$^{-2}$~sr$^{-1}$. Contours of the PAH 7.7~$\mu$m flux are shown in black as in Figure~\ref{irs maps}. The position of SVS~3 is shown as a white cross. North is up and east is to the left. Pixels below the 3~$\sigma$ noise level of each respective map are shown in black. Axes are given in right ascension and declination (J2000).}
\label{FIR maps}
\end{figure}

\subsection{FIFI--LS Maps}
\label{FIFI results}

In Figure \ref{FIR maps}, we show the FIFI-LS spectral maps of the three FIR cooling line fluxes. Each of these maps is convolved to the PSF of the [\CII] 158~$\mu$m observations and regridded to the PACS 160~$\mu$m map grid (see Section~\ref{FIFI measure}). In each case, these emission lines show a simple concentric morphology with emission peaking near the illuminating source. However, the [\OI] 146~$\mu$m peak intensity is offset from the source position (to the southeast) similar to the 7.7~$\mu$m PAH peak emission, whereas [\OI] 63 and [\CII] 158~$\mu$m peaks are nearly co--spatial with the star. The [\CII] emission is slightly displaced from the central star to the northeast and exhibits significant emission to the north and north east of the star. 
Another notable difference is the [\OI] 63~$\mu$m is significantly more centrally concentrated relative to the other FIR cooling lines which cannot be discounted because this map has been convolved to match the [\CII] 158~$\mu$m resolution.

\subsection{PACS Maps}
\label{PACS results}

\begin{figure}
\begin{center}
\includegraphics[clip,trim =0.5cm 0.cm 0.cm 1cm,width=8.4cm]{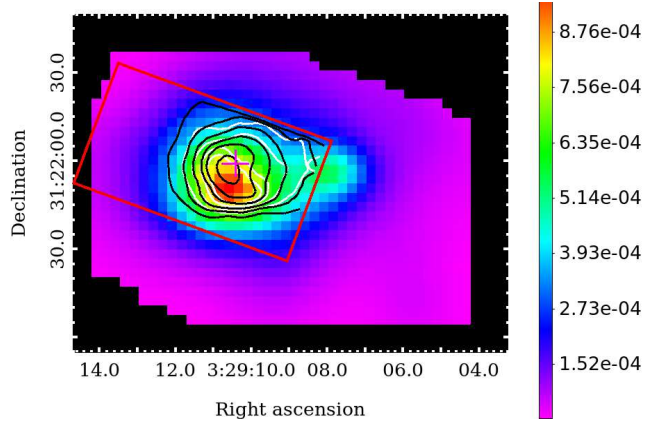}
\end{center}
\caption{The FIR dust continuum in units of W~m$^{-2}$~sr$^{-1}$. The IRS~SL aperture is shown in red. Contours of the 11.2 and 7.7~$\mu$m emission are shown respectively in white and black as in Figure~\ref{irs maps}. The position of SVS~3 is shown as a magenta cross. North is up and east is to the left. Pixels below the 3~$\sigma$ noise level are shown in black. Axes are given in right ascension and declination (J2000).}
\label{PACS FIR map}
\end{figure}

In Figure \ref{NGC1333 PACS}, we show the PACS 70, 100, 160~$\mu$m images in blue, green and red respectively. These maps show a very similar concentric morphology within the RN with peak emission found to the south of SVS~3 in each case. They also show emission extending to the west of SVS~3 with a secondary peak corresponding to the YSO IRAS~03260+3111E. In Figure \ref{PACS FIR map}, we show the FIR dust continuum emission map derived from the PACS images as detailed in Section \ref{SED}. This map is essentially identical in spatial morphology to the individual PACS maps.   

\subsection{PDR Modelling}
\label{PDR Models}

 We make a simple estimate for the FUV radiation field strength (between 6 and 13.6~eV) using the FIR dust continuum emission:
\begin{equation}
 G_{0} \simeq \frac{1}{2} \frac{I_{\textrm{FIR}}}{1.3 \times 10^{-7}}
 \end{equation}
 
 \noindent where we assume a face--on PDR morphology \citep[][]{hol97} and take G$_{0}$ in units of the Habing field \citep[1.3~$\times$~10$^{-7}$~W~m$^{-2}$~sr$^{-1}$;][]{hab68}.
 We determine the gas density and temperature by using i) this FIR dust continuum derived G$_{0}$ along with the FIR cooling lines [\OI] 63 and [\CII] 158~$\mu$m and ii) the [\OI] 63/ [\CII] 158 ratio and the gas temperature PDR model grids as a function of gas density and FUV radiation field strength from the Photo Dissociation Region Toolbox (PDRT) software \citep{kau06,pou08}. Several of the chemical and thermal processes in the model have been updated with recent rates. In particular, the photo-dissociation and photo-ionization rates of \cite{hea17} and the collisional excitation rates of [\OI] from \cite{liq18}  and Lique (private communication) are used. These model grids have resolution elements of 0.125 in log~G$_{0}$ and n$_{\textrm{H}}$ and cover a range of -0.5--6.5 in log G$_{0}$ and a range of 1--7~cm$^{-3}$ in log~n$_{\textrm{H}}$. We observe the median [\OI] 146/63 ratio to be 0.154 (in pixels where the [\OI] 63~$\mu$m line has a 3~$\sigma$ detection), which is higher than the 0.1 value suggested by \cite{oss15} as a sign for self-shielding in the [\OI] 63~$\mu$m line. We therefore multiply our [\OI] 63~$\mu$m observations by a factor of 2 \citep[e.g.][]{sch18}. We compare the observed 2 $\times [\OI]$ 63/ [\CII] 158 ratio and G$_{0}$ in each pixel with the [\OI] 63/ [\CII] 158 model grid in G$_{0}$ and n$_{\textrm{H}}$ to predict the gas density, n$_{\textrm{H}}$, based on where the observed quantities intersect the grid. We use the derived densities along with the G$_{0}$ values to derive a gas temperature by comparing our predicted values with a model grid for gas temperature, T, as a function of gas density and FUV radiation field strength. We then convert the gas density to electron density assuming that all free electrons result from the photo--ionization of carbon and all gas--phase carbon is ionized \citep[e.g.][]{gal08}: n$_{\textrm{e}}$~$\simeq$~(C/H)~n$_{\textrm{H}}$~$\simeq$~1.6~$\times$~10$^{-4}$~n$_{\textrm{H}}$, where 1.6~$\times$~10$^{-4}$ is the interstellar gas--phase carbon abundance \citep{sofia04}. From this, we derive the PAH ionization parameter \citep[$\gamma$~=~G$_{0}$~T$^{0.5}$~/~n$_{\textrm{e}}$;][]{bak94}.
 
 We find that the derived G$_{0}$, n$_{\textrm{H}}$, and T are of the same order of magnitude as the results of the FIR analysis performed by \cite{you02} for this region: 10$^{3}$, 10$^{4}$~cm$^{-3}$, and 10$^{2}$~K respectively. Slight discrepancies in the absolute values found between both studies may be attributed to a difference in spatial resolution. Indeed, \cite{you02} consider a single pointing for each line with beam sizes $>$~40$^{\prime\prime}$ whereas our maps have been convolved to the [\CII] 158~$\mu$m PSF of 15.9$^{\prime\prime}$ and resampled to the PACS 160~$\mu$m grid with a pixel scale of 3.2$^{\prime\prime}$.

\begin{figure*}
\begin{center}
\begin{tabular}{cc}
\includegraphics[clip,trim =0.2cm 0.cm 0.cm .5cm,width=7.6cm]{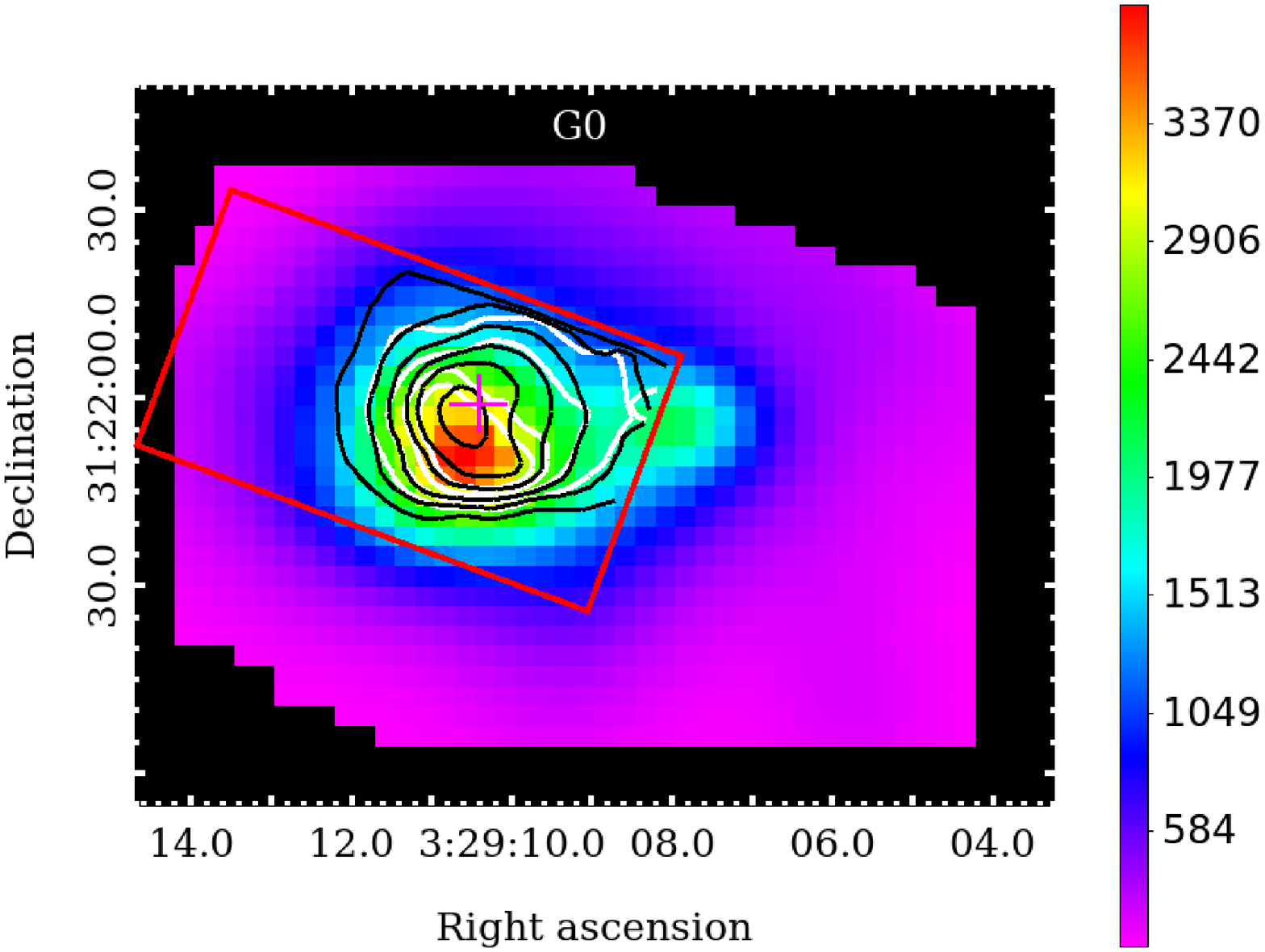} &
\includegraphics[clip,trim =0.8cm 0.cm 0.cm .5cm,width=7.6cm]{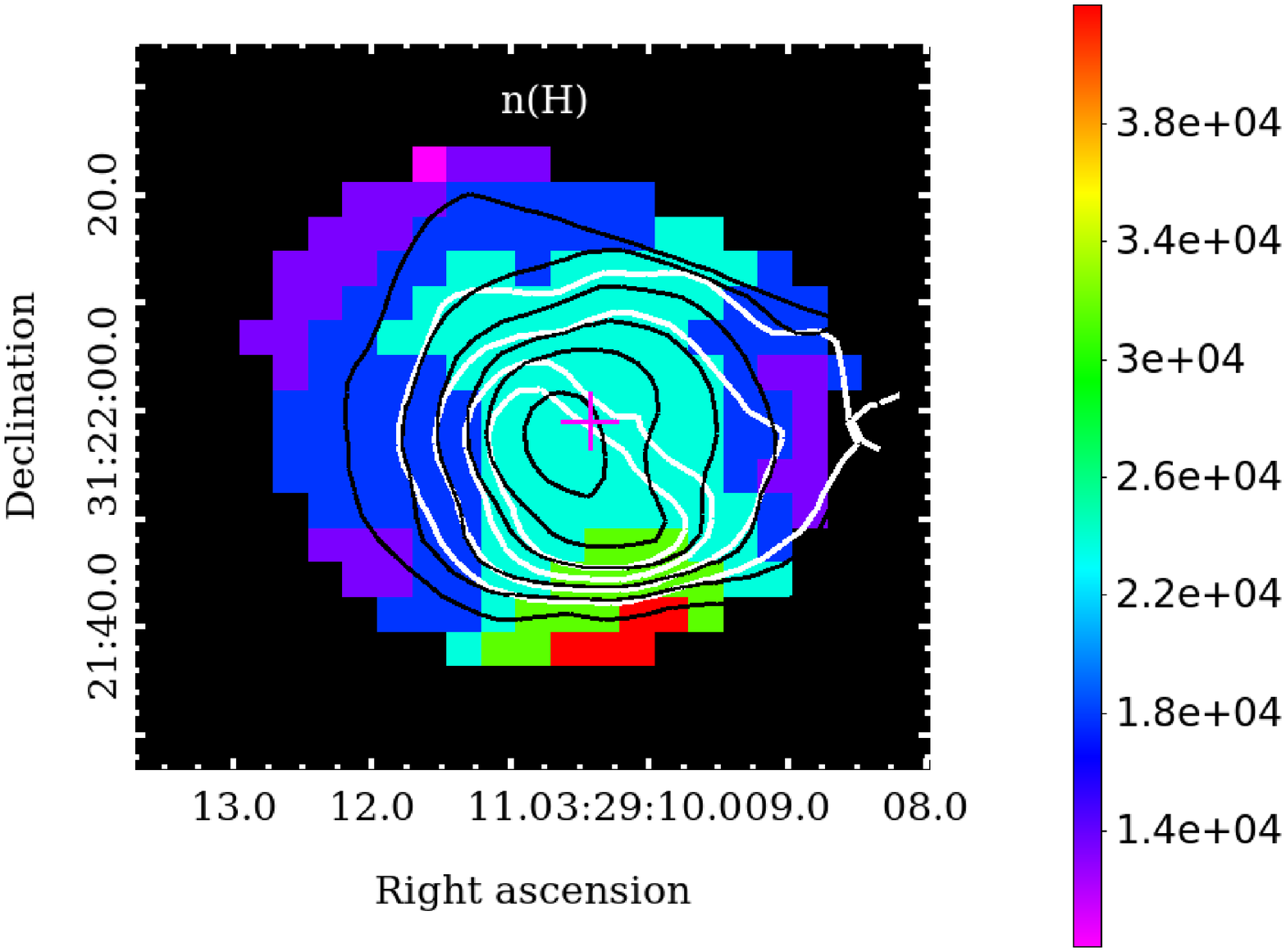} \\
\includegraphics[clip,trim =0.8cm 0.cm 0.cm .5cm,width=7.6cm]{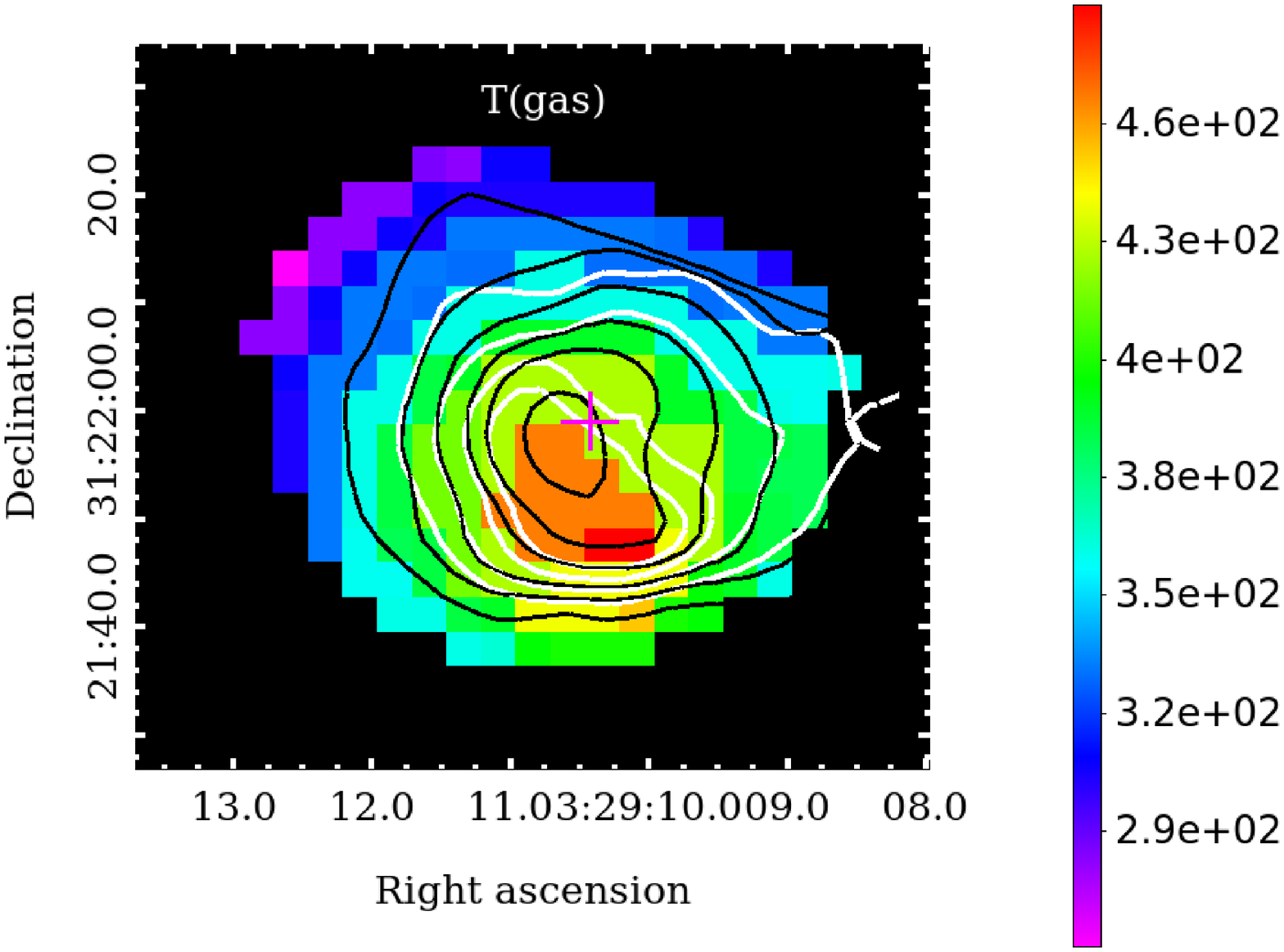} &
\includegraphics[clip,trim =0.8cm 0.cm 0.cm .5cm,width=7.6cm]{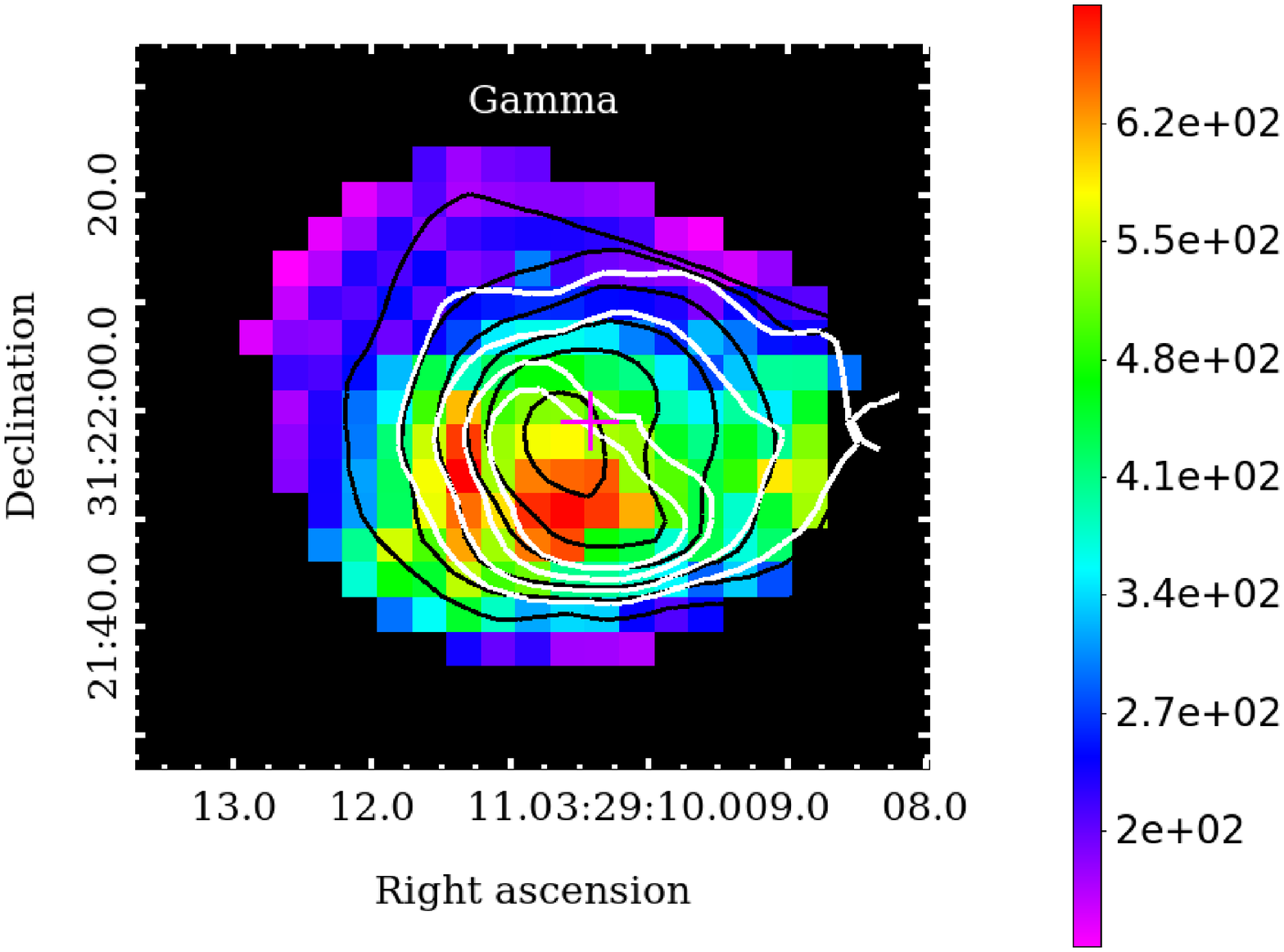} \\
\end{tabular}
\end{center}
\caption{The physical conditions in NGC~1333 derived from FIR data and PDR models. From top left to bottom right: the FUV radiation field strength, G$_{0}$, in units of the Habing field, with the IRS~SL aperture shown in red; the gas density, n$_{\textrm{H}}$ (cm$^{-3}$); the gas temperature, T (K); and the ionization parameter, $\gamma$~=~G$_{0}$~T$^{0.5}$~/~n$_{\textrm{e}}$. Contours of the 11.2 and 7.7~$\mu$m emission are shown respectively in white and black as in Figure~\ref{irs maps}. Pixels where [\OI] 63~$\mu$m emission line has a SNR~$<$~3 are shown in black all panels except G$_{0}$. The position of SVS~3 is indicated by a magenta cross. North is up and east is to the left. Axes are given in right ascension and declination (J2000).}
\label{PDR maps}
\end{figure*}

The resulting maps of the PDR conditions are shown in Figure \ref{PDR maps}. Except for the G$_0$ map, these maps are limited to the [\OI] 63~$\mu$m 3~$\sigma$ detection limit. The G$_{0}$ and T maps show a high degree of spherical symmetry emanating from their nearly co--spatial peaks. 
Likewise, the morphology of the PAH ionization parameter is similar to that of G$_{0}$.  Each of these three parameters (G$_{0}$, T, $\gamma$) peak to the south of SVS~3 where they almost overlap with the PAH 11.2~$\mu$m emission peak. In contrast, the gas density map shows very little variation, with a large plateau of $\sim$~2.3~$\times$~10$^{4}$~cm$^{-3}$ over much of the nebula and a substantial rise towards the southern edge of the map beyond the PDR front as traced by H$_{2}$ S(3) 9.7~$\mu$m (Figure \ref{irs maps}).

\begin{figure*} 
\begin{center}
\resizebox{\hsize}{!}{%
\includegraphics[clip,trim =.3cm 1.3cm 1cm 3.2cm,width=0.5\textwidth]{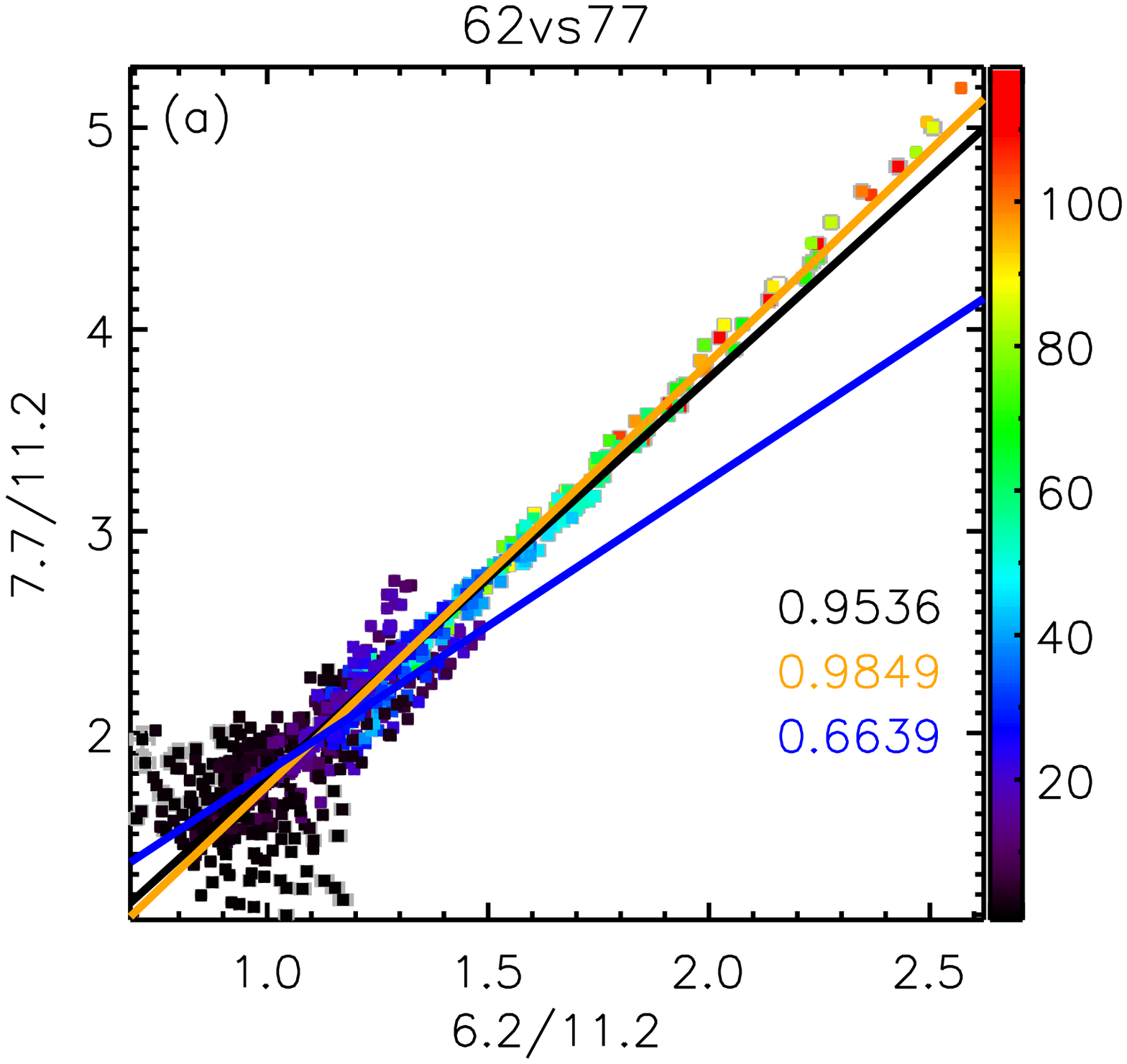} 
\includegraphics[clip,trim =.3cm 1.3cm 1cm 3.2cm,width=0.5\textwidth]{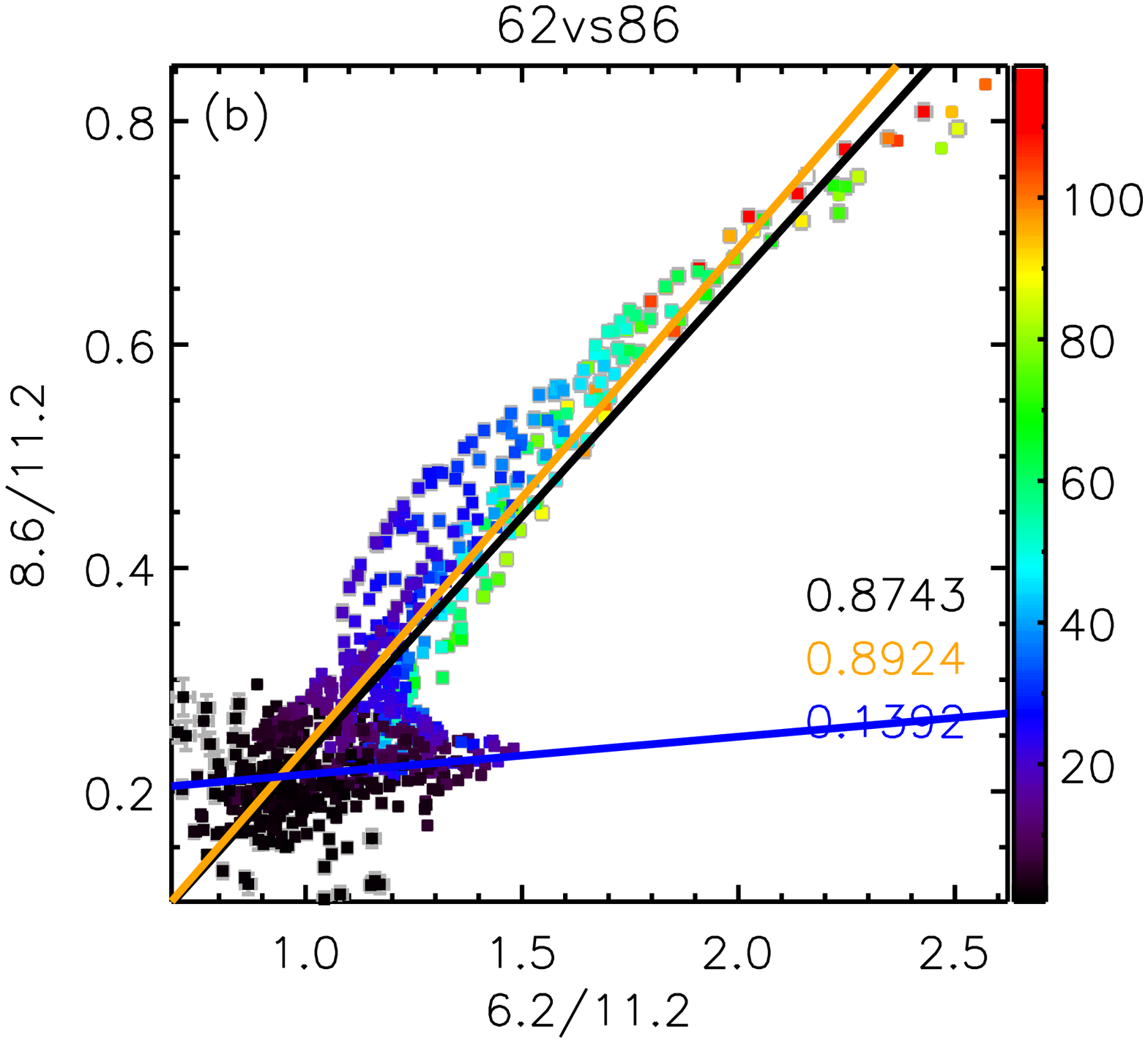} 
\includegraphics[clip,trim =.3cm 1.3cm 1cm 3.2cm,width=0.5\textwidth]{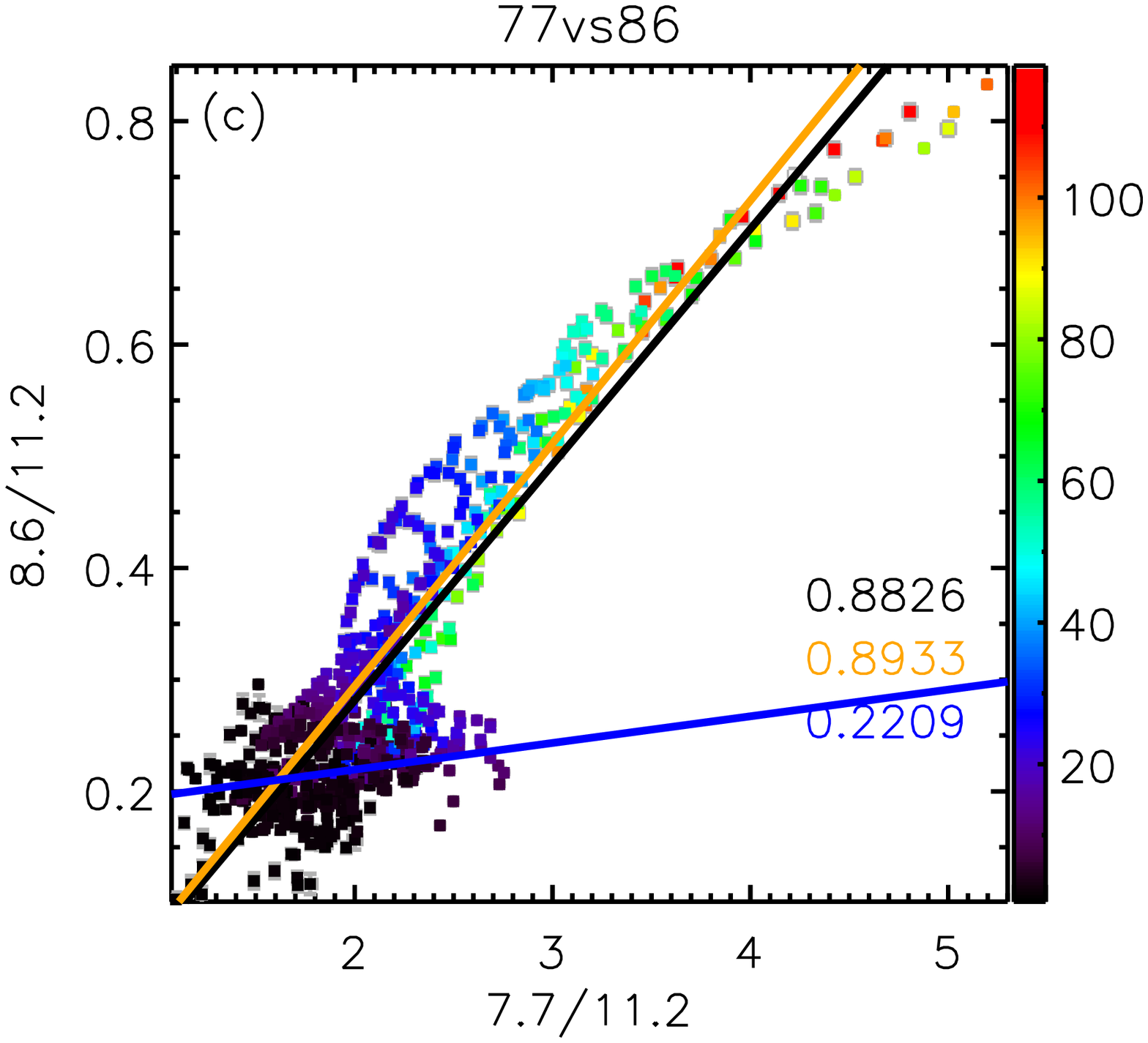}}
\resizebox{\hsize}{!}{%
\includegraphics[clip,trim =.3cm 1.3cm 1cm 3.2cm,width=0.5\textwidth]{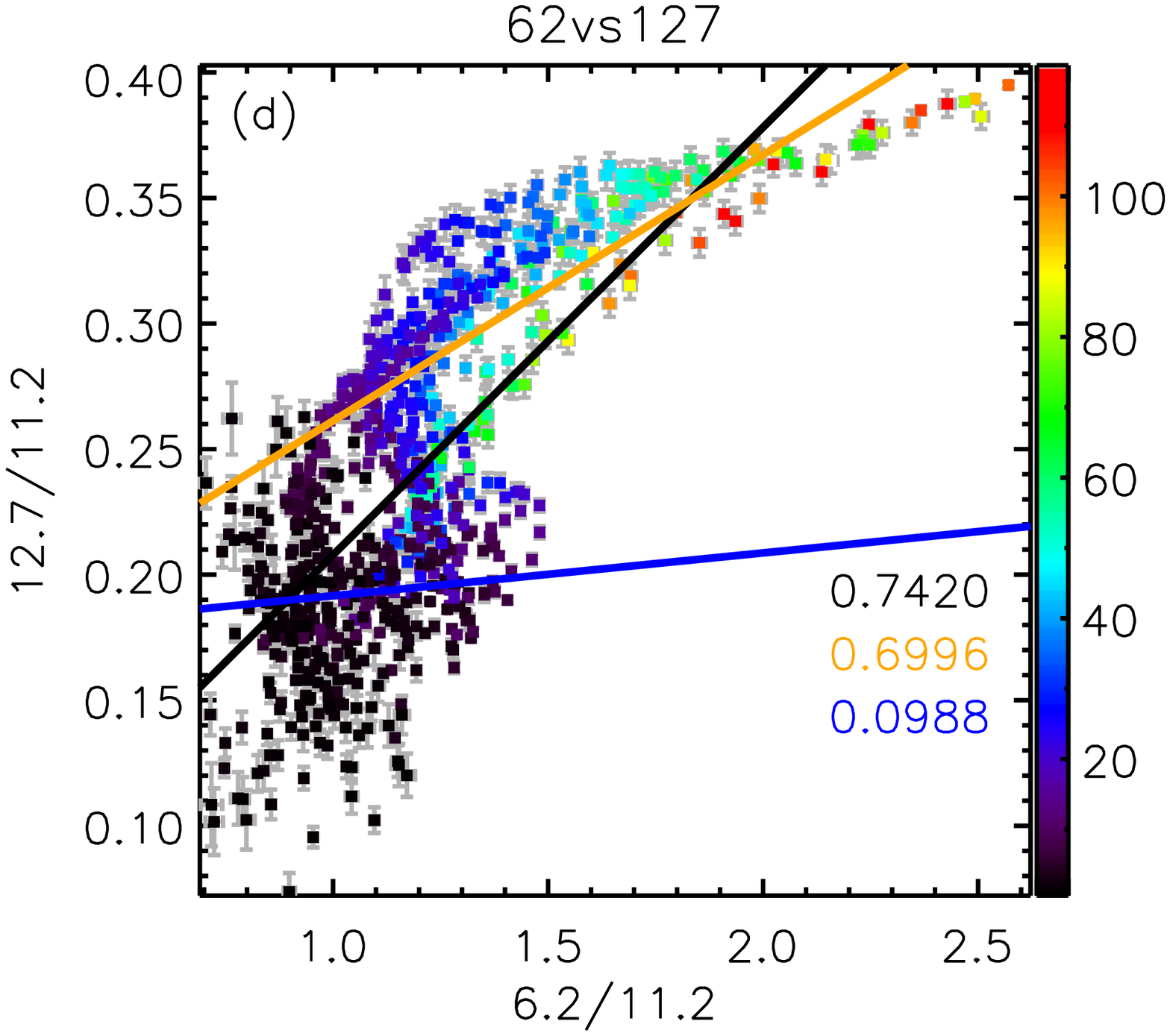}
\includegraphics[clip,trim =.3cm 1.3cm 1cm 3.2cm,width=0.5\textwidth]{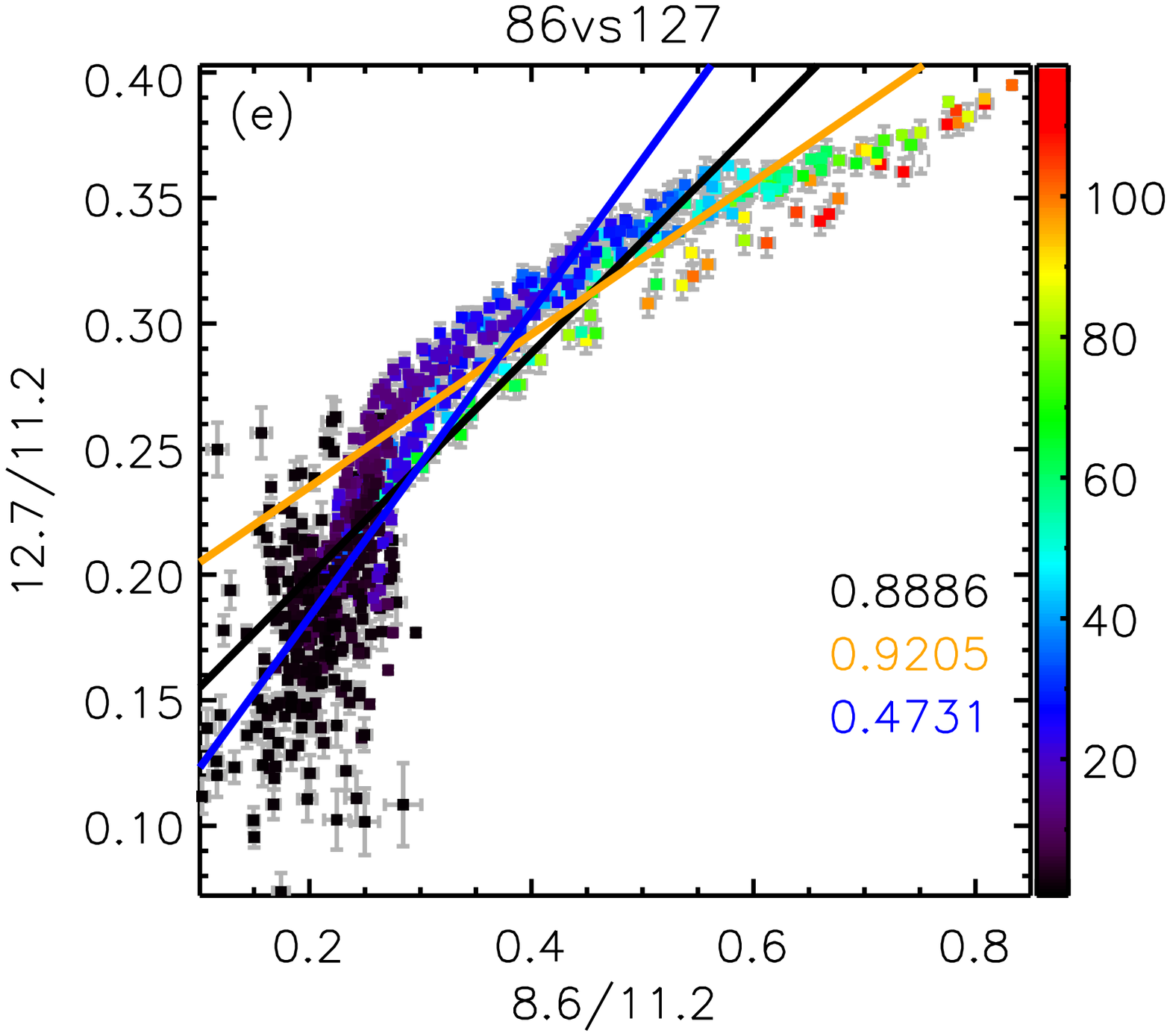}
\includegraphics[clip,trim =.3cm 1.3cm 1cm 3.2cm,width=0.5\textwidth]{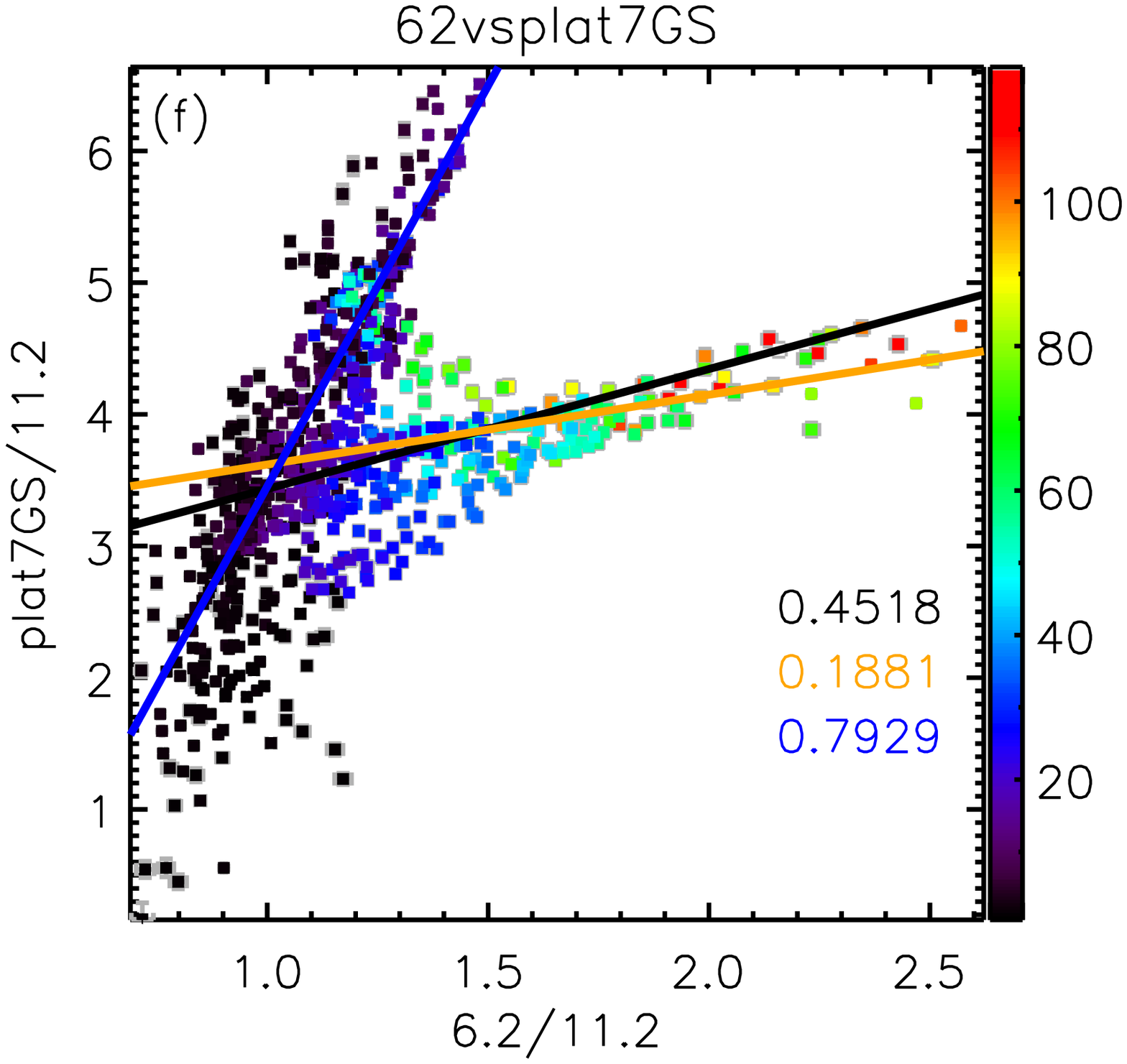}}

\resizebox{\hsize}{!}{%
\includegraphics[clip,trim =.3cm 1.3cm 1cm 3.2cm,width=0.5\textwidth]{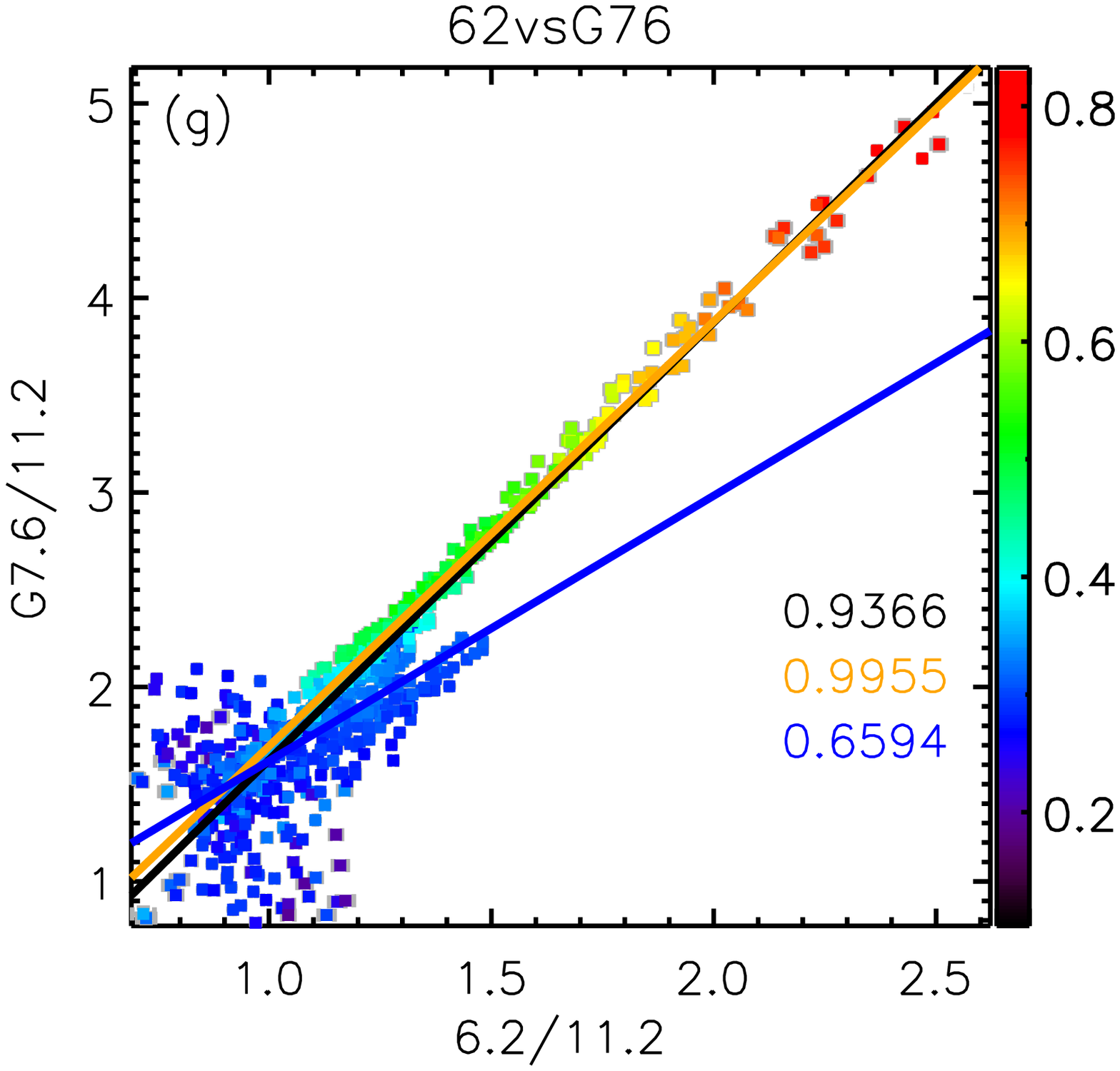} 
\includegraphics[clip,trim =.3cm 1.3cm 1cm 3.2cm,width=0.5\textwidth]{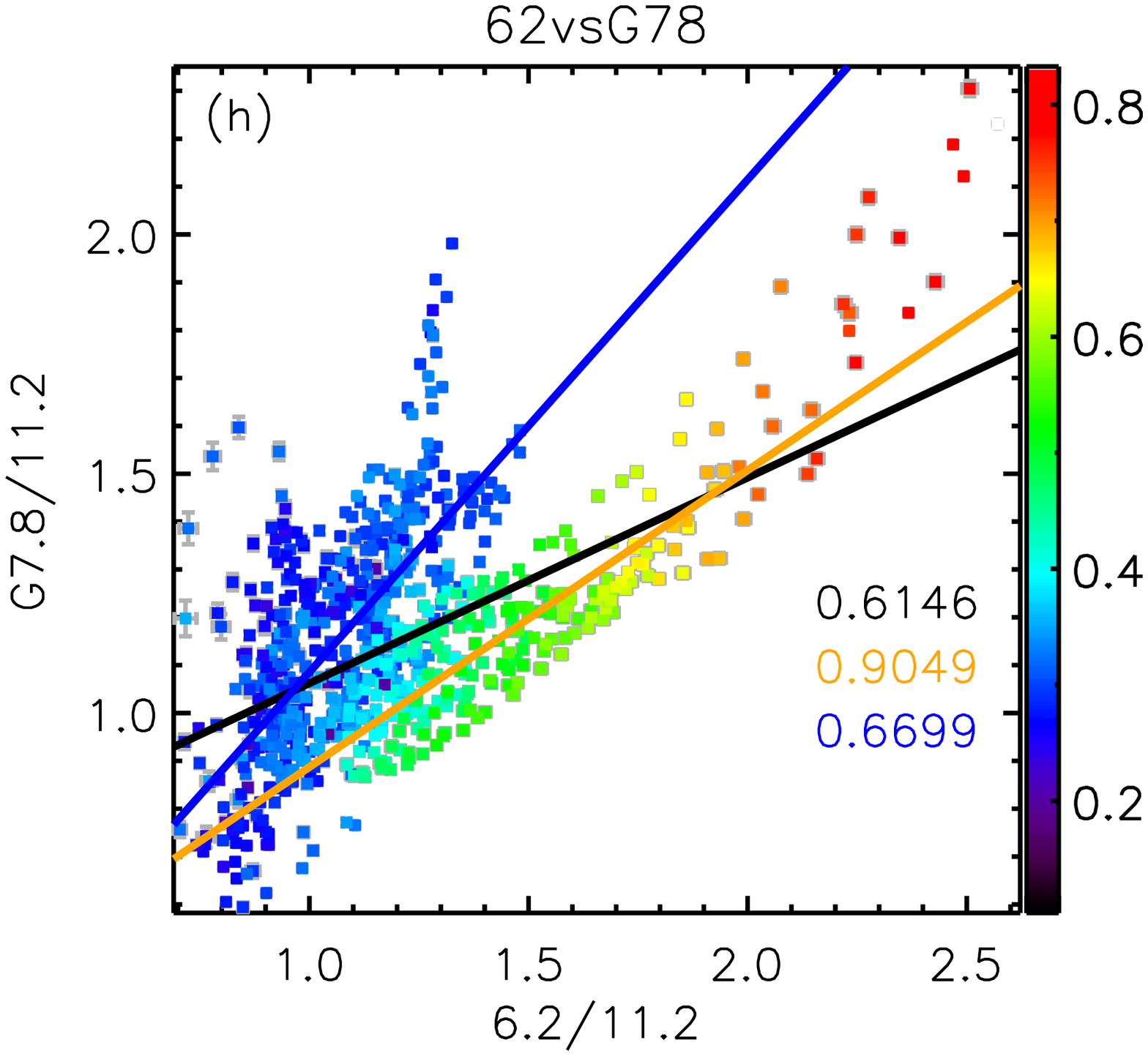}
\includegraphics[clip,trim =.3cm 1.3cm 1cm 3.2cm,width=0.5\textwidth]{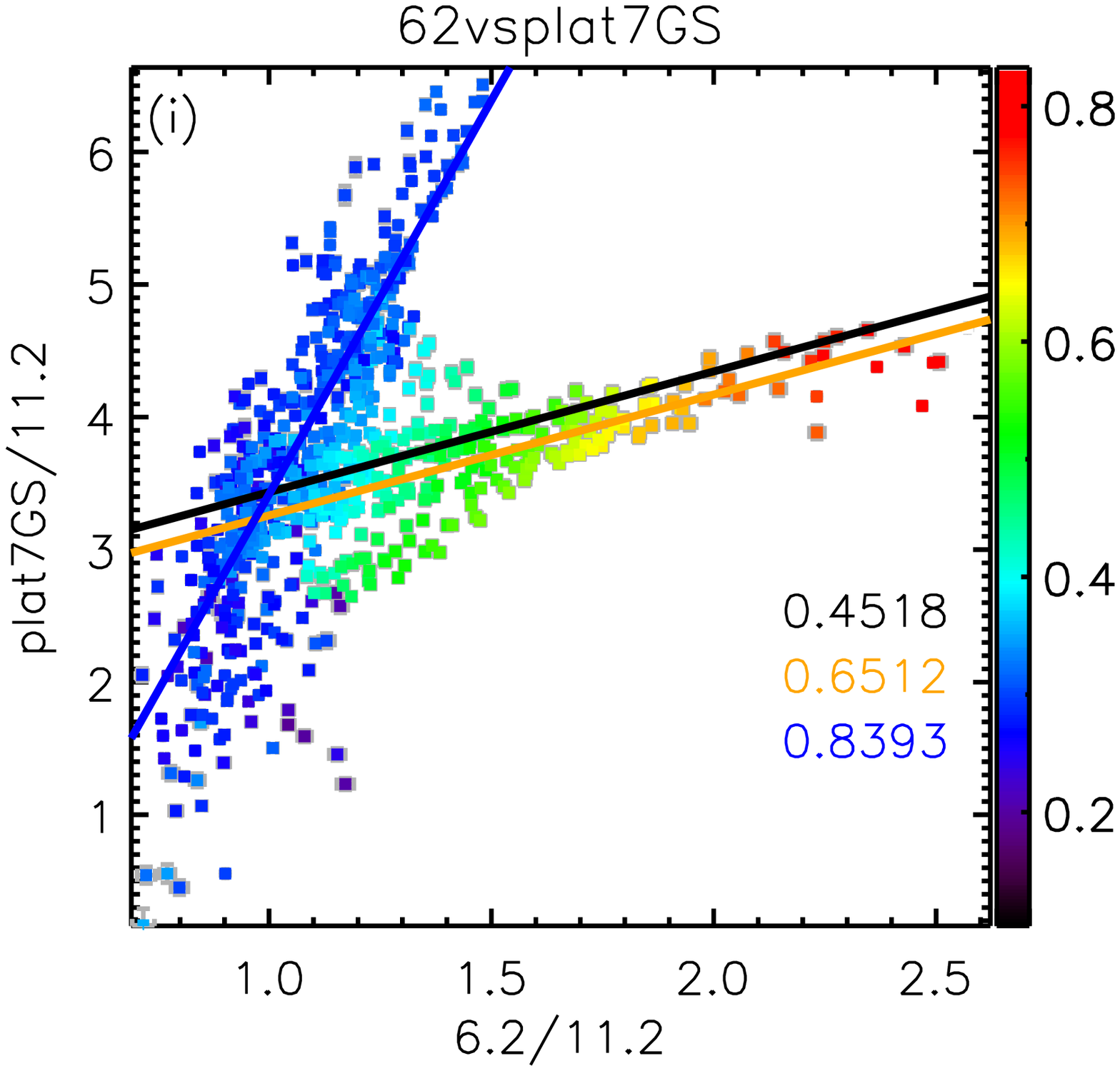}}
\resizebox{\hsize}{!}{%
\includegraphics[clip,trim =.3cm 1.3cm 1cm 3.2cm,width=0.5\textwidth]{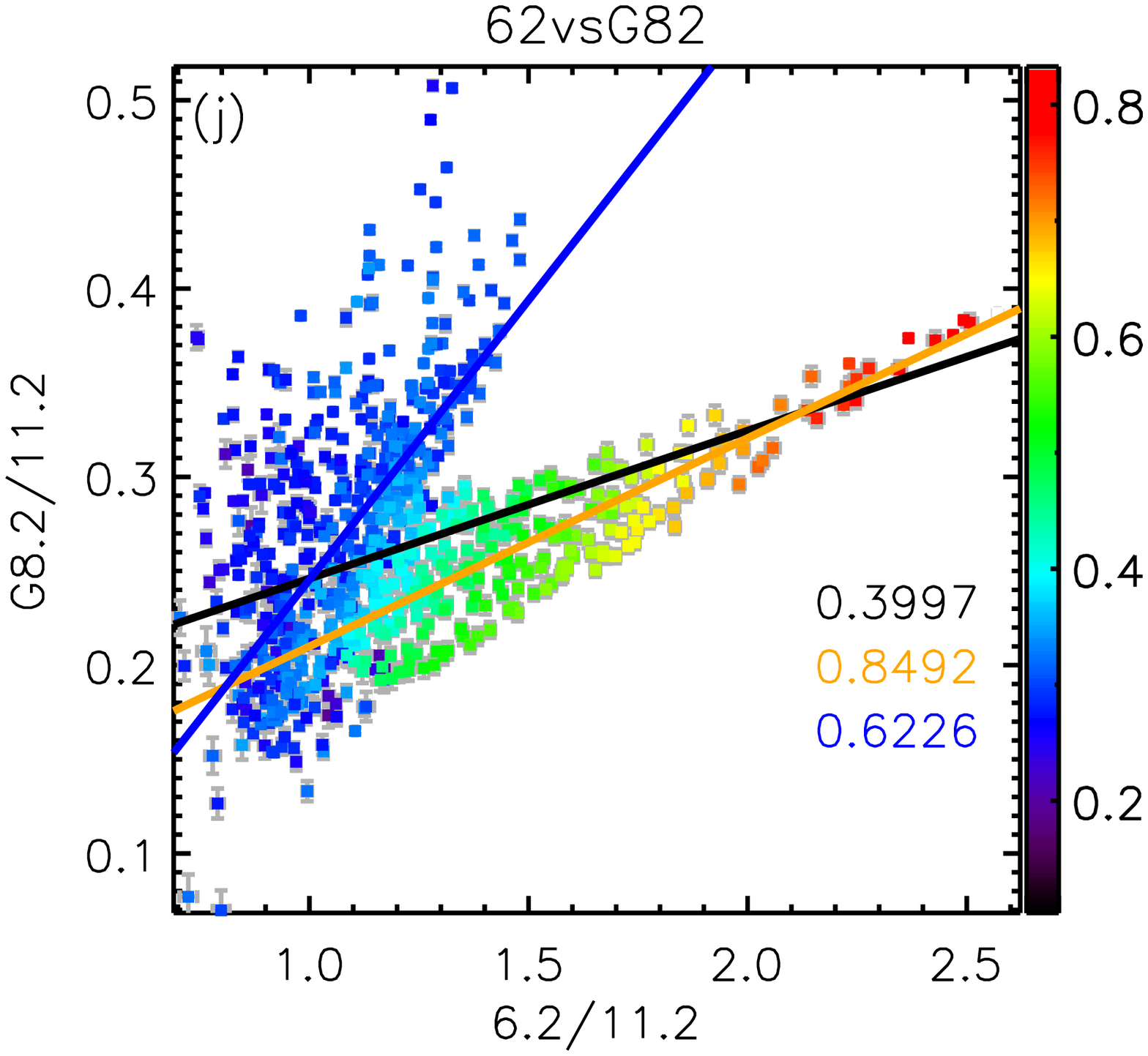}
\includegraphics[clip,trim =.3cm 1.3cm 1cm 3.2cm,width=0.5\textwidth]{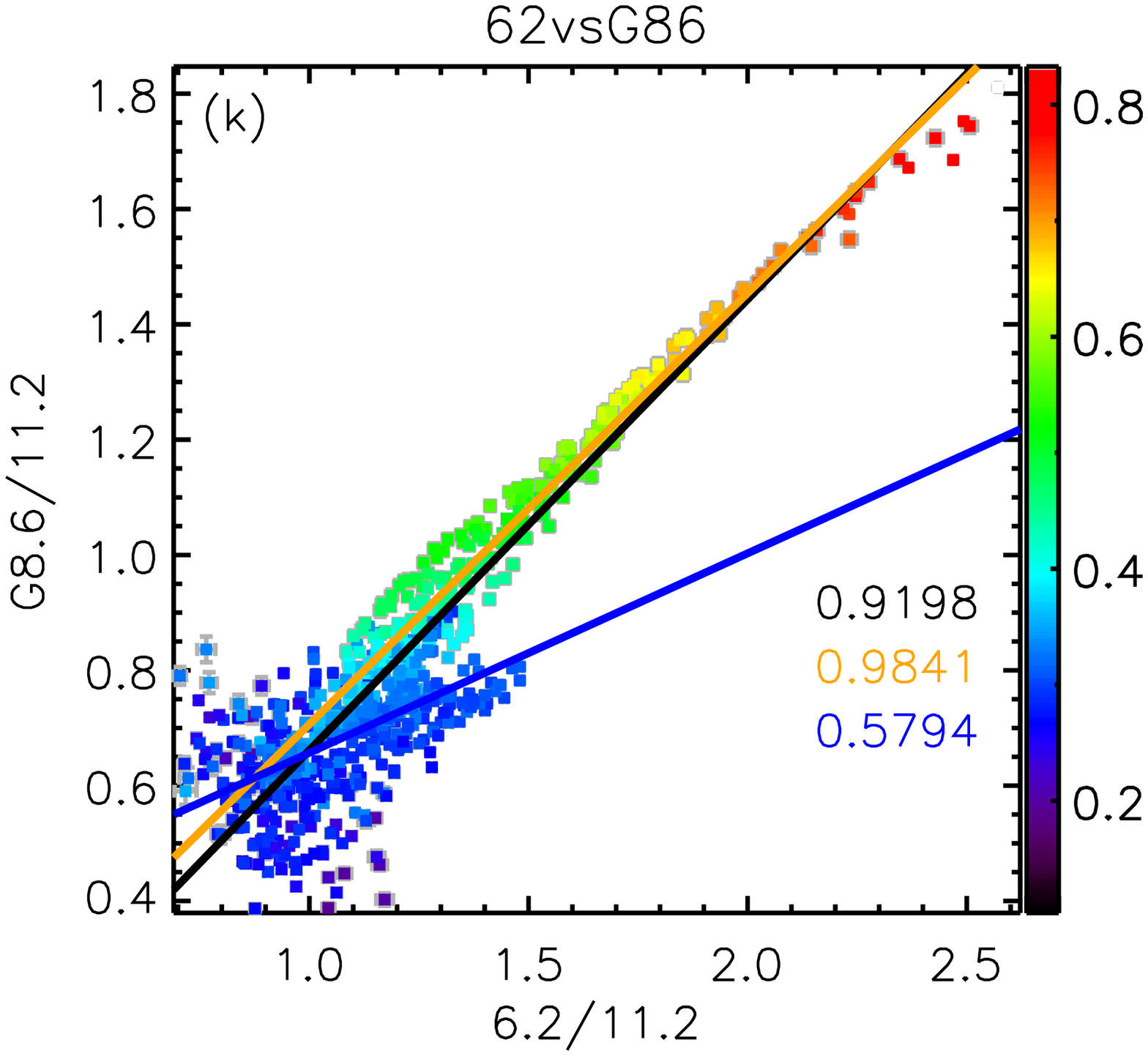}
\includegraphics[clip,trim =.3cm 1.3cm 1cm 3.2cm,width=0.5\textwidth]{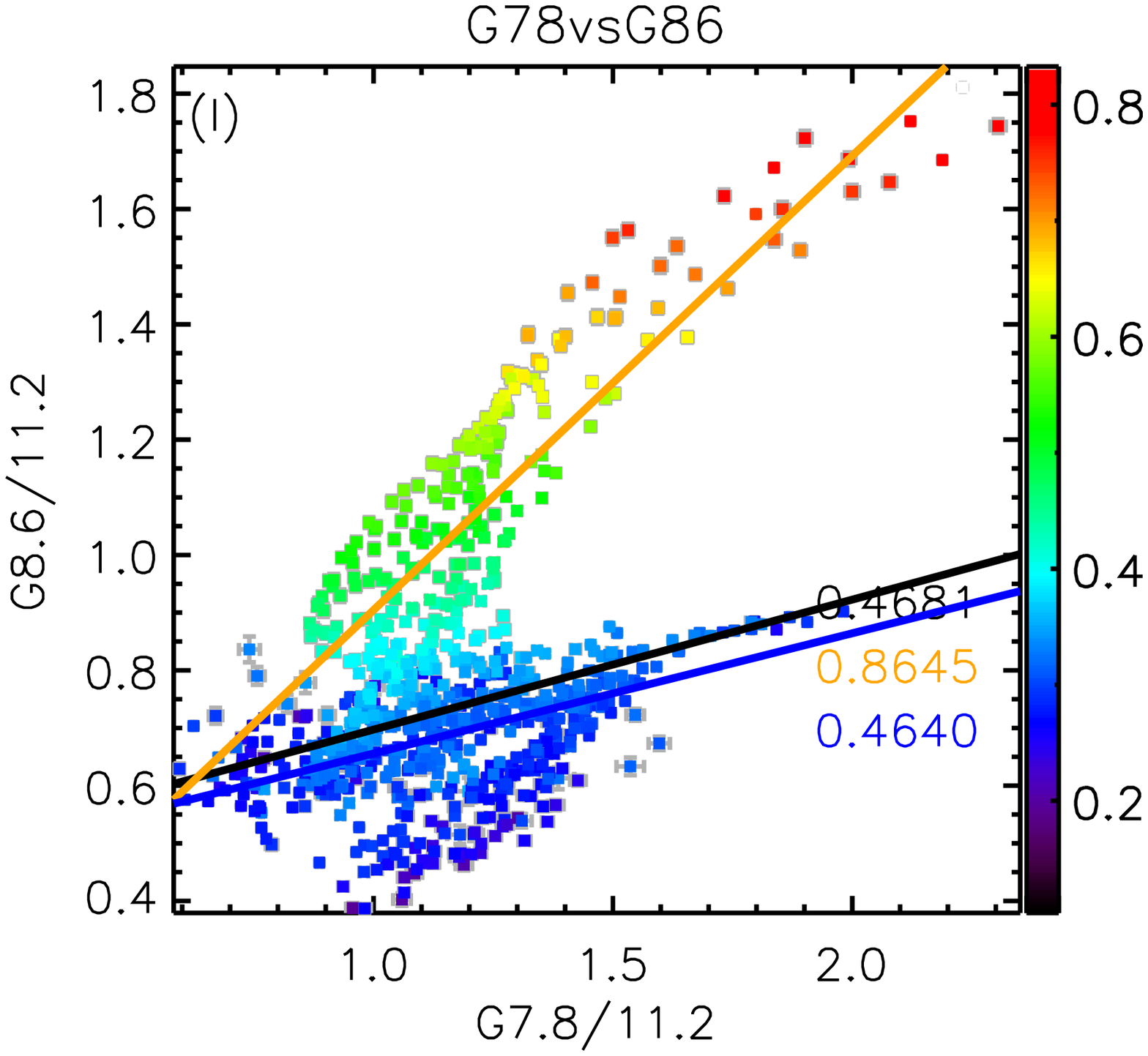}
}
\end{center}

\caption{Correlation plots between PAH features in NGC~1333. The data is color--coded to the PAH~7.7~LS~$\mu$m integrated intensity (in units of 10$^{-6}$~W~m$^{-2}$~sr$^{-1}$; top two rows) and to the PAH~8.6~LS/11.2 ratio (bottom two rows). Correlation coefficients are given in each panel for the entire map (black) and for the pixels where the 7.7~$\mu$m~LS flux $>$~10$^{-5}$~W~m$^{-2}$~sr$^{-1}$ (orange) and $\leq$~10$^{-5}$~W~m$^{-2}$~sr$^{-1}$ (blue, top two rows) and for the pixels where the  8.6/11.2 ratio $>$~0.3 (orange) and $\leq$~0.3 (blue, bottom two rows). Linear fits are shown in the same respective colors as for the regimes defined for the correlation coefficients.
}
\label{irs_corr_77rainbow}
\end{figure*}

\section{Discussion}
\label{discussion}

In this section, we investigate potential drivers of the PAH characteristics reported in Section~\ref{results} such as the environmental conditions and properties of the underlying PAH populations. Specifically, we investigate the origin of the non-linearity behavior in the reported correlations and relate these to the environment (diffuse ISM versus irradiated PDR) in Section~\ref{PAH dis}. We discuss the presence of multiple PAH sub-populations in Section~\ref{spat seq}. In Section~\ref{PAHs and PDRs}, we explore the diagnostic power of PAHs as PDR tracers.   

\subsection{PAH emission in the diffuse ISM versus irradiated PDR}
\label{PAH dis}

While we report significant correlations between various PAH features, clear deviations from the tight correlations and anomalous groups of pixels exist within our FOV (see Figure \ref{irs_corr}) that deserve a more in--depth investigation. We have illustrated that in particular for the 11.0~$\mu$m band, the degree of extinction introduces such an anomalous group of pixels. Clearly, due to its weakness and being located in the wing of the silicate absorption feature, the 11.0~$\mu$m band is more affected by extinction (correction) in comparison to the major PAH bands. However, significant scatter remains when we only account for the low extinction regime. Here we apply other weighting schemes in order to better characterize these discrepancies. In Figure \ref{irs_corr_77rainbow}, we show a selection of the correlation plots shown in Figure \ref{irs_corr} where we apply a color to each pixel based on the PAH~7.7~LS~$\mu$m flux (top two rows) and the 8.6/11.2 ratio (bottom two rows). We use the same color scale as shown for the 7.7~$\mu$m map in Figure \ref{irs maps} and for the 8.6/11.2~$\mu$m map in Figure \ref{irs_ratios} for easy comparison.

In panels (a), (b), and (c) of Figure \ref{irs_corr_77rainbow}, we compare the correlations between the major PAH bands in the 6--9~$\mu$m range. When only considering positions where the 7.7~$\mu$m flux is $>$~10$^{-5}$~W~m$^{-2}$~sr$^{-1}$, the correlations are tighter than in the case of the low silicate regime of Figure \ref{irs_corr}. This cut-off value for 7.7~$\mu$m flux distinguishes between the nebula itself and the outskirts in the FOV which correspond to the more diffuse ISM (see Figure \ref{irs maps}). Thus, these bands are much better correlated in the irradiated PDR than in the diffuse ISM. In addition, when the 8.6~$\mu$m band is considered, a stumpier second correlation is observed for locations where the 7.7~$\mu$m flux is $\leq$~10$^{-5}$~W~m$^{-2}$~sr$^{-1}$, i.e. the diffuse ISM. This second branch exhibits little change in the 8.6~$\mu$m band intensity while a larger change is seen in the 6.2 or 7.7~$\mu$m band intensity. Another notable anomaly is the `arc' found in the correlation of the 6.2 versus 8.6~$\mu$m bands and the 7.7 versus 8.6~$\mu$m bands while it is absent in the correlation of the 6.2 versus 7.7~$\mu$m bands. The majority of the pixels in the arc exhibit a low degree of extinction thereby excluding extinction as its origin. However, the pixels in the arc are located in the western edge of the FOV and thus the PAH emission may be affected by the nearby YSO. In addition, it is only when considering the 8.6~$\mu$m band that at high ratio values the data levels off. We note that the arc disappears and the degree of levelling off decreases significantly when considering the G8.6 component instead of the 8.6~$\mu$m band (which were obtained with different underlying continua). 

Concerning correlations with the 12.7~$\mu$m band (Figure~\ref{irs_corr_77rainbow}~(d),~(e)), two major regimes are present and characterized by high and low 12.7/11.2 (i.e. $>$ 0.25 or $<$ 0.25 respectively). This boundary corresponds approximately to the region between the two lowest contours in our 7.7~$\mu$m map (Figure \ref{irs maps}), representing the transition from the nebula to the diffuse ISM. In the lower flux regimes, there is significant scatter with a much wider range in 12.7/11.2 compared to those found in the other ionic bands normalized to the 11.2~$\mu$m band. In particular, the 6.2 or 7.7 versus 12.7 relation exhibit two separate arcs in this low 12.7/11.2 regime that seem to make up opposite halves of a circular trend. Based on the 7.7~$\mu$m band intensity, the right arc is largely located in the transition region between the nebula and the diffuse ISM, the left arc is relegated to the diffuse emission outside of the PDR while the nebula represents the high 12.7/11.2 regime. 

A similar behaviour with pixel location is seen for the strong bi-linear trend observed between the 6.2~$\mu$m band and 5--10~$\mu$m plateau (Figure~\ref{irs_corr_77rainbow}~(f)). The branch representing a large change in the plateau emission relative to the 11.2~$\mu$m band has a high correlation coefficient and low ionization degree (as traced by 8.6/11.2 intensity ratio in Figure \ref{irs_corr_77rainbow} (i)). This branch arises from the diffuse ISM. In contrast, the second branch exhibiting a smaller range in plateau intensity normalized to the 11.2~$\mu$m band represents the nebula where a medium to high degree of ionization is found. 
Bi-linear trends arising from different regions are also found between the 6.2~$\mu$m band and the four Gaussian components (Figure \ref{irs_corr_77rainbow}~(g),~(h),~(j),~(k),~(l)), in particular for the G8.2 and G7.8 components. An 8.6/11.2 value of about 0.3 distinguishes the nebula from the diffuse ISM and represents the transition between both (Figure \ref{irs_ratios}). Using this boundary condition, one branch arises from the nebula and exhibits a strong correlation while the other branch arises from the diffuse ISM and the transition region between the nebula and the diffuse ISM and shows more scatter. The diffuse ISM branch is very pronounced and distinct for the G8.2 component and subsequently the G7.8 component while for the G7.6 and G8.6 this branch is closer to the nebula branch and shows much less variation. In addition, the G7.8 and G8.2 components normalized to the 11.2~$\mu$m band show a stronger increase for a given increase in 6.2/11.2 for the diffuse ISM branch than the branch arising in the nebula while the opposite holds for the G7.6 and G8.6 components.

Finally, we note the existence of a third branch for relations between the G7.8 component and the G7.6 component, G8.6 component, or 8~$\mu$m bump (Figure ~\ref{irs_corr_77rainbow} (l) shows the second correlation). Specifically, the relationship between this set of bands is different in the nebula, the diffuse ISM as well as the transition region between both with the third branch representing the latter. Moreover, the diffuse ISM branch and the branch representing the nebula have roughly similar slopes. 
The common factors in this set of relations is the G7.8 component and the almost `identical' characteristics of the G7.6 and G8.6 components (cf. these bands exhibit the strongest correlation between any PAH bands). As the relation of the G7.8 with the G8.2 component exhibits a single linear correlation, albeit with significant scatter, and the G8.2 component exhibit bi-linear trends with the G7.6 and G8.6 components, the existence of the third branch confirms that the G7.8 and G8.2, as defined here, do not completely arise from the same PAH population. \\

Multi-linear trends in the relation between various PAH ratios have already been reported in the literature. Specifically, \cite{sto16} demonstrated that correlations between the 12.7 and 7.7~$\mu$m bands, normalized to the 11.2~$\mu$m band, showed different linear trends with respect to the different physical environments they probed from most quiescent environments to harsher environments associated with \HII\, regions and PDRs characteristic of RNe in between. Using the same sample, \cite{sto17} reported a bifurcation in the correlations between the G7--9~$\mu$m Gaussian components involving G7.8 and G8.2~$\mu$m, normalized to the 11.2~$\mu$m band intensity. These authors associated the G7.8~$\mu$m with diffuse regions and the G7.6~$\mu$m with irradiated PDRs, consistent with the results reported above. 
The presented analysis here and by these authors thus clearly demonstrates that the relative behaviour of the PAH bands (i.e. how the different band intensities normalized to the 11.2~$\mu$m band intensity relate to each other) depends on the type of environment in which they reside, the diffuse ISM, the irradiated PDR or the transition region between these environments.  Note that we detect a discrete set of trends or branches (e.g. one, two or three) and not a continuous distribution.  

This implies that (some of) the PAH bands arise from multiple PAH subpopulations or from multiple grand-PAHs \citep{and15} that i) each have distinct relative PAH intrinsic intensities, and ii) have different relative abundances in these different environments. For instance, the fact that we only see one branch or linear relationship between the G7.6 and G8.6 components and between the 6.2 and 7.7~$\mu$m bands indicates that both sets of bands arises from either a single PAH subpopulations or from PAH subpopulations which are co-located across the entire FOV (i.e. for all three environments: the diffuse ISM, the irradiated PDR, and the transition region). In contrast, when we observe two linear relationships or branches for a set of PAH bands, e.g. the 6.2 and the 8.6~$\mu$m band, this indicates these bands arise from at least two distinct subpopulations that have relative abundances dependent on the environment in which they reside. As a consequence, this then indicates that, similar to the 7.7~$\mu$m band, the 6.2~$\mu$m band arises from at least two distinct PAH subpopulations.  The latter was already implied by the fact that the 6.2~$\mu$m band correlates very strongly with the 7.7~$\mu$m band which has been found to arise from at least two subpopulations \citep[][this paper]{pee17}. Thus, we present support of this hypothesis by the unique behaviour of the PAH emission in the diffuse ISM versus the nebula.

\subsection{Spatial Sequence}
\label{spat seq}

While NGC~1333 has a concentric ovular morphology and is not observed as an edge-on PDR, our results reveal distinct spatial behaviour of the PAH emission components (Sections~\ref{slmaps} and \ref{irs lp}). In particular, with increasing distance from SVS~3 (in the S direction), we observe the peak emission of first the ionic bands (i.e. 6.2, 7.7, 8.6, and the 11.0~$\mu$m bands), followed by the neutral band (11.2~$\mu$m), and subsequently by H$_2$. Such a pattern has been reported numerous times and reflects that PAH charge is the primary driver of relative PAH intensities \citep[e.g.][]{job96b, slo99, ber07, gal08, boe13, pee17, sid20}. As for two other reflection nebulae \citep[NGC~2023 and NGC~7023;][]{pee12, sha16}, the 12.7~$\mu$m band peaks between the neutral and ionic peak emission, reflecting either an origin in both neutral and cationic PAHs, or that conditions conducive for changes in the charge balance are also conducive for changes in the molecular edge structure, or both.

Recent analysis of Spitzer-IRS spectral maps showed distinct behaviour of the ionic bands, which indicates we do probe changes in the PAH family beyond cationic versus neutral PAHs \citep{pee17, sid20, Knight:orion}. Specifically, these studies report that the ionic bands behave as roughly two distinct groups: one being the 6.2 and 7.7~$\mu$m bands and the other being the 8.6 and 11.0~$\mu$m bands, with the latter being less tight than the former potentially reflecting the need for three instead of two groups. Such a division in two groups is present but less clear in NGC~1333. The 11.0~$\mu$m band peaks closest to the star while the 6.2, 7.7, and 8.6~$\mu$m bands peak slightly further away (Figure~\ref{ngc1333_nsline1}~(a)). However, the 8.6~$\mu$m emission drops faster going south of the star compared to the 6.2 and 7.7~$\mu$m emission (Figure~\ref{ngc1333_nsline1}~(a)) and the 6.2, 7.7, and 8.6~$\mu$m ratios indicate that the 8.6~$\mu$m band is clearly distinct from the 6.2 and 7.7~$\mu$m bands (Figure \ref{ngc1333_nsline2}~(b)). Moreover, 
amongst the 6.2, 7.7, and 8.6~$\mu$m bands, the 8.6~$\mu$m band exhibits the strongest correlation with the 11.0~$\mu$m band (within the low silicate extinction regime; Figures~\ref{slcorr}~(r),~(s),~(t)). 
While this is not clear evidence for the 8.6 and 11.0~$\mu$m bands belonging to a single group, it does show that both the 8.6 and the 11.0 $\mu$m bands are distinct from the 6.2 and 7.7~$\mu$m bands. Thus, we infer there may be at minimum two and potentially three distinct PAH sub-populations responsible for the ionic bands.

Different sub-components of the 7--9~$\mu$m PAH emission in NGC~1333 exhibit distinct morphologies (Figure \ref{irsGS}). Specifically, the G7.6 and G8.6 components have the same spatial characteristics (seen in spatial maps and correlation plots) which is significantly different from the spatial characteristics of the G7.8 and G8.2 components, which are also similar, yet clearly not the same. This is consistent with the results of \citet{pee17} and \citet{sto17} for RNe.  In contrast to NGC~2023, the G7.8 and G8.2 show distinct morphology compared to that of the plateaus, which in turn, is distinct from those of the individual bands perched on top of them. The origin of these differences may originate from different band assignments for these components (and thus different structure and size of their carriers), different sizes of their carriers (including PAH, PAH clusters, and VSGs), and/or different charge states \citep[see][for a detailed discussion]{pee17}. Clearly, the applied decomposition in sub-components is simple in nature and will need to be refined when JWST observations of PDRs are available. Nevertheless, this analysis points to the contribution of multiple PAH sub-populations to the 7--9~$\mu$m PAH emission. \\

Summarizing, we have found that the PAH emission bands and their associated ratios qualitatively depend on their environment. Probing the relationships between PAH emission ratios and the physical conditions quantitatively could provide insight into a novel way to use PAH emission as PDR diagnostics. We further explore this in the next Section.

\begin{figure*}
\begin{center}
\resizebox{\hsize}{!}{%
\includegraphics[clip,trim =0.cm 0.3cm 1cm 1.2cm,width=3.5cm]{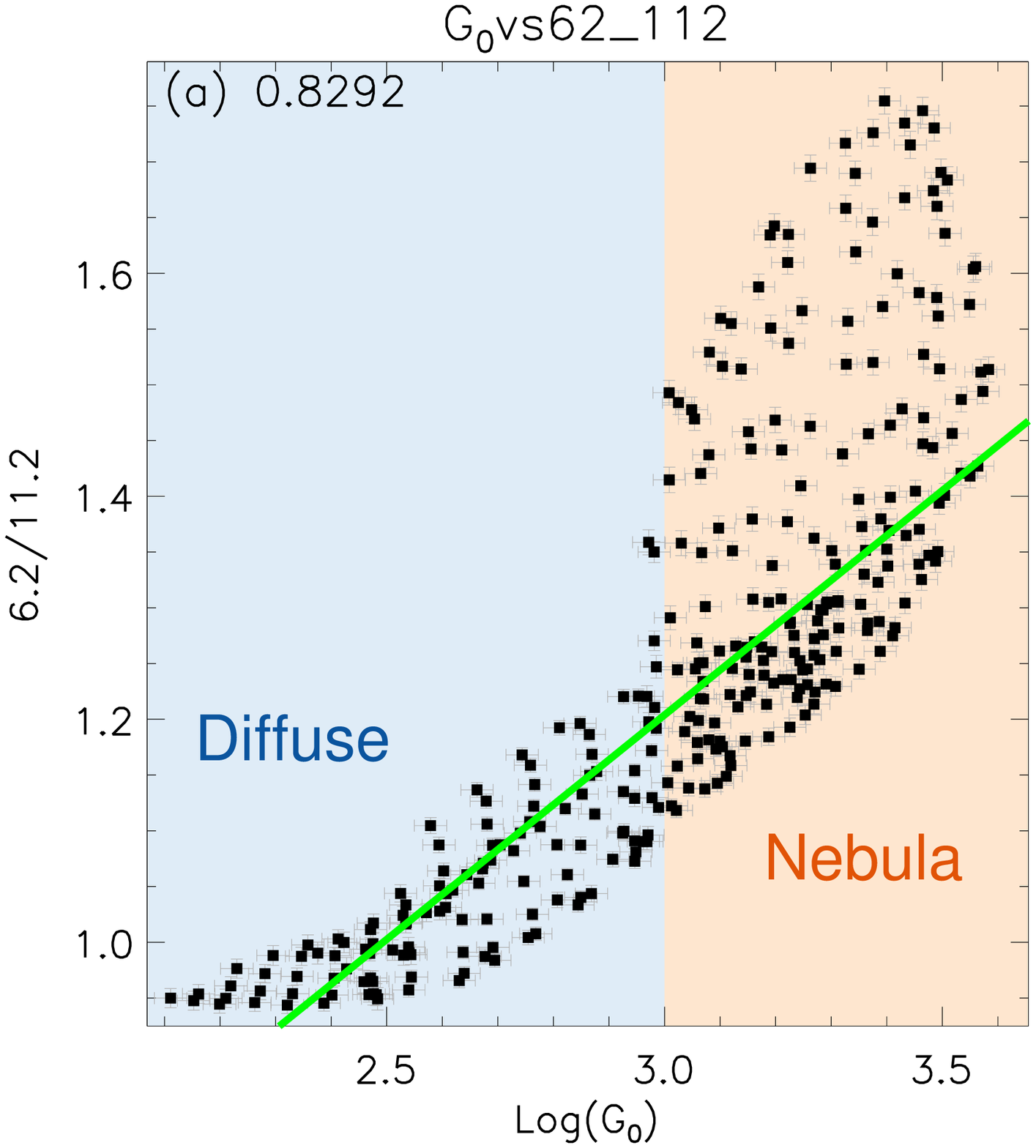}
\includegraphics[clip,trim =0.cm 0.3cm 1cm 1.2cm,width=3.5cm]{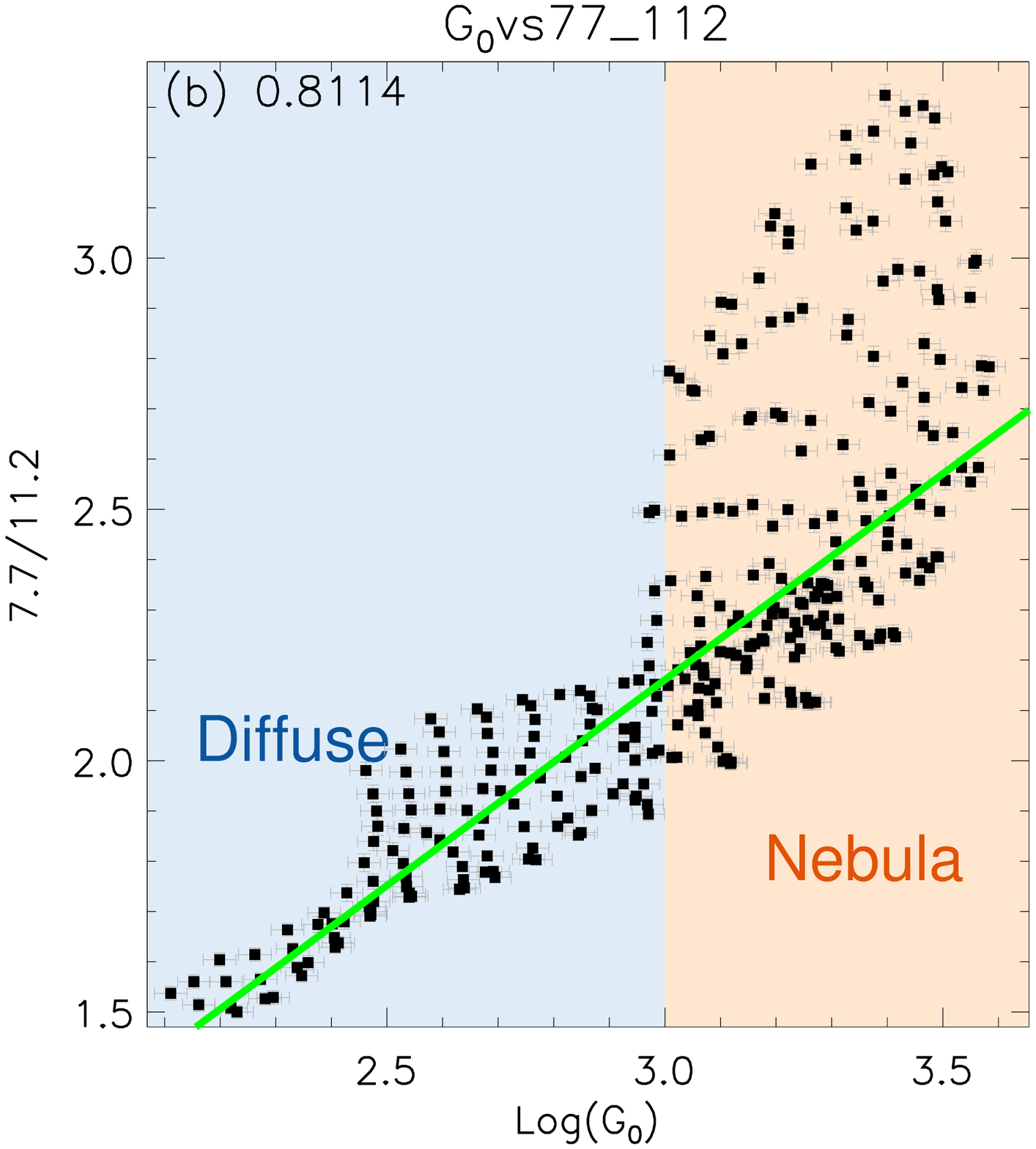}
\includegraphics[clip,trim =0.cm 0.3cm 1cm 1.2cm,width=3.5cm]{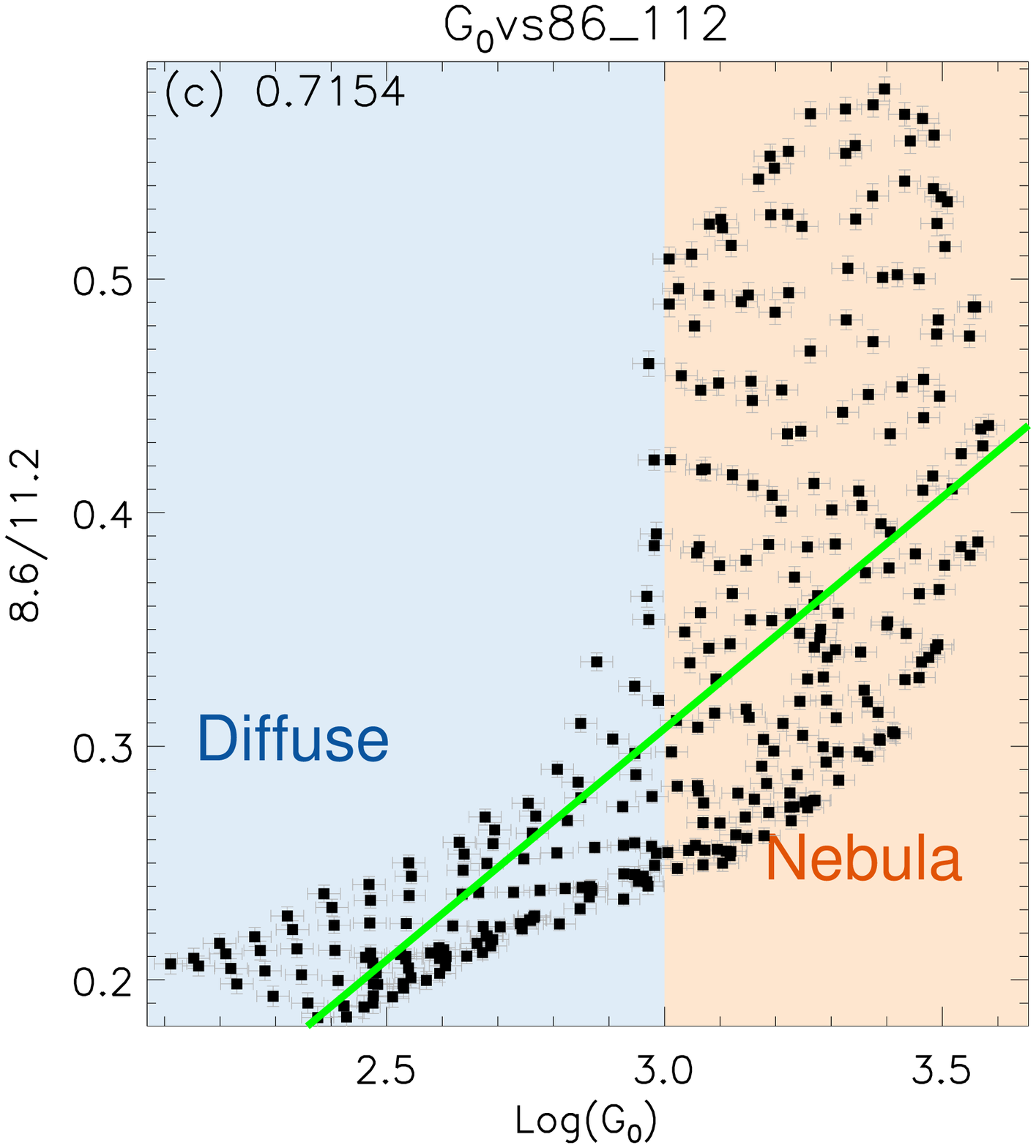}}
\resizebox{\hsize}{!}{%
\includegraphics[clip,trim =0.cm 0.3cm 1cm 1.2cm,width=3.5cm]{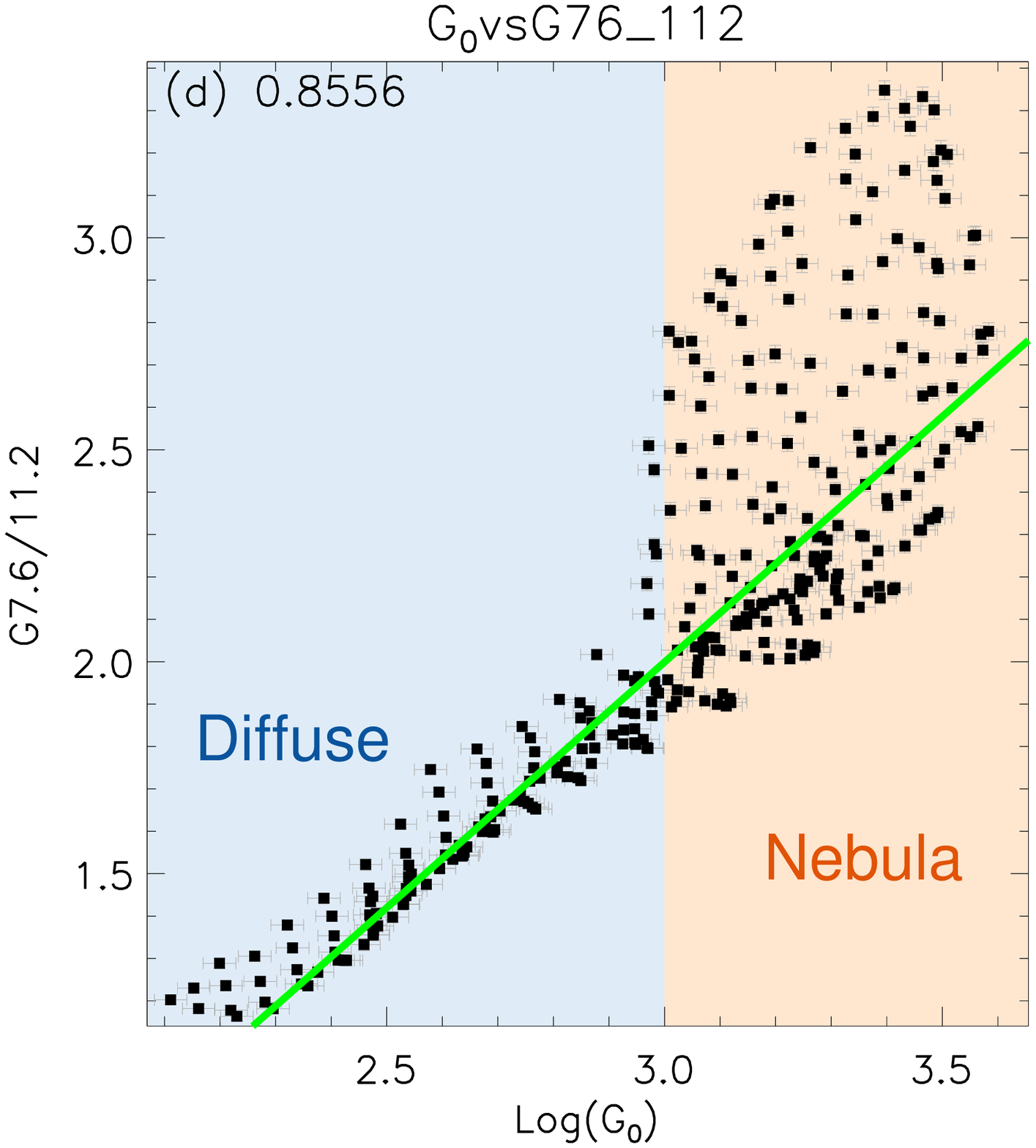}
\includegraphics[clip,trim =0.cm 0.3cm 1cm 1.2cm,width=3.5cm]{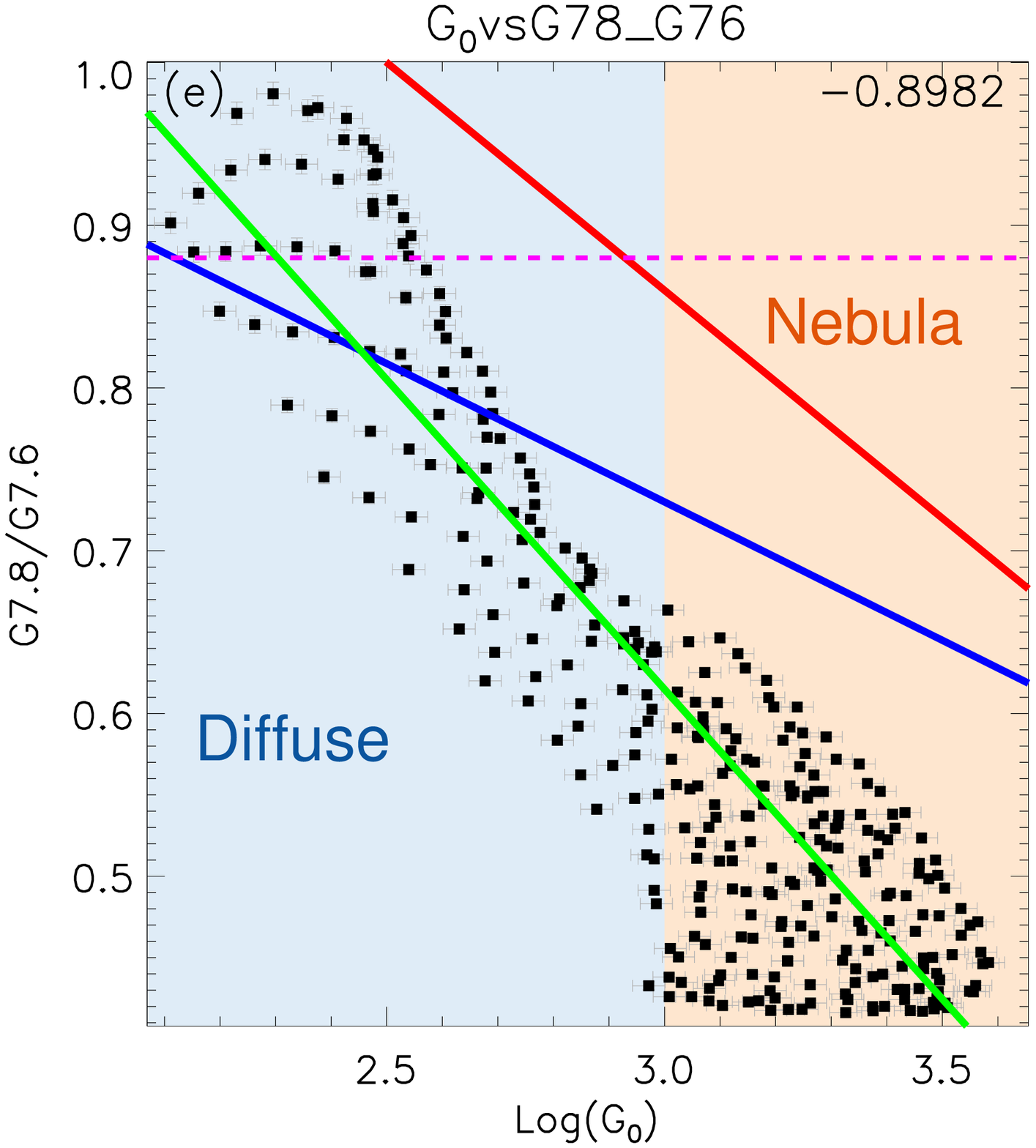}
\includegraphics[clip,trim =0.cm 0.3cm 1cm 1.2cm,width=3.5cm]{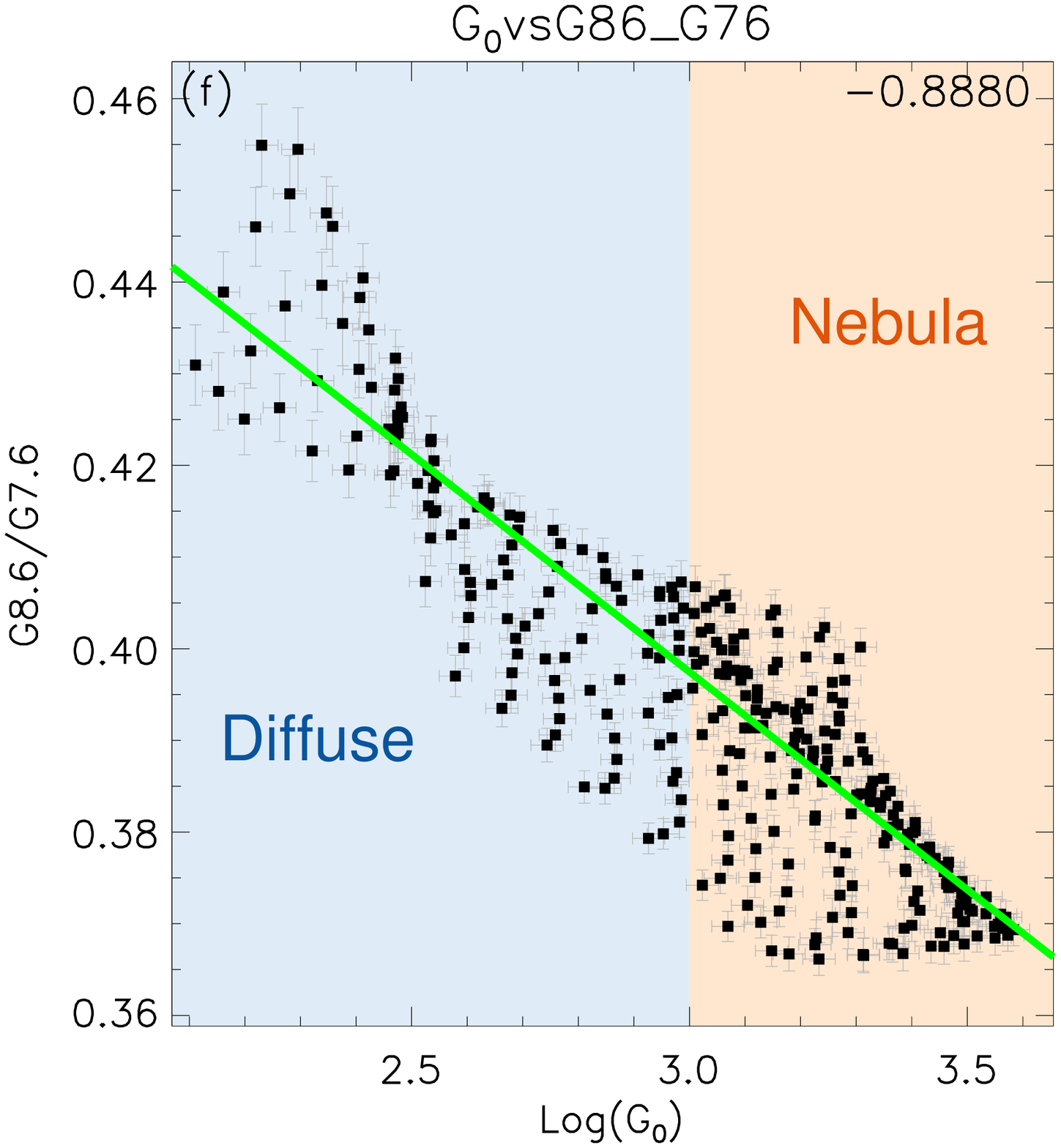}
}
\end{center}
\caption{Correlation plots between the strength of the FUV radiation field, G$_{0}$, and PAH emission ratios in NGC~1333. Correlation coefficients are given in each panel. Linear fits between log(G$_{0}$) and each PAH ratio is shown by a green line. We divide each plot into two regimes based on the relative strength of the FUV radiation field: log(G$_{0}$) $\geq$ 3 in orange corresponds to the irradiated PDR and log(G$_{0}$) $\le$ 3 in blue corresponds to the diffuse ISM. In the bottom center panel, we show linear fits of \citet{sto17}, with (blue) and without (red) inclusion of low G$_0$ regions. In this panel, the maximum G7.8/G7.6 ratio of 0.88 found in the diffuse outskirts of W49A by \citet{sto17} is shown as a magenta horizontal dashed line. }
\label{G0 corr}
\end{figure*} 

\begin{figure*}
\begin{center}
\resizebox{\hsize}{!}{%
\includegraphics[clip,trim =0.cm 0.3cm 1cm 1.15cm,width=3.5cm]{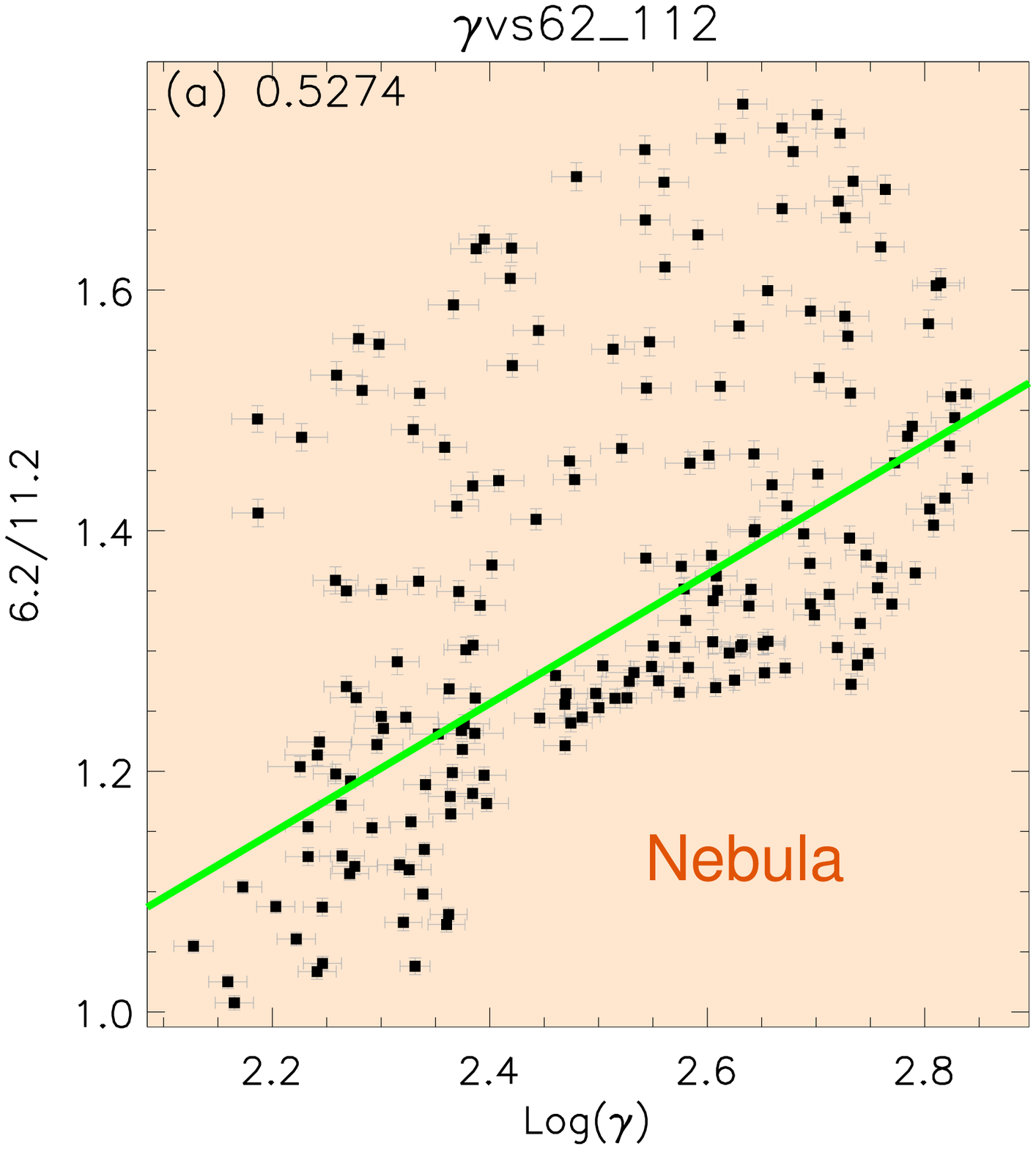}
\includegraphics[clip,trim =0.cm 0.3cm 1cm 1.15cm,width=3.5cm]{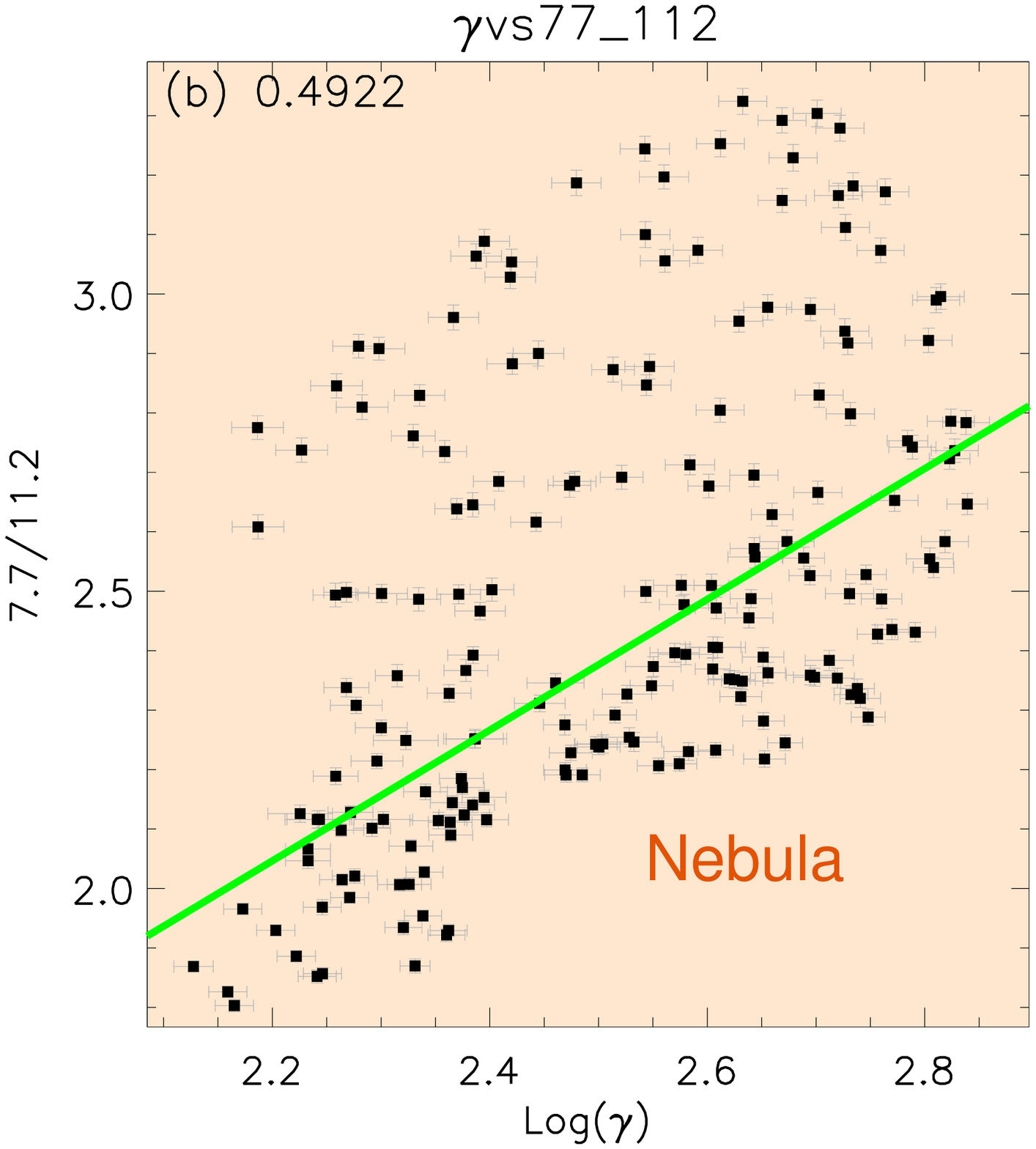}
\includegraphics[clip,trim =0.cm 0.3cm 1cm 1.15cm,width=3.5cm]{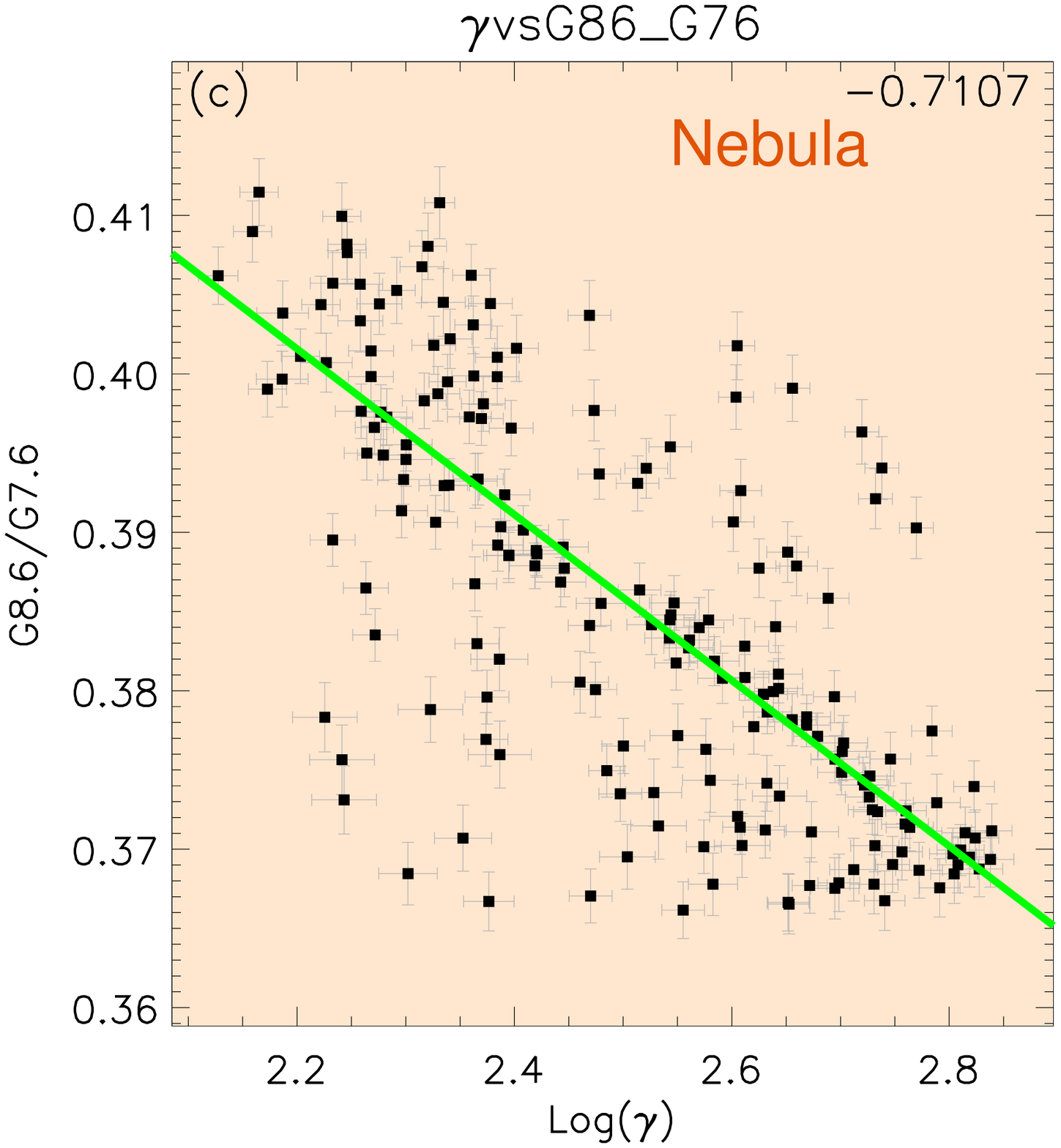}}
\end{center}
\caption{Relationship between the PAH ionization parameter,  $\gamma$, and PAH emission ratios in NGC~1333. The correlation coefficient are given in each panel. Linear fits between log($\gamma$) and each PAH ratio are shown by a green line.}
\label{Gamma corr}
\end{figure*} 

\subsection{PAHs as PDR Diagnostics}
\label{PAHs and PDRs}

In Figures~\ref{G0 corr}~and~\ref{G0 corr2}, we compare the strength of the FUV radiation field (G$_{0}$; as derived from the FIR dust continuum) with selected PAH ratios at the [\CII] 158~$\mu$m resolution. The three major PAH bands in the 6--9~$\mu$m range as well the G7.6~$\mu$m component, normalized to the 11.2~$\mu$m band, show very similar, positive, trends with G$_{0}$ (Figure~\ref{G0 corr}~(a),~(b),~(c)). The G7.6/11.2 intensity ratio has the best positive correlation with G$_{0}$ of all PAH (and PAH-related) bands (Figure~\ref{G0 corr}~(d)). These positive correlations reflect the fact that all four ratios trace the PAH ionization state which is driven, in part, by a stronger radiation field. We observe two separate regimes: a tight correlation between each ratio and G$_{0}$ for lower ranges of each quantity, while, for G$_{0}$ $>$ 10$^{3}$, we find substantial scatter for all four PAH ratios. In this regime, the observed relation rather gives a lower boundary to the considered PAH ratio. These stronger radiation field strengths are found within the nebula (see Figure~\ref{PDR maps}). The increased range in these PAH ratios for a given FUV radiation strength across the nebula likely reflects the larger variation present in the electron density and gas temperature across the nebula compared to the outskirts as these parameters also determine the PAH ionization parameter and thus the PAH ionization fraction. Specifically, the spatial distribution of the 6.2/11.2 emission ratio is approximately spherical (similarly for 7.7/11.2 and 8.6/11.2; see Figure~\ref{irs_ratios}). In contrast, the G0 map has a much larger divergence from spherical symmetry (Figure~\ref{PDR maps}). This discrepancy in the 2--dimensional morphology results in scatter seen in Figure~\ref{G0 corr} (See Appendix \ref{PAH Scatter} for more details).

A strong anti--correlation exists between G7.8/G7.6 and G$_{0}$ (Figure~\ref{G0 corr}~(e)) of the form:
\begin{equation}
I_{7.8}/I_{7.6} = 1.7575 \  -  \ 0.3797 \  \textrm{log}\ G_{0} 
\label{eq_78_76}
\end{equation}
In contrast with the ratios discussed above, this relationship does not show increased scatter for strong FUV radiation fields (G$_0 > 10^3$). This dependence of G7.8/G7.6 on G$_{0}$ strongly supports the results of \cite{sto17}. These authors found a strong anti--correlation of G7.8/G7.6 with G$_{0}$ for a sample of integrated PDRs over a wide range of physical conditions. Here we show that this also holds true within a spatially resolved source. However, we find that the slope of our fit (equation \ref{eq_78_76}) is significantly larger than the slope of both fits given by \citet[][see Figure \ref{G0 corr}~(e)]{sto17}. This contradicts their prediction that the inclusion of diffuse ISM regions or regions with a weak FUV radiation field will cause a shallower correlation between G7.8/G7.6 and G$_{0}$. One explanation for this discrepancy may be the lack of data at low G$_{0}$ in the \cite{sto17} study and the large uncertainties involved in estimating G$_{0}$ as mentioned in Section \ref{intro}. In particular, we only employed a single method to derive G$_{0}$ values at the same spatial scale as the PAH emission while \cite{sto17} resorted to multiple methods and considered the spatially integrated values of G$_{0}$ and G7.8/G7.6 in a sample of PDRs, which combined resulted in considerable uncertainties in G$_{0}$.
While we do not probe G$_{0}$ values as low as the Ophiuchus positions in \cite{sto17}, we do encounter G7.8/G7.6 ratios greater than those found in the diffuse outskirts of W49A, which has been argued to be representative of PAH spectra for diffuse ISM sightlines \citep{sto14}. In fact, pixels where G7.8/G7.6 $>$ 0.88 seem to fall in between the two correlations from \cite{sto17}, suggesting their predictions are not too far off for the diffuse ISM. We note that a similar increased slope was also found by \citet{Knight:orion} across the Orion Bar (also when employing a single method to determine G$_0$). 

We find another strong anti-correlation between G8.6/G7.6 and G$_{0}$ (Figure~\ref{G0 corr}~(f)). Given that the G7.6 and G8.6 components exhibit the strongest correlation of all PAH bands (when normalized to the 11.2~$\mu$m band) for NGC~1333 (Figure \ref{irs_corr}) and for NGC~2023  \citep{pee17}, this is somewhat surprising. However, we note that the range of variation in G8.6/G7.6 is quite small (of $\sim$~0.1) while G$_{0}$ varies over almost two orders of magnitude. Nevertheless, the observed anti-correlation with G8.6/G7.6 suggests that our simple four Gaussian decomposition needs further improvement to disentangle the components contributing to the 7 to 9~$\mu$m emission. Alternatively, or in addition, it may reflect the different band assignments of the 7.6 and 8.6~$\mu$m PAH emission. Indeed, while the G7.6~$\mu$m component is attributed to compact symmetric cationic PAH species with sizes ranging from 50~--~100~carbon atoms, the G8.6~$\mu$m PAH band is only detected in larger, highly symmetric, PAHs  \citep[, with a few exceptions,  in the size range from 96~--~150~carbon atoms, e.g.][]{bau08, bau09, ric12, pee17} or in less symmetric PAHs with straight edges, containing 50~carbon atoms or more \citep{ric18}. This tight correlation with G$_{0}$ may be suggestive of photo--processing of the compact, less symmetric PAHs with straight edges to compact, highly symmetric PAHs as they are increasingly exposed to the radiation field.

In Figure \ref{Gamma corr}, we compare the ionization parameter, $\gamma$, derived from the PDR models, with a few relevant PAH ratios. We note that convolving the IRS spectral maps to the PSF of the [\CII] 158~$\mu$m observation and accounting for the 3~$\sigma$ detection limits of the [\OI] 63~$\mu$m map in our PDR models restricts this comparison primarily to pixels within the nebula as shown in Figure \ref{PDR maps}. We do not find a significant correlation between 6.2/11.2 and 7.7/11.2 with $\gamma$ (Figure~\ref{Gamma corr}~(a),~(b)). Interestingly, no low PAH ratios are found for medium to high values of the PAH ionization parameter. Here as well, the observed relation (fit) gives a rough lower boundary to the considered PAH ratio. This is in contrast with the strong correlation between $\gamma$ and 6.2/11.2 found by \cite{gal08}, \cite{boe15}, and \cite{Knight:orion}. We note the different regimes in $\gamma$ considered in these studies: these studies consider regions where log($\gamma$)~$>$~4 with the addition of only 2 data points at lower $\gamma$ values in \citet{gal08} and \cite{Knight:orion}. Similarly as for the observed scatter in the 6.2/11.2-G$_0$ correlation, the observed scatter likely reflects the complex line-of-sight contribution of this spherical nebula (see also Appendix~\ref{PAH Scatter}). Moreover, NGC~1333 is not viewed as an edge-on PDR while the other studies were largely based on edge-on PDRs: the NW PDR in NGC~7023 in \cite{boe15} and the Orion Bar in \cite{gal08} and \cite{Knight:orion}. Additionally, the different methodology used to derive the PDR conditions may contribute to the observed discrepancy between the studies: \cite{gal08} assumed a fixed n$_{\textrm{H}}$ and T for the Orion Bar whereas we derived both n$_{\textrm{H}}$ and T using PDR models. Moreover, these models may not be sensitive enough to probe the PDR at the same resolution as the MIR and the FIR maps. Finally, we see a modest anti--correlation between $\gamma$ and G8.6/G7.6 (Figure~\ref{Gamma corr}~(c)).  However, there is significant scatter and the range in G8.6/G7.6 is very small ($\sim$~0.05). Hence, this may as well reflect the imperfectness of the four Gaussian decomposition.

In summary, from this comparison between PAH emission ratios and PDR conditions in and around the RN NGC~1333 there is clearly evidence to support the case for the emission features as viable PDR diagnostics in particular with respect to G$_{0}$.

\section{Conclusion}
\label{conclusion}

We characterize the PAH emission features and investigate their dependence on the physical conditions for the RN NGC~1333 by using a {\it Spitzer}--IRS spectral map from 5--15 $\mu$m, {\it SOFIA} FIFI--LS spectral maps of the FIR cooling lines [\OI] at 63 and 146~$\mu$m and [\CII] at 158~$\mu$m and {\it Herschel} PACS images at 70, 100 and 160~$\mu$m. We derive maps for the PDR's physical conditions using the FIR cooling lines and the FIR dust continuum emission determined from SED fitting of the PACS images in combination with the PDR models of \cite{kau06} and \cite{pou08}. The derived physical conditions are in agreement with previous PDR modelling efforts of \cite{you02}. Subsequently, we compare the MIR PAH emission with the PDR conditions determined from the FIR data using matching apertures and spatial resolution for the MIR and FIR data. 

Within the nebula, we find a distinction in the behaviour between the PAH emission bands in the 6--9~$\mu$m range and the 11.2~$\mu$m band along with the underlying plateaus between 5--10 and 10--15~$\mu$m with respect to the distance from the illuminating source, SVS~3. Namely, the 6--9~$\mu$m PAH bands are much more condensed within the nebula relative the 11.2~$\mu$m band and the plateaus. This dichotomy is also reflected in the 7--9~$\mu$m Gaussian components where the G7.6 and G8.6~$\mu$m components show a similar structure to the 6--9~$\mu$m bands whereas the G7.8 and G8.2~$\mu$m components are more comparable to the 11.2~$\mu$m band group.

More precise detail emerges upon considering emission cross cuts from SVS~3 in the N-S direction. In particular, the peak emission of the various bands exhibit a clear stratification with respect to distance from the star. The 6.2 and 7.7~$\mu$m band show similar spatial behaviour that is distinct from both the 8.6 and 11.0~$\mu$m bands, which are also distinct from one another. The 7--9~$\mu$m emission has contributions from at least two sub-populations with distinct morphology and the spatial characteristics of the plateaus are distinct from that of the features perched on top. We attribute these to the effects of photoprocessing on the implicit PAH population characterized through spatial variations in the molecular properties including PAH charge, size, and structure. 

In addition, we find the PAH characteristics of the nebula and the diffuse outskirts to be distinct from each other. Specifically, the PAH emission features within the nebula tend to be stronger and more tightly correlated with respect to the PAH features from the diffuse regions. In several cases, we find separate linear correlations for the diffuse emission and the nebula suggesting their PAH population consist of distinct sub-populations with different underlying properties. This supports previous evidence of multi-linear trends discovered between PAH ratios with respect to different physical environments.

We investigate previously reported relationships between PAH emission ratios and the physical condition within the relatively simple PDR geometry of the RN NGC~1333. We find strong correlations between the ratio of the 6--9~$\mu$m bands to the 11.2~$\mu$m band and the FUV radiation field strength particularly within the diffuse ISM. We also find strong negative correlations between ratios of the 7--9~$\mu$m Gaussian components, G7.8/G7.6 and G8.6/G7.6, with respect to G$_{0}$. This supports a similar relationship found by \cite{sto17} between G7.8/G7.6 and G$_{0}$ for a sample of PDRs covering 3 orders of magnitude in G$_{0}$. We did not find strong correlations between the ionization parameter and the 6.2/11.2 contradicting the previously established relationships of \cite{gal08} and \cite{boe15}. However, we did find there is a promising correlation between G8.6/G7.6 and $\gamma$ that warrants additional investigation.

To conclude, PAHs have much potential as PDR diagnostics. Further refinement of these correlations between PAH emission and PDR conditions will solidify the value of the PAH emission features in PDR studies. A robust survey of PAH and FIR observations in spatially resolved PDR sources within the Galaxy is the next step to accomplish this goal. With the imminent launch of JWST and the FIR access of SOFIA, a new golden age of astronomy will allow astronomers to characterize PDR environments and their PAH emission in our own Milky Way as well as in extragalactic sources.

\section*{Acknowledgements}
The authors thank Dario Fadda for his helpful suggestions for using his SOSPEX software. C.K. acknowledges support from an Ontario Graduate Scholarship (OGS). E.P. acknowledges support from an NSERC Discovery Grant and a SOFIA grant. Based [in part] on observations made with the NASA-DLR Stratospheric Observatory for Infrared Astronomy (SOFIA). SOFIA is jointly operated by the Universities Space Research Association, Inc. (USRA), under NASA contract NAS2-97001, and the Deutsches SOFIA Institut (DSI) under DLR contract 50 OK 0901 to the University of Stuttgart. This work is based [in part] on observations made with the Spitzer Space Telescope, which is operated by the Jet Propulsion Laboratory, California Institute of Technology under a contract with NASA. {\it Herschel} is an ESA space observatory with science instruments provided by European-led Principal Investigator consortia and with important participation from NASA. PACS has been developed by a consortium of institutes led by MPE (Germany) and including UVIE (Austria); KU Leuven, CSL, IMEC (Belgium); CEA, LAM (France); MPIA (Germany); INAF-IFSI/OAA/OAP/OAT, LENS, SISSA (Italy); IAC (Spain). This development has been supported by the funding agencies BMVIT (Austria), ESA-PRODEX (Belgium), CEA/CNES (France), DLR (Germany), ASI/INAF (Italy), and CICYT/MCYT (Spain).

\section*{Data Availability}

The data underlying this article will be shared on reasonable request to the corresponding author.




\bibliographystyle{mnras}
\bibliography{mainbib}



\appendix

\section{NGC~1333 Spitzer IRS~SL Cross Cut Data}
\label{ngc1333 irs components}

In Table \ref{table:1}, we summarize the prominent emission components found in the NGC~1333 Spitzer IRS~SL cross cuts as discussed in Section~\ref{irs lp}. We use the same nomenclature used to refer to each feature in Figure \ref{ngc1333_nsline1} and \ref{ngc1333_nsline2}. Furthermore, in Tables \ref{table:2} and \ref{table:3}, we provide the normalization factors for each cross cut shown in this work.

\begin{table}
\begin{center}
\caption{\label{table:1}{}NGC~1333 IRS~SL Spectral Emission Components.}
\begin{tabular}{ c c }
\hline\hline
Feature Tag$^{1}$ &  Emission Description  \\

\hline\\[-2pt]
\multicolumn{2}{c}{\bf Dust Emission}\\[1pt]
cont.~10.3 & Dust continuum emission at  10.3~\mum\, \\
cont.~13.9 & Dust continuum emission at  13.9~\mum\, \\
A$_{k}$ & Silicate extinction in units of A$_{k}$\\
\hline\\[-10pt]
\multicolumn{2}{c}{\bf PAH--Related Emission}\\[2pt]
6.2 & PAH 6.2~\mum\, band  \\
7.7 & PAH 7.7~\mum\, band  \\
8.6 & PAH 8.6~\mum\, band  \\
G7.6 & PAH G7.6~\mum\, sub--component  \\
G7.8 & PAH G7.8~\mum\, sub--component  \\
G8.2 & PAH G8.2~\mum\, sub--component  \\
G8.6 & PAH G8.2~\mum\, sub--component  \\
11.0 & PAH 11.0~\mum\, band \\
11.2 & PAH 11.2~\mum\, band  \\
12.7 & PAH 12.7~\mum\, band   \\
8~Bump & 8~\mum\, bump PAH plateau  \\
5--10 plat & 5--10~\mum\, PAH plateau  \\
10--13 plat & 10--13~\mum\, PAH plateau \\
\hline\\[-10pt]
\multicolumn{2}{c}{\bf Molecular Hydrogen Lines}\\[2pt]
H$_{2}$ 9.7 & H$_{2}$ 9.7~\mum\, emission line \\
\hline\\[-5pt]

\hline\\[-10pt]

\end{tabular}
\end{center}
$^1$ Shorthand used to refer to individual features in Figures \ref{ngc1333_nsline1} and \ref{ngc1333_nsline2}.
\end{table}

\renewcommand{\arraystretch}{1.5}
\begin{table}
\caption{\label{table:2}{}NGC~1333 emission cross cut normalization factors.}
\begin{center}
\begin{tabular}{ p{2.5cm} p{2 cm} }
\hline\hline
Emission Feature ($\mu$m) &  Normalization Factor$^1$  \\
\hline\\[-5pt]
cont.~10.3 & 3.72 (-5) \\
cont.~13.9 & 3.33 (-5)  \\
A$_{k}$ & 1.15 \\
6.2 & 5.71 (-5)  \\
7.7 & 1.10 (-4) \\
8.6 & 1.96 (-5) \\
G7.6 & 1.15 (-4) \\
G7.8 & 4.12 (-5) \\
G8.2 & 9.08 (-6) \\
G8.6 & 4.15 (-5)\\
11.0 & 3.88 (-6)\\
11.2 & 3.18 (-5)\\
12.7 & 1.01 (-5)\\
8~Bump  & 7.58 (-5)\\
5--10 plat & 1.32 (-4)\\ 
10--13 plat & 3.71 (-5)\\
H$_{2}$ 9.7 & 9.26 (-7)\\
\hline\\[-10pt]

\end{tabular}

\end{center}
$^1$ Normalization factors in units W~m$^{-2}$~sr$^{-1}$ (i.e multiply by these values to get the original values). Emission strength given in format: 3 significant digits (Order of magnitude). 
\end{table}

\renewcommand{\arraystretch}{1.5}
\begin{table}
\caption{\label{table:3}{} NGC~1333 emission ratio cross cut normalization factors.}
\begin{center}
\begin{tabular}{ p{2.5cm} p{2 cm} }
\hline\hline
Emission Ratio ($\mu$m) & Normalization factor$^{1}$ \\

\hline\\[-5pt]

6.2/7.7 & 0.636  \\
8.6/7.7  & 0.185 \\
8.6/6.2  & 0.343 \\
6.2/11.2  & 2.54 \\
11.0/11.2 & 0.173 \\
7.7/11.2 & 5.12 \\
8.6/11.2 & 0.820 \\
12.7/7.7 & 0.131 \\ 
12.7/11.2 & 0.392 \\
12.7/6.2 & 0.244 \\
12.7/8.6 & 0.884 \\
11.0/12.7 & 0.447 \\
11.0/6.2 & 0.0723 \\
11.0/7.7 & 0.0366 \\
11.0/8.6 & 0.218 \\
G7.8/G7.6 & 0.669 \\
G7.8/G8.2 & 5.91 \\
G8.2/G8.6  & 0.509\\
G8.6/G7.6 & 0.402 \\
(G7.6 +G7.8)/7.7 & 1.56 \\
\hline\\[-10pt]
\end{tabular}
\end{center}
$^{1}$ Multiply emission ratio by respective normalization factors to get original ratio.
\end{table}

\section{Additional Considerations for PAHs as PDR Diagnostics}
\label{PAHs and PDRs 2}

\begin{figure*}
\begin{center}
\resizebox{\hsize}{!}{%
\includegraphics[clip,trim =0.cm 0.3cm 1cm 1.2cm,width=3.5cm]{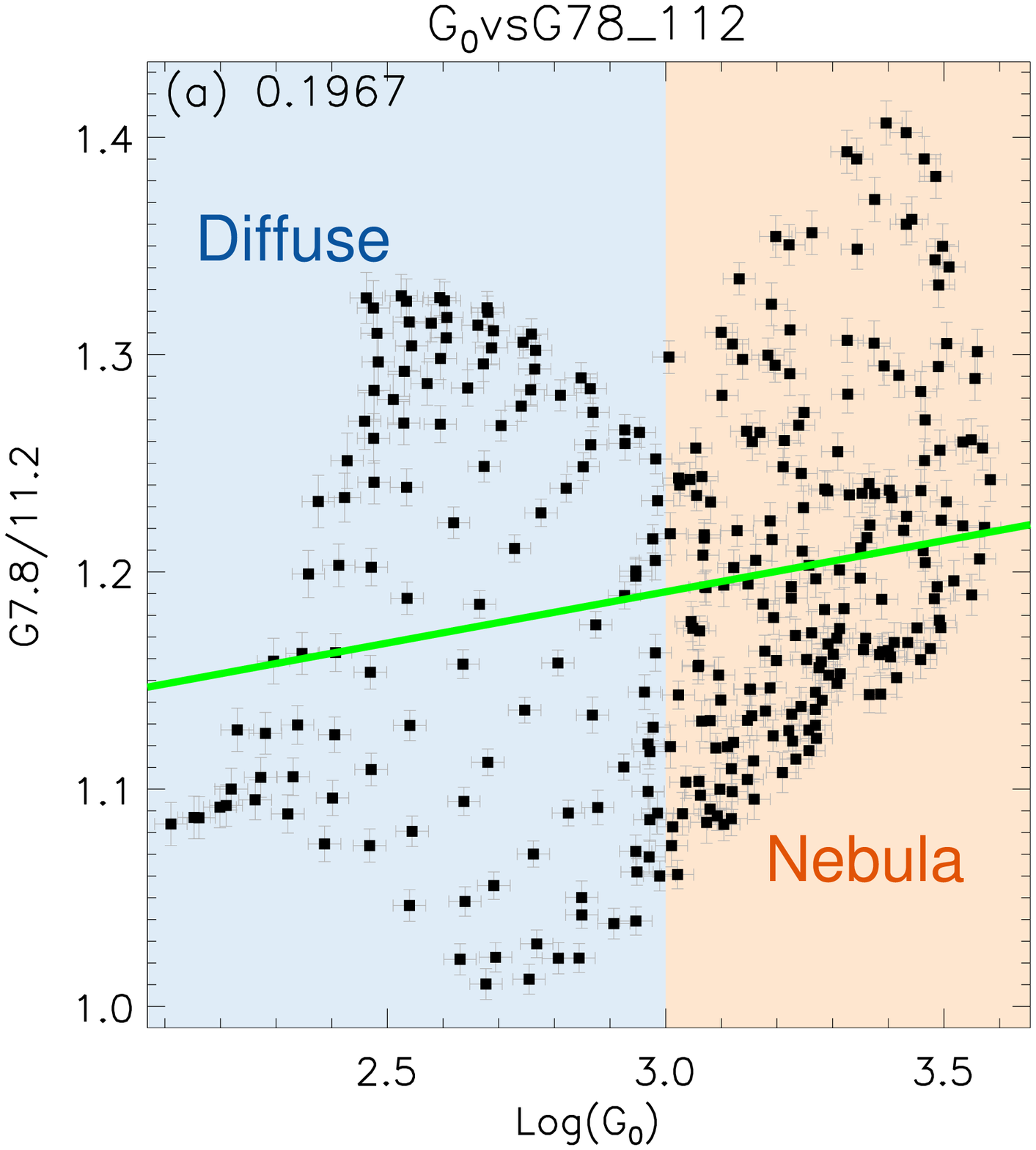}
\includegraphics[clip,trim =0.cm 0.3cm 1cm 1.2cm,width=3.5cm]{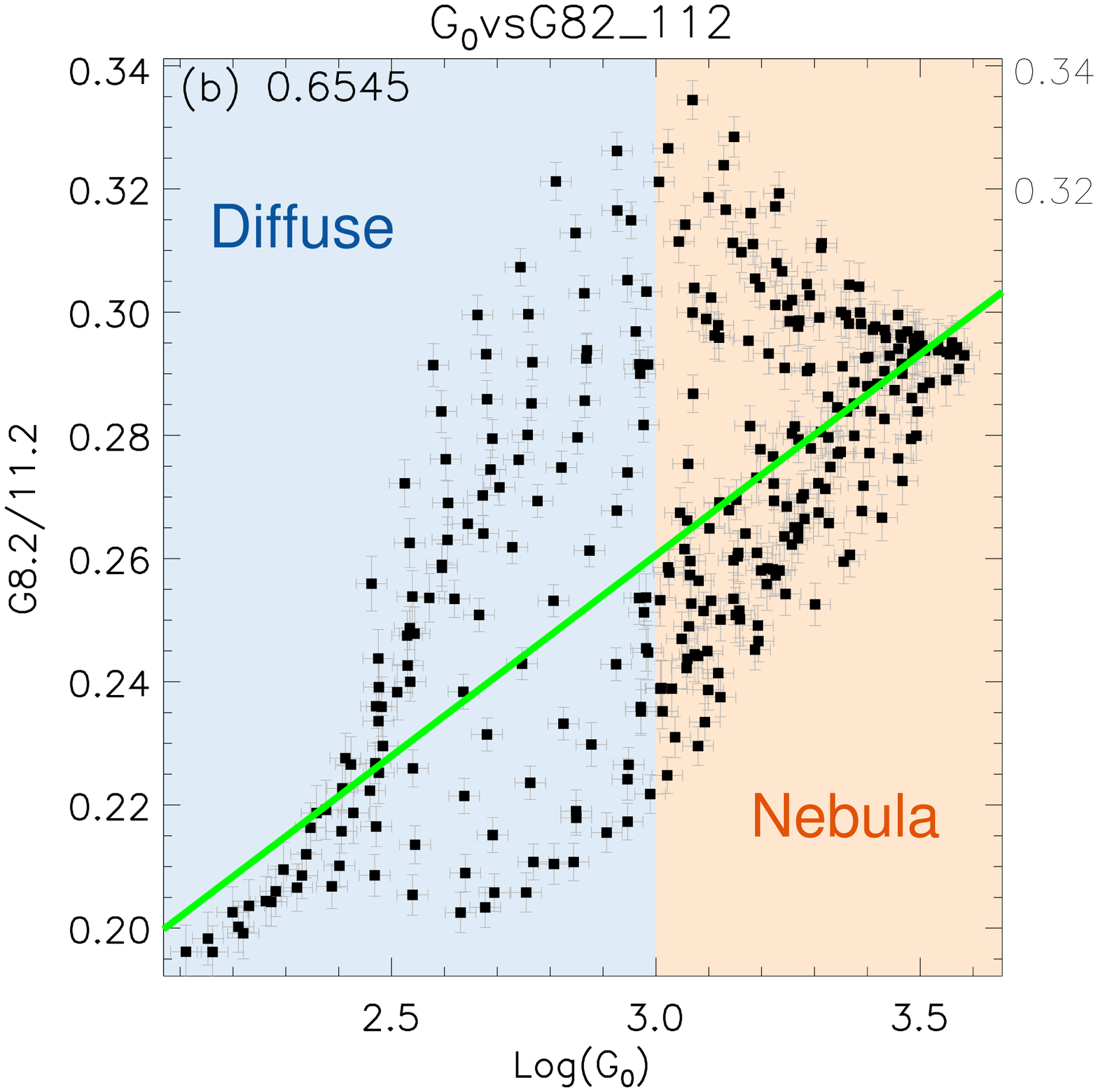}
\includegraphics[clip,trim =0.cm 0.3cm 1cm 1.2cm,width=3.5cm]{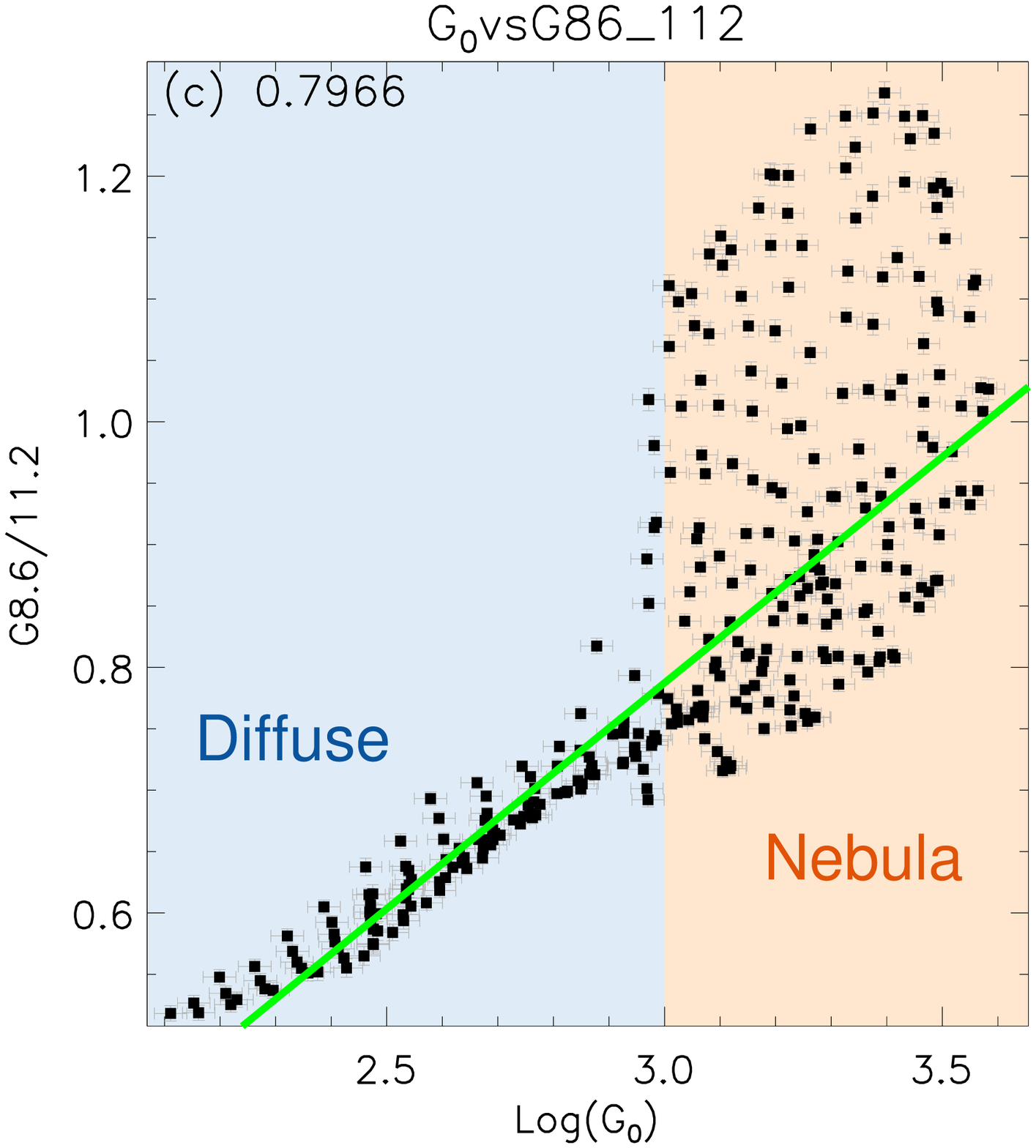}}
\end{center}
\caption{Correlation plots between the strength of the FUV radiation field, G$_{0}$, and PAH emission ratios in NGC~1333. Correlation coefficients are given in each panel. Linear fits between log(G$_{0}$) and each PAH ratio is shown by a green line. We divide each plot into two regimes based on the relative strength of the FUV radiation field: log(G$_{0}$) $\geq$ 3 in orange corresponds to the irradiated PDR and log(G$_{0}$) $\le$ 3 in blue corresponds to the diffuse ISM.}
\label{G0 corr2}
\end{figure*}

In this section, we present additional correlation plots between the FUV radiation field strength and other PAH ratios of interest (Figure \ref{G0 corr2}). Notably, in panels (a) and (b), we do not see an appreciable correlation between G$_{0}$ and G7.8/11.2 or G8.2/11.2 respectively. As the G7.8 and G8.2~components may arise from larger and/or neutral species \citep[i.e.][]{pee17}, both the G7.8/11.2 and G8.2/11.2 emission ratios are not expected to depend on the radiation field strength. Hence the relationship between these emission ratios and G$_{0}$ is more complex in comparison to traditional tracers of PAH ionization.
In panel (c), we find that G8.6/11.2 shows a relatively strong correlation with G$_{0}$, with a similar higher scatter within the nebula such as was found in Figure \ref{G0 corr} panels (a)--(d). This is not surprising since G7.6 and G8.6 have a very tight correlation (Figure \ref{irs_corr} (i)).

\section{Addressing the Scatter in PAH Diagnostics}
\label{PAH Scatter}

\begin{figure}
\begin{center}
\resizebox{\hsize}{!}{%
\includegraphics[clip,trim =0.cm 0.3cm 1cm 3.3cm,width=3.5cm]{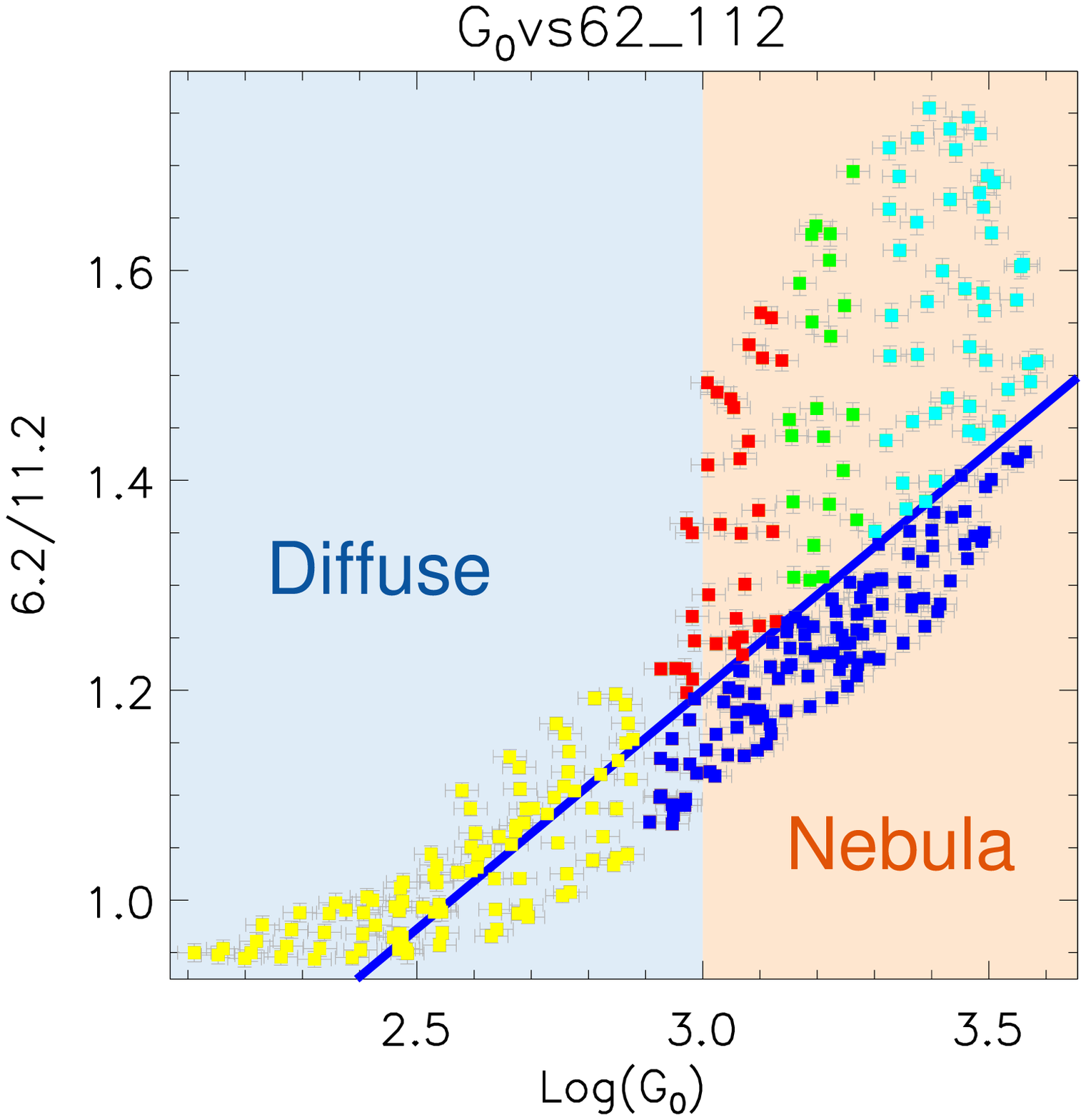}}
\end{center}
\caption{Correlation plot between the strength of the FUV radiation field, G$_{0}$, and the 6.2/11.2 PAH emission ratio in NGC~1333 showing five distinct groups of data--points. The line of best fit between log(G$_{0}$) and 6.2/11.2 is shown by a blue line. We divide each plot into two regimes based on the relative strength of the FUV radiation field: log(G$_{0}$) $\geq$ 3 in orange corresponds to the irradiated PDR and log(G$_{0}$) $\le$ 3 in blue corresponds to the diffuse ISM.}
\label{G0 corr3}
\end{figure}

\begin{figure}
\begin{center}
\includegraphics[clip,trim =0.5cm 0.cm 0.cm 1cm,width=8.4cm]{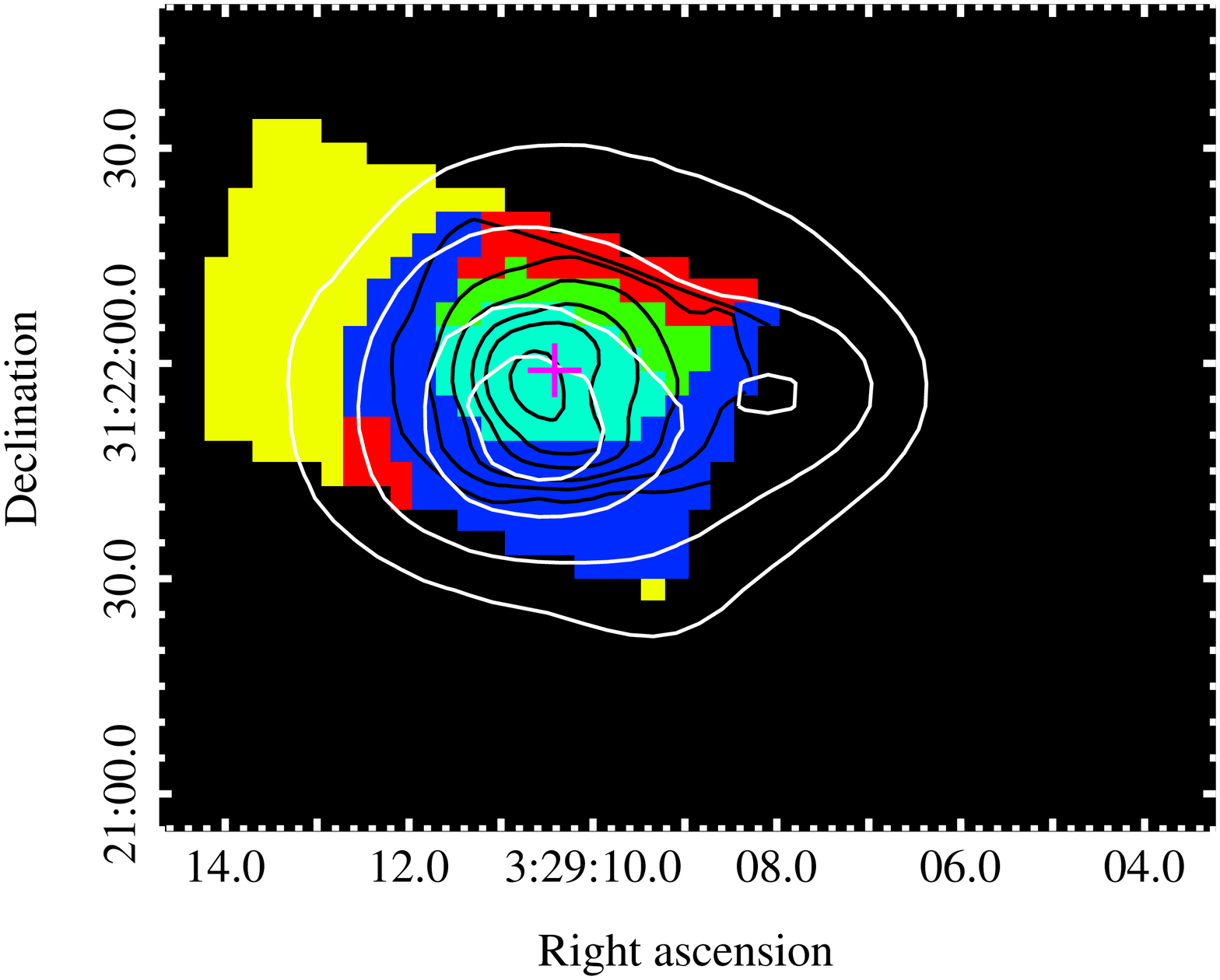}
\end{center}
\caption{Spatial location of the five different color-coded groups of data points selected in Figure~\ref{G0 corr3}. Contours shown in white are G0 (500, 1000, 2000, 3000) in units of of the Habing field and in black the PAH 7.7 \mum\ emission at the native IRS~SL resolution as in Figure~\ref{irs maps}. The position of SVS~3 is shown as a magenta cross. North is up and east is to the left. Pixels below the 3~$\sigma$ noise level are shown in black. Axes are given in right ascension and declination (J2000).} 
\label{loc G0vs62_112}
\end{figure}

In Section~\ref{PAHs and PDRs}, we find significant scatter in the relationships between the derived PDR physical conditions and relevant PAH emission ratios, particularly within the nebula (Figures~\ref{G0 corr} and \ref{Gamma corr}). To further illustrate this, we investigated the spatial location of various groups of data points in the scatter for the G0 vs 6.2/11.2 plot. As can be seen from the spatial location of the different-colored groups (Figure~\ref{G0 corr3}), data points located above the best-fit line in the correlation plot (see Figure~\ref{loc G0vs62_112} for color--coding) originate from positions near the center and northern edge of the nebula where the emission drops less steep than at southern edge (which coincides with increased extinction).

\begin{figure}
\begin{center}
\resizebox{\hsize}{!}{%
\includegraphics[clip,trim =0.cm 0.3cm 1cm 3.3cm,width=3.5cm]{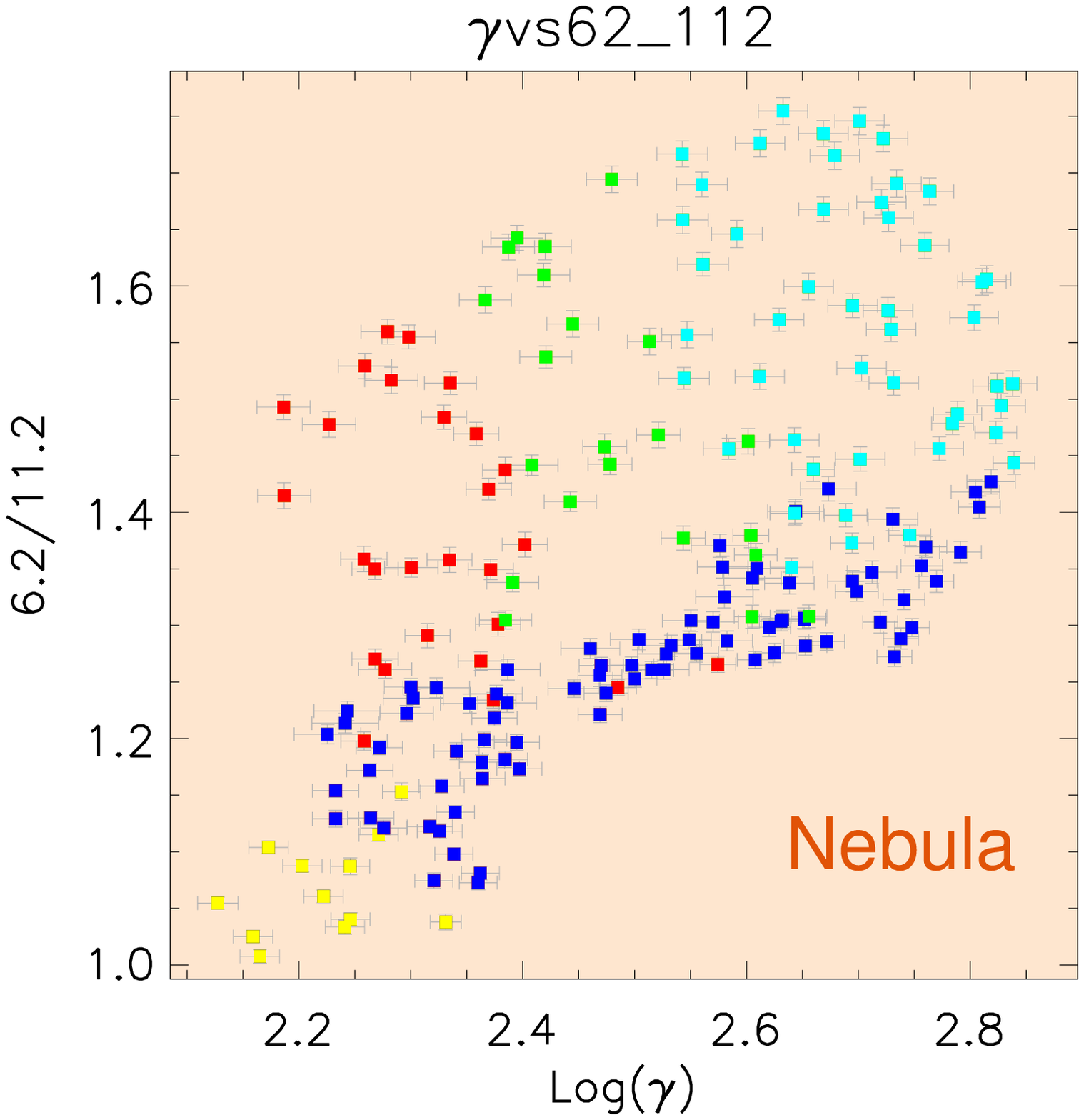}}

\end{center}
\caption{Relationship between the PAH ionization parameter, $\gamma$, and 6.2/11.2 emission ratio in NGC~1333. The color-coding shown here is the same as in Figures~\ref{G0 corr3} and ~\ref{loc G0vs62_112}.  }
\label{Gamma corr2}
\end{figure}

Using the color-coding of Figure~\ref{loc G0vs62_112}, we explore how these color-coded groups behave in the gamma vs 6.2/11.2 space (Figure~\ref{Gamma corr2}). Overall, we find the same trends between $\gamma$ vs 6.2/11.2 as we find for G$_{0}$ vs 6.2/11.2 albeit with higher scatter.

\bsp	
\label{lastpage}
\end{document}